\documentclass[12pt,prd,nofootinbib,superscriptaddress]{revtex4-1}
\usepackage[utf8]{inputenc}

\usepackage{xparse, physics, graphicx, dcolumn, bm, amsmath, amssymb, mathtools, cancel, dsfont, enumitem, braket, tikz, color}
\usepackage[com pat = 1.1.0]{tikz-feynman} 
\usepackage[linktocpage=true]{hyperref}
\usepackage[all]{hypcap}
\usepackage{cleveref}
\crefname{equation}{Eq.}{Eqs.}
\crefname{figure}{Fig.}{Figs.}
\crefname{tabular}{Tab.}{Tabs.}
\crefname{section}{Sec.}{Secs.}
\crefname{appendix}{App.}{Apps.}
\usepackage{color}
\usepackage{multirow}
\usepackage{siunitx}
\usepackage{mdframed}
\usepackage{rotating}
\usepackage{setspace}

\newcommand{\Lag}[1][]{\mathcal{L}_\text{#1}}
\newenvironment{feynd}{\begin{tikzpicture}[baseline=(current bounding box.center)] \begin{feynman}[small]}{\end{feynman} \end{tikzpicture}}
\newcommand{\dbar}{{\mathchar'26\mkern-12mu \dd}}
\newcommand{\iso}[1]{\mathcal{P}_{#1}}
\newcommand{\defeq}{\equiv}

\newcommand{\U}[1]{\textup{U}(#1)}

\newcommand{\cmt}[1]{}
\newcommand{\ii}{\mathrm{i}}
\newmdenv[backgroundcolor=gray!15,%
          skipabove=0pt,%
          skipbelow=5pt,%
          leftmargin=0pt,%
          rightmargin=0pt,%
          innertopmargin=-5pt,%
          innerbottommargin=7pt,%
          innerleftmargin=2pt,%
          innerrightmargin=2pt,%
          splittopskip=0pt,%
          splitbottomskip=0pt,%
          linewidth=0pt,%
          nobreak=true]%
          {keyeqn}
\renewcommand{\vev}{\ev}
\newcommand{\MPl}{M_\text{Pl}}
\definecolor{maroon}{rgb}{0.5,0,0}
\newcommand{\red}[1]{{\color{maroon} #1}}
\definecolor{deepblue}{rgb}{0.14,0.36,0.64}
\newcommand{\blue}[1]{{\color{deepblue} #1}}
\definecolor{brightyellow}{rgb}{0.87,0.55,0.073}
\newcommand{\yellow}[1]{{\color{brightyellow} #1}}
\newcommand{\W}{\mathrm{W}}

\newcommand{\beq}{\begin{equation}}
\newcommand{\eeq}{\end{equation}}
\newcommand{\bal}{\begin{align}}
\newcommand{\eal}{\end{align}}
\newcommand{\bit}{\begin{itemize}}
\newcommand{\eit}{\end{itemize}}
\newcommand{\ben}{\begin{enumerate}}
\newcommand{\een}{\end{enumerate}}
\newcommand{\f}{\frac}
\newcommand{\hb}[1]{\hat{\vb #1}}

\let\deltafunc\delta
\DeclareDocumentCommand\delta{}{\trigbraces{\deltafunc}}

\begin{document}
	
\title{Anatomy of Parity-violating Trispectra in Galaxy Surveys}

\author{Yunjia Bao}
\email{yunjia.bao@uchicago.edu}
\affiliation{Department of Physics, University of Chicago, Chicago, IL, 60637, USA}
\affiliation{Enrico Fermi Institute, University of Chicago, Chicago, IL, 60637, USA}
\affiliation{Kavli Institute for Cosmological Physics, University of Chicago, Chicago, IL 60637, USA}

\author{Lian-Tao Wang}
\email{liantaow@uchicago.edu}
\affiliation{Department of Physics, University of Chicago, Chicago, IL, 60637, USA}
\affiliation{Enrico Fermi Institute, University of Chicago, Chicago, IL, 60637, USA}
\affiliation{Kavli Institute for Cosmological Physics, University of Chicago, Chicago, IL 60637, USA}

\author{Zhong-Zhi Xianyu}
\email{zxianyu@tsinghua.edu.cn}
\affiliation{Department of Physics, Tsinghua University, Beijing 100084, China}
\affiliation{Peng Huanwu Center for Fundamental Theory, Hefei, Anhui 230026, China}

\author{Yi-Ming Zhong}
\email{yiming.zhong@cityu.edu.hk}
\affiliation{Department of Physics, City University of Hong Kong, Kowloon, Hong Kong SAR, China}

\begin{abstract}
	Parity-violating interactions are ubiquitous phenomena in particle physics. If they are significant during cosmic inflation, they can leave imprints on primordial perturbations and be observed in correlation functions of galaxy surveys. Importantly, parity-violating signals in the four-point correlation functions (4PCFs) cannot be generated by Einstein gravity in the late universe on large scales, making them unique and powerful probes of high-energy physics during inflation. However, the complex structure of the 4PCF poses challenges in diagnosing the underlying properties of parity-violating interactions from observational data. In this work, we introduce a general framework that provides a streamlined pipeline directly from a particle model in inflation to galaxy 4PCFs in position space. We demonstrate this framework with a series of toy models, effective-field-theory-like models, and full models featuring tree-level exchange-type processes with chemical-potential-induced parity violation. We further showed the detection sensitivity of these models from BOSS data and highlighted potential challenges in data interpretation and model prediction.
\end{abstract}

\maketitle
\begingroup
\setstretch{0.9}
\tableofcontents
\endgroup

\section{Introduction}
\label{sec:intro}
Parity violation plays an important role in fundamental physics. Since Wu \textit{et al.}'s experiment in the 1950s \cite{Wu:1957my, Garwin:1957hc}, we have learned that parity is violated in the electroweak sector \cite{Lee:1956qn}. Further, the observed baryon asymmetry of the universe strongly suggests some parity-violating process in the early universe \cite{Sakharov:1967dj} that is beyond the Standard Model (SM) of particle physics. While many parity-violating electroweak processes have been well measured and extensively studied, the parity-violating processes, such as those producing the baryon asymmetry, may have happened at a much higher energy scale, which is hard to probe directly. In this regard,  cosmological observations can provide valuable insights into parity violations in the early universe. 

Parity-violating signatures in cosmology can be generated in certain inflationary scenarios. Since the typical energy scale of inflation, the inflationary Hubble parameter, $H$, is probably much higher%
\footnote{Current observation only obtains an upper bound on the inflationary Hubble scale due to the absence of an observed primordial tensor perturbation. The current bound finds $H \lesssim 4.5\times 10^{13} \;\text{GeV}$ from BICEP-Keck 2018 Observation \cite{BICEP:2021xfz}. Future tensor-to-scalar ratio measurement with Simons Observatory will be able to reduce this limit by another factor of $2$ \cite{SimonsObservatory:2018koc}.}
than that has been probed by collider experiments, the inflaton field that drives the cosmic inflation may couple to other heavy fields through beyond-SM interactions, which can distort its almost Gaussian distribution. Decoding the properties of these heavy particles and their interactions from inflaton correlation functions (such as bispectrum and trispectrum)  is known as ``cosmological collider'' physics \cite{Chen:2009we, Chen:2009zp, Arkani-Hamed:2015bza}. Roughly, if the correlation function of the inflaton involves the exchange of a particle of mass $m$, this correlation function in momentum space (a.k.a., the spectrum) is expected to scale as $\sim \exp(-m/H) \sin( m/H \ln(k/k_r) + \phi_r)$ in a soft limit, in which $k$ denotes the comoving momentum that is reaching the soft limit $k\to 0$, $k_r$ denotes the reference momentum, and $\phi_r$ denotes the reference phase offset \cite{Wang:2021qez,Qin:2022lva}. Thus, identifying the oscillation frequency in log-$k$ space of the inflaton spectra determines the mass of the exchanged particle in the unit of $H$, and much more information about the particle, such as its spin and interactions, can be extracted by measuring other parameters, such as the angular dependences and the phase shift $\phi_r$. 

However, this schematic formula also shows a conundrum of cosmological collider physics. On the one hand, a large mass $m \gg H$ leads to a more conspicuous oscillation in the inflaton spectrum but is severely suppressed due to the Boltzmann suppression factor, $\exp(-m/H)$. On the other hand, a small mass $m \ll H$ results in a spectrum with a large amplitude but a very low-frequency oscillation in log-$k$ space, demanding ambitious observation efforts across many orders of magnitude of momentum scale. While calling for better observation techniques, this conundrum is solved in several theoretically well-motivated mechanisms, such as the introduction of a chemical potential $\mu$ that enhances the spectrum by $\sim \exp(\mu/H)$ \cite{Chen:2018xck, Wang:2019gbi, Wang:2020ioa, Bodas:2020yho}. Interestingly, many known chemical potential models are helical and naturally possess parity-violating interactions. Therefore, this class of models could be promising targets for probing both heavy-particle oscillations and parity violations in the universe.

Parity violations may show up in various observables. With primordial tensor modes, parity violation can imprint two-point and three-point correlators. (See, e.g., \cite{Lue:1998mq,Alexander:2004us,Tong:2022cdz}). However, if we only focus on scalar mode, the parity violation appears first at the four-point level, which is easy to understand: Without breaking the spatial rotation symmetries, a parity-violating scalar observable is constructed by contracting the Levi-Civita (antisymmetric) tensor $\epsilon_{ijk}$ with three vectors. This contraction is nonvanishing only when the three vectors are not coplanar, which requires at least four points. In this study, we only consider correlation functions of galaxy number overdensities, i.e., a correlation function of scalar fluctuations, and thus focus on the four-point correlation function (4PCF) in position space.

While the 4-point scattering amplitude is a familiar concept in Minkowski spacetime, the 4PCF in cosmology presents additional complexity due to the effects of the cosmic background. A relativistically invariant four-point scattering amplitude depends on only two free parameters. This simplicity arises because it is defined by four on-shell 3-momentum vectors, which initially have 12 degrees of freedom. Energy and momentum conservation, along with three boosts and three rotations, eliminate 10 redundant degrees of freedom, leaving two independent parameters.  In contrast,  the 4PCF in cosmology generally requires {six} free parameters to fully specify. This difference arises because the cosmic expansion and inflaton rolling break the Lorentz invariance,  leaving only three spatial translations and three rotations to reduce the parameter space. Consequently, the 4PCF retains $12 - 6 = 6$ free parameters. One can further reduce this number to {five} in the presence of (approximate) scale invariance.

Although the momentum-space trispectrum is natural to compute theoretically, direct inference of this quantity is hampered by linear growth, survey geometries, and window-function effects \cite{Feldman:1993ky, Tegmark:1997yq, Bond:1998zw, Yamamoto:2005dz, Smith:2015uia, Slepian:2025fdx}. While such a procedure for deconvolving the window effect has been developed for the spectrum \cite{Oh:1998sr, Hamilton:2005kz, Hamilton:2005ma, Hand:2017irw, Bianchi:2015oia, Philcox:2020vbm} and bispectrum \cite{Scoccimarro:2000sp, Smith:2006ud, Gil-Marin:2014sta, Slepian:2015hca, Philcox:2021ukg}, its implementation for higher-point statistics remains challenging. On the other hand, position-space higher-point correlation functions can be measured directly \cite{Guo:2014nka, Slepian:2016kfz, Philcox:2021hbm, Hou:2022wfj, Philcox:2022hkh}, albeit computationally challenging to handle \cite{Philcox:2021bwo}. However, these observables are difficult to compare with theories. This study bridges the two aspects by translating the momentum-space trispectra into position-space 4PCFs, while taking into account linear structure growth and galaxy clustering effects. 

From the observational perspective, focusing on the scalar 4PCF, such as the 4PCF of galaxy number overdensities, is timely and well-motivated. We are entering an era with abundant galaxy survey data from programs such as DESI~\cite{DESI:2016fyo}, SPHEREx~\cite{SPHEREx:2014bgr}, Rubin~\cite{LSST:2008ijt}, Roman~\cite{Spergel:2015sza}, DESI-II~\cite{DESI:2022lza}, Spec-S5~\cite{Sailer:2021yzm, DESI:2022lza, Schlegel:2022vrv}.  Notably, Refs.~\cite{Hou:2022wfj} and \cite{Philcox:2022hkh} recently reported the discovery of parity violation in the 4PCF of galaxy number overdensities using BOSS data, a phenomenon unlikely to be caused by gravitational or baryonic processes on small scales.%
\footnote{See, however, Refs.~\cite{Krolewski:2024paz, Philcox:2024mmz} for a potential mismatch bias in the mock-based model-independent inference of parity violation from BOSS.}

Since these initial findings, the community has made significant progress in understanding parity violations in the large-scale structure (LSS). 
On the theoretical side, an extension to the no-go theorem in Ref.~\cite{Liu:2019fag} was proposed, along with several exemplary models that produce parity-violating trispectra~\cite{Cabass:2022rhr}. Using the wavefunction formalism, new factorization properties were identified to simplify the computation of parity-violating trispectra  \cite{Stefanyszyn:2023qov, Stefanyszyn:2024msm,Lee:2023jby}, and a more formal perspective was proposed to extend the no-go theorem beyond tree level using a de Sitter version of the CPT theorem \cite{Hewson:2024rnb, Thavanesan:2025kyc}. A similar but slightly different model with an axion in a sinusoidal potential that couples to a \U{1} gauge field and the inflaton was also studied, and the study suggests a larger parity-violating signal relative to the parity-preserving signal \cite{Fujita:2023inz}. Other studies investigated parity-violating trispectra of axion inflation \cite{Niu:2022fki, Reinhard:2024evr} and in Chern-Simons gravity \cite{Creque-Sarbinowski:2023wmb}. Recent studies also investigated the possibility that redshift-space distortion may introduce an apparent parity violation in the galaxy redshift survey data \cite{Paul:2024uim, Paul:2024xff}. On the observational side, a procedure was proposed to compress the six-dimensional parity-violating part of the trispectrum to one-dimensional power-spectrum-like functions \cite{Jamieson:2024mau}. Constraints from the cosmic microwave background (CMB) showed no significant parity violation with a model-independent limit \cite{Philcox:2023ffy, Philcox:2023ypl} and with model template fitting \cite{Philcox:2025wts}. There has also been interest in the role of parity-violating correlations for baryon acoustic oscillation observations~\cite{Hou:2024udn}, as well as smaller-scale constraints using $N$-body simulations~\cite{Coulton:2023oug} or machine-learning-based techniques ~\cite{Taylor:2023deh, Craigie:2024bhk, Hewson:2024rnb}.

An important open question is which model best accounts for the parity-violating signals observed by BOSS and what constraints can be derived from the data. Some of these were investigated in Refs.~\cite{Philcox:2022hkh, Cabass:2022oap, Cabass:2024wob}. However, extracting information from the primordial (parity-violating) process from the late-time observations remains challenging. Theoretical developments in cosmological collider physics mainly focused on shapes of curvature perturbation correlators in momentum space, whereas galaxy surveys document galaxy number overdensities in position space. Certain limits in momentum space (such as squeezed or collapsed limits) are frequently taken when discussing momentum-space spectra for clarity. However, all momentum modes contribute to the position-space correlation functions, and it is unclear whether spectral features under those limits are manifested in the position space. Furthermore, while curvature perturbations provide a qualitative picture of galaxy overdensities, they miss essential details, such as transfer functions linking the initial curvature perturbations to the matter overdensities,  galaxy bias that associates matter with galaxy distributions,  and projection effects such as redshift space distortion (see review, e.g.,~\cite{Desjacques:2016bnm, Hamilton:1997zq}).

The no-go theorems found in previous works \cite{Liu:2019fag, Cabass:2022rhr} show that the parity-violating 4PCF of curvature perturbations cannot be generated by local contact diagrams in inflation under a few very mild assumptions. These no-go theorems thus point to exchange-type diagrams for realistic model building of parity-violating signals. In other words, \emph{the parity violation in 4PCF is naturally associated with nonlocal propagation effects}. This fact has important consequences. On the one hand, a propagating degree typically carries some spatial angular momentum, so that the corresponding parity-violating signal typically spreads over a range of azimuthal quantum number $\ell$. This is in contrast with the parity-violating signal from a diagram from contact interactions that are local, where the signal is typically peaked at a fixed $\ell$. We will see these structures explicitly from the results below. On the other hand, the presence of intermediate particles necessarily complicates the computation, due to the increased layers of time integrals and the non-factorizable integrand when performing the Fourier transform.

Thus, the principal challenge in translating from momentum space to position space lies in the high dimensionality of the Fourier transform, the highly oscillatory momentum spectra such as the $e^{\ii k \tau}$ terms in the inflaton propagator, the implementation of special functions appearing in the integrand, etc. A key strategy for taming this complexity is to exploit the \emph{factorizability} of the trispectra. We show that for a tree-level exchange-type cosmological collider process, the Schwinger-Keldysh formalism \cite{Chen:2017ryl} naturally brings out this factorizability. Building on these ideas, we develop a complete pipeline from a generic momentum-space trispectrum to position-space 4PCF of galaxy survey, with careful attention to interface the numerical routine that maximizes factorizability. Using this pipeline, we compute position-space templates for several trispectra, focusing on those generated by the parity-violating chemical potential models \cite{Chen:2018xck, Wang:2019gbi, Wang:2020ioa, Tong:2022cdz}. We then identify qualitative features in the angular and radial distributions of the resulting 4PCF that can be tried back to their momentum-space counterparts. We then compare their 4PCF templates with BOSS data without restricting them to particular soft limits, such as the squeezed or collapsed limit.

The paper is organized as follows. In \cref{sec:theoreticalModels}, we discuss general features of parity-violating trispectrum. \cref{sec:canonicalModels} surveys canonical trispectrum models, and \cref{sec:largeMassModels} introduces a new contact-interaction-like trispectrum from an EFT perspective.  In \cref{sec:fullModels}, we go beyond contact interactions by presenting two trispectra without taking the EFT limit. Translating these trispectra to position space 4PCF is challenging, \cref{sec:4PCFGeneralStrategy} outlines a numerically feasible strategy that applies broadly to exchange-type trispectra. Results of this study are presented in \cref{sec:results}. \cref{sec:templateFeatures} discusses the feature of the numerically computed position-space templates and \cref{sec:compareWithBOSS} compares them with BOSS data. 
We conclude in \cref{sec:conclusion}. The appendices provide further details on the 4PCF computation (\cref{app:detail4PCF}), important numerical technicalities (\cref{app:numTechnicalities}), the factorizable trispectrum for spin-2 exchange under the large mass limit (\cref{app:spin2LargeMass}), and the full 4PCF templates of several models (\cref{app:fullresults}). 

\section{Parity-violating Trispectra: Toy Shapes and Full Models \label{sec:theoreticalModels}}
We now move on to providing a general characterization of parity-violating momentum trispectra of primordial curvature fluctuation $\zeta$. We take a generic inflation model, where the curvature fluctuation $\zeta$ is generated by a nearly massless inflaton fluctuation $\phi$. The two fluctuations are related by 
\beq
\zeta=-\frac{H}{\dot\phi_0}\phi
\eeq
at the Gaussian level, where $\dot\phi_0$ is the rolling speed of the inflaton background. In this setup, the primordial trispectra are essentially 4PCF of inflaton fluctuations $\ev{\phi^4}$. When computing the 4PCF, we normally work in the 3-momentum space. Then, a parity-violating trispectrum, or the parity-violating part of a trispectrum, corresponds to the imaginary part of a trispectrum:  
\begin{equation}
    \text{Parity-violating part}\{ \ev{\phi_{\vb{k}_1}\phi_{\vb{k}_2}\phi_{\vb{k}_3}\phi_{\vb{k}_4}}' \} = \ii \Im\{ \ev{\phi_{\vb{k}_1}\phi_{\vb{k}_2}\phi_{\vb{k}_3}\phi_{\vb{k}_4}}' \}, 
\end{equation}
due to the nature of the Fourier transform. 

In this work, we always assume a weakly interacting theory during inflation so that we can evaluate trispectra with diagrammatic expansion using the Schwinger-Keldysh formalism \cite{Chen:2017ryl}. It has been shown in Refs.~\cite{Liu:2019fag,Cabass:2022rhr} that tree diagrams from contact interactions of inflaton fluctuations do not produce any parity-violating trispectrum under certain mild assumptions. Thus, the simplest nontrivial example is a trispectrum mediated by a single bulk field at the tree level. For such a process, the trispectrum can be written as: 
\begin{equation}
\label{eq_phi4toJK}
    \begin{multlined}
        \ev{\phi_{\vb{k}_1}\phi_{\vb{k}_2}\phi_{\vb{k}_3}\phi_{\vb{k}_4}}' 
        = \begin{feynd}
				\vertex (i) at (-1,0);
				\vertex (f) at (2,0);
				\node[square dot, fill = none] (i1) at (-2,-1);
				\node[square dot, fill = none] (i2) at (-2,1);
				\node[square dot, fill = none] (f1) at (3,-1);
				\node[square dot, fill = none] (f2) at (3,1);
				\diagram*{
					(i) -- [scalar, momentum' = \(\vb{k}_{1,i}\)] (i2);
					(i1) -- [scalar, reversed momentum' = \(\vb{k}_2\)] (i);
					(f) -- [scalar, momentum = \(\vb{k}_{3,j}\)] (f2);
					(f) -- [scalar, momentum' = \(\vb{k}_4\)] (f1);
					(i) -- [photon, edge label = \( \Pi_{ij} \supset \mp \ii \epsilon_{ijl} {\hat{k}_s}^l \), momentum' = \( \vb{k}_s \)] (f);
				};
			\end{feynd} \\
    = \mathcal{J}(k_1, \ldots, k_4, k_s) \mathcal{K}(\hat{\vb{k}}_1, \ldots, \hat{\vb{k}}_4, \hat{\vb{k}}_s) + \text{perms},
    \end{multlined}
\end{equation}
in which $\vb{k}_1, \ldots, \vb{k}_4$ denote external momenta, $k_i\equiv |\vb{k}_i|$, $\hb k_i\equiv \vb{k}_i/k_i$, and $\vb{k}_s=\vb{k}_1+\vb{k}_2$ is the $s$-channel momentum, and represents the momentum of the exchanged particle. The permutation is over channels, namely $\vb{k}_s\to \vb{k}_t\equiv \vb{k}_1+\vb{k}_4$  and $\vb{k}_s\to \vb{k}_u\equiv\vb{k}_1+\vb{k}_3$. Here, we separate the angular dependence to $\mathcal{K}$ (also called the kinematic part of the trispectrum) from the radial dependence $\mathcal{J}$ (called the dynamical part of the trispectrum) for later convenience. 

Let us first look at the radial-dependent factor $\mathcal{J}(k_1, \ldots, k_4, k_s)$. Eventually, we will express everything in position space, which means that we need to perform the inverse Fourier transform to the momentum-space 4PCF. Ideally, if one can express the radial factor $\mathcal{J}(k_1, \ldots, k_4,k_s)$ in the Fourier integrand as a factorized form, $\mathcal{J} = \prod_{i=1}^{4} \mathcal{J}_i(k_i)$, then the computation is efficient since a factorized integrand essentially reduces a higher-dimensional integral to a product of lower-dimensional integrals \cite{Regan:2010cn, Funakoshi:2012ms, Smith:2015uia, Lee:2020ebj}. This, unfortunately, is not always guaranteed. However, an interesting feature of the Schwinger-Keldysh formalism is that one may partition the computation into subdiagrams with each interaction vertex having a temporal integral. For instance, for the single exchange in the $s$-channel, one can partition the integral into 

\begin{equation}
    \begin{aligned}
    &\ev{\phi_{\vb{k}_1}\phi_{\vb{k}_2}\phi_{\vb{k}_3}\phi_{\vb{k}_4}}_s' \sim \int \dd \tau_L \dd \tau_R\; 
        \begin{feynd}
            \vertex (v1) at (0,0); \node at (0,0) [yshift = 10] {\( \tau_L \)}; 
            \vertex (v2) at (1,0); \node at (1,0) [yshift = 10] {\( \tau_R \)};
            \node[square dot, fill = white] (i1) at (-0.71,0.71);
            \node[square dot, fill = white] (i2) at (-0.71,-0.71);
            \node[square dot, fill = white] (f1) at (1.71,0.71);
            \node[square dot, fill = white] (f2) at (1.71,-0.71);
            \diagram{
                {(i1), (i2)} -- [very thick, color=maroon] (v1) -- [photon, very thick, color=brightyellow, momentum' = {[arrow style=brightyellow]\(\vb{k}_s\)}] (v2) -- [very thick, color=deepblue] {(f1), (f2)};
            };
        \end{feynd} \\
        \supset& \int \dd \tau_L \dd \tau_R\; \red{\mathcal{J}_L(k_1, k_2, k_s, \tau_L)} \yellow{\Pi(k_s, \tau_L, \tau_R)} \blue{\mathcal{J}_R(k_3, k_4,k_s, \tau_R)}.
	   \label{eqn:dynamical}
    \end{aligned}
\end{equation}
This ensures that the momentum-space trispectrum is at least partially factorizable at the cost of introducing some time integral. In other words, the integrand can be factorized into three parts: (1) $\mathcal{J}_L$ that only depends on $k_1$, $k_2$, and $k_s$, (2) $\mathcal{J}_R$ only on $k_3$, $k_4$, and $k_s$, and (3) $\Pi$ only on $k_s$.

The vertex term $\mathcal{J}_{L/R}$ usually involves complex exponentials or Hankel functions, and the propagator term $\Pi$ usually contains products of Hankel functions or Whittaker functions and Heaviside step functions. Then, analytically studying these nested temporal integrals tends to be challenging as the temporal integral creates complicated special functions in $\{k_i\}$. Considerable progress has been made in this direction \cite{Arkani-Hamed:2018kmz,Baumann:2019oyu,Qin:2022fbv,Qin:2023ejc,Xianyu:2023ytd,Fan:2024iek,Liu:2024xyi}, but the resulting expression for $\mathcal{J}(\{k_i\})$ is usually complicated. Consequently, one has to perform a high-dimensional Fourier transform that tends to be prohibitively expensive computationally. However, if the time integral is not performed analytically first and is delayed after a numerical integral of $\{k_i\}$ is performed, the computation load is significantly reduced as the $k_3$ and $k_4$ integral on $\mathcal{J}_L(k_1, k_2, k_s)$ is trivial. This achieves a partially factorizable trispectrum. Therefore, in the following subsections, we will report the radial-dependent factor of the trispectrum without performing the time integral in anticipation of performing such integral numerically. 

Next we consider the angular part $\mathcal{K}(\hat{\vb{k}}_1, \ldots, \hat{\vb{k}}_4, \hat{\vb{k}}_s)$ in \cref{eq_phi4toJK}. Since we are considering the trispectrum of a scalar field, $\mathcal{K}(\{ \hat{\vb{k}}_i \})$ must also be a scalar, in the sense that it is invariant under a co-rotation of all momenta. Also, the parity-violating part of $\mathcal{K}(\{ \hat{\vb{k}}_i \})$ must be a pseudoscalar, and contains a factor of Levi-Civita symbol $\epsilon_{ijk}$. To utilize the properties of $\mathcal{K}(\{ \hat{\vb{k}}_i \})$ under rotations and space inversion, it is convenient to expand it in terms of a set of basis functions, known as $N$-point isotropic basis functions \cite{Cahn:2020axu}. For concreteness, we show explicit formulae of 3-point and 4-point isotropic basis functions here:
\begin{gather}
	\iso{\ell_1, \ell_2, \ell_3}(\hat{\vb{k}}_1,\hat{\vb{k}}_2,\hat{\vb{k}}_3) 
	\defeq \!\!\! \sum_{m_1, m_2, m_3} \mqty(\ell_1 & \ell_2 & \ell_3 \\ m_1 & m_2 & m_3) Y_{\ell_1}^{m_1*}(\hat{\vb{k}}_1) Y_{\ell_2}^{m_2*}(\hat{\vb{k}}_2) Y_{\ell_3}^{m_3*}(\hat{\vb{k}}_3),
 \label{eq:3pt}\\
 \begin{aligned}
	\iso{\ell_1, \ell_2, (\ell'), \ell_3, \ell_4}(\hat{\vb{k}}_1,\hat{\vb{k}}_2,\hat{\vb{k}}_3, \hat{\vb{k}}_4) 
	\defeq &\!\!\!\!\!\! \sum_{m_1, m_2, m_3, m_4}\!\!\! \!\!\!  \sqrt{2\ell'+1} \sum_{m'} (-1)^{\ell'-m'}\mqty(\ell_1 & \ell_2 & \ell' \\ m_1 & m_2 & -m')\mqty(\ell_3 & \ell_4 & \ell' \\ m_3 & m_4 & m')\\
    & \times Y_{\ell_1}^{m_1*}(\hat{\vb{k}}_1) Y_{\ell_2}^{m_2*}(\hat{\vb{k}}_2) Y_{\ell_3}^{m_3*}(\hat{\vb{k}}_3)Y_{\ell_4}^{m_4*}(\hat{\vb{k}}_4),
    \end{aligned}
 \label{eq:4pt}
\end{gather}
in which big parentheses denote Wigner's 3$j$ symbols, $Y_{\ell}^{m}(\hat{\vb{k}})$ represents spherical harmonics, and $*$ means complex conjugation. Note that, among the subscripts of $\mathcal{P}_{\ell_1, \ell_2, (\ell'), \ell_3, \ell_4}$, $\ell_1,\cdots,\ell_4$ outside the parentheses label external angular momenta (named primaries in Ref.~\cite{Cahn:2020axu}), while $(\ell')$ denotes the intermediate angular momentum label (named intermediates in Ref.~\cite{Cahn:2020axu}) that is subject to the triangle inequality $|\ell_1 - \ell_2| \leq \ell' \leq \ell_1+\ell_2$ as well as $|\ell_3-\ell_4|\leq \ell'\leq \ell_3+\ell_4$. Usually, $\mathcal{K}(\{\hat{\vb{k}}_i\})$ comes from tensor structures expressed in the Cartesian basis, but the conversion to the spherical basis is straightforward. Some of these conversions are given in Appendix A of Ref.~\cite{Cahn:2020axu}. As examples, 
\begin{align}
    \iso{111}(\hb k_1, \hb k_3, \hb k_s)={}&-\f{3\ii}{ 2^{1/2}(4\pi)^{3/2}} (\hb k_1 \times \hb k_3)\cdot \hb k_s = -\frac{\sqrt{4\pi} k_2}{k_s} \iso{11(1)10}(\hat{\vb{k}}_1, \hat{\vb{k}}_2, \hat{\vb{k}}_3, \hat{\vb{k}}_4),\label{eq:P111}\\
    \iso{212}(\hb k_1, \hb k_3, \hb k_s) ={}& - \frac{3\sqrt{5}\ii}{2^{1/2}(4\pi)^{3/2}} \hb k_1 \cdot (\hb k_3 \times \hb k_s) (\hb k_1\cdot \hb k_s),\\
\iso{333}(\hb k_1, \hb k_3, \hb k_s) ={}& \frac{175\ii}{2\sqrt{6}(4\pi)^{3/2}} \hb k_1 \cdot (\hb k_3 \times \hb k_s) \left[\frac{2}{25}+(\hb k_1 \cdot \hb k_3)(\hb k_1 \cdot \hb k_s)(\hb k_3 \cdot \hb k_s) \right. \nonumber\\
&\left.  -\frac{1}{5}\left((\hb k_1 \cdot \hb k_3)^2+(\hb k_1 \cdot \hb k_s)^2 + (\hb k_3 \cdot \hb k_s)^2\right)\right],
\end{align}
where we have applied $\hb k_s = (\vb k_1 + \vb k_2)/k_s$ and $\iso{\ell_1, \ell_2(\ell_3)\ell_3,0} = \iso{\ell_1, \ell_2, \ell_3}/\sqrt{4\pi}$ given $Y_{0}^0 = {1}/{\sqrt{4\pi}}$ in the derivation of~\cref{eq:P111}.

We will express $\mathcal{K}(\{\hat{\vb{k}}_i\})$ in the isotropic basis functions, in anticipation of using identities or relations of spherical harmonics to simplify some of the angular integrations when performing the Fourier transform. A useful property for the upcoming discussion is that a parity-violating isotropic function should have an odd total angular momentum label. To be sepecific, for 3-point isotropic function $\iso{\ell_1, \ell_2, \ell_3}$, this means that $\ell_1 + \ell_2 + \ell_3$ is an odd number. For 4-point isotropic function $\iso{\ell_1, \ell_2, (\ell'), \ell_3, \ell_4}$, this means that $\ell_1 + \ell_2 + \ell_3 + \ell_4$ is an odd number. 

\subsection{Canonical Toy Shapes: Local and Equilateral \label{sec:canonicalModels}}
First, we survey some toy models of trispectra that are frequently used when analyzing non-Gaussianities and higher-point statistics in cosmology. These toy shapes take simple analytical forms, thus easier to evaluate numerically as well. There are usually two possible spectral shapes, local or equilateral, for the trispectrum. We start with toy models constructed from the local trispectrum. One of the local trispectra is \cite{Lee:2020ebj}:
\begin{align}
\vev{\zeta^4}'_{\tau_\text{NL}} =&~ \tau_\text{NL} P_\zeta(k_1) P_\zeta(k_2) P_\zeta (k_s) + \text{11 perms},
\end{align}
in which $P_\zeta(k)$ denotes the curvature power spectrum. Here, we report the curvature trispectrum 
\beq
\ev{\zeta^4}' = \qty(\frac{H}{\dot{\phi}_0})^4 \ev{\phi^4}'
\eeq
as it is more convenient to work with this dimensionless quantity and convert it to the galaxy statistics in the next section. The local shape may arise from introducing additional nonlinear local field redefinition of inflaton fluctuation on the late-time boundary, i.e., $\phi(x) \to \phi(x) + \sqrt{3\tau_\text{NL}/10} (\phi^2(x) - \ev{\phi^2(x)}) + \ldots$, and hence the name of local shape. Due to the smallness of measured large-scale non-Gaussianity, one may consider the nonlinear term in the field redefinition as small perturbations on top of the Gaussian inflaton fluctuation. In momentum space, this shape peaks in the squeezed limit (e.g., $k_1 \ll k_2 \simeq k_3 \simeq k_4$) as one of the external momenta is significantly smaller than the rest. This shape is parity-preserving as the field redefinition is parity invariant. To construct a parity-violating local-shape-like trispectrum, we multiply $\vev{\zeta^4}'_{\tau_\text{NL}}$ by a parity-violating isotropic function, e.g., $\iso{111}(\hat{\vb{k}}_1, \hat{\vb{k}}_3, \hat{\vb{k}}_s)$. The resulting trispectrum is given by, e.g.,
\beq
\tau_\text{NL}^{-1} \vev{\zeta^4}'_\text{toy,local} = P_\zeta(k_1) P_\zeta(k_3) P_\zeta (k_s) \mathcal P_{111} (\hb k_1, \hb k_3, \hb k_s) + \text{23 perms}.
\label{eq:local-1}
\eeq
We note that this is an \emph{ad hoc} shape which serves as a toy example, and is not from a model with a concrete bulk Lagrangian. Here, we factored out an overall magnitude $\mu_{A,\text{ Local}} \defeq \tau_\text{NL}$ so that we may focus on the shape of the trispectrum.

Next, we consider the equilateral shape~\cite{Smith:2015uia, Lee:2020ebj}:
\begin{align}
	\vev{\zeta^4}'_{g_\text{NL}^{\text{eq},1}} ={}& \f{221184}{25} g_\text{NL}^{\text{eq},1}  \f{1}{k_1k_2k_3k_4 k_{1234}^5},
	\label{eq:toy_model}
\end{align}
where $k_{1234} = k_1+k_2+k_3+k_4$. The equilateral shape is obtained by considering derivatively-coupled higher-point inflaton contact interaction, such as $\phi'^4/\Lambda^4$. Due to the flatness of the slow-roll inflaton potential, inflaton and its fluctuation may enjoy a shift symmetry, which motivates these derivatively-coupled terms in the Lagrangian. In momentum space, this spectrum peaks in the equilateral limit ($k_1 \simeq k_2 \simeq k_3 \simeq k_4$) when all external momenta are of comparable scale. This shape is still mediated by a contact interaction, hence parity-preserving. Similar to the local-shape-like toy shapes, we construct parity-violating equilateral-shape-like toy shapes by multiplying the equilateral shape by a parity-violating isotropic function. The resulting trispectrum is given by, e.g.,
\begin{align}
\qty(\frac{9216}{25}g_\text{NL}^{\text{eq},1})^{-1} \vev{\zeta^4}'_\text{toy,eq} ={}& \f{24}{k_1k_2k_3k_4 k_{1234}^5} \mathcal P_{111} (\hb k_1, \hb k_3, \hb k_s) + \text{23 perms} \nonumber\\
={}& \int^{\infty}_0 dt_E \; t_E^4 \left(\prod_{i=1}^4 \frac{e^{-k_i t_E}}{k_i}\right) \mathcal P_{111} (\hb k_1, \hb k_3, \hb k_s) + \text{23 perms},
\label{eq:model_toyE1}
\end{align}
where in the second equality, we perform the Schwinger parametrization that makes the equilateral shape factorizable up to introducing an auxiliary integral. Note that using Schwinger parameterization is essentially restoring the temporal integral introduced by the $\phi'^4/\Lambda^4$ vertex. Also, a coupling-dependent overall magnitude $\mu_{A,\text{ Equil.}} \defeq 9216 g_\text{NL}^{\text{eq},1}/25$ is pulled out of the expression.

\subsection{EFT-like Toy Shapes: Large-mass Limit \label{sec:largeMassModels}}
Beyond the canonical toy shapes for the trispectrum, we propose a procedure to generate new parity-violating contact-like shapes. As an example, we will consider exchanging a massive spin-1 particle with a chemical potential \cite{Liu:2019fag}, and the contact-like toy shape comes from taking a particular \emph{large-mass limit}, much like how heavy particle exchanges are treated in the effective field theory (EFT).%
\footnote{An example for the massive spin-2 exchange enhanced by a chemical potential is given in \cref{app:spin2LargeMass}. See also Ref.~\cite{Tong:2022cdz} for a detailed discussion.} 
This shape also illustrates how a parity-violating angular-dependent factor $\mathcal{K}(\{\hat{\vb{k}}_i\})$ enters the trispectrum through realistic models in inflation.

Let us consider an Abelian Higgs model whose gauge boson couples to a rolling homogeneous background $\theta(t)$ as a function of time $t$ via an axion-like coupling \cite{Wang:2020ioa}:
\begin{equation}
    \Lag = - \sqrt{-g} \qty[\frac{1}{4} F_{\mu\nu} F^{\mu\nu} + D_\mu \Sigma^* D^{\mu} \Sigma - m_H^2 \abs{\Sigma}^2 + \lambda \abs{\Sigma}^4] - \frac{c_0 \theta(t)}{4} \epsilon^{\mu\nu\rho\sigma} F_{\mu\nu} F_{\rho\sigma} 
    \label{eqn:lagSpin1}
\end{equation}
in which $F^{\mu\nu}$ and $\Sigma$ are the gauge boson and the Higgs fields, respectively, and $\epsilon_{\mu\nu\rho\sigma}$ denotes the 4-dimensional Levi-Civita symbol.%
\footnote{
Following Ref.~\cite{Liu:2019fag}, we will use the convention $\sqrt{-g} = a^4(\tau) = \qty(-H\tau)^{-4}$ in which $\tau$ denotes the conformal time with $a(\tau) \dd \tau = \dd t$ and $a(\tau)$ is the scale factor. By introducing the conformal time, all spacetime indices can be raised and lowered with $\eta^{\mu\nu}$ in the mostly plus signature. For the time derivatives, we introduce the notation 
\[
    \dot{f} \defeq \dv{f}{t}, \quad f' \defeq \dv{f}{\tau}. 
\]
} 
The rolling of $\theta(t)$ introduces a chemical potential on the gauge boson controlled by a dimensionless parameter \beq
c \defeq c_0 \dot{\theta}/H\eeq
in which $\dot{\theta}$ denotes the time derivative. Also, the 4D Levi-Civita tensor becomes a 3D Levi-Civita tensor after performing an integration by parts on the last term since $\theta(t)$ is a homogeneous background. This shows how the chemical potential term in a spin-1 model provides parity violation. 

To make further progress, we need the propagator of the spin-1 particle. Instead of following the standard Schwinger-Keldysh formalism, we solve the propagator as the Green's function to a simplified equation of motion directly. This is because time-ordering Heaviside functions $\Theta(\tau_L, \tau_R)$ from the Schwinger-Keldysh formalism introduces extra layers of time integrals, which we would like to avoid. (The full Schwinger-Keldysh propagator for a spin-1 exchange with a chemical potential is given in \cref{eq:spin1Prop,eq:spin1TOrdering}.)  Ideally, simplifying this time ordering to $\delta(\tau_L - \tau_R)$ would eliminate a layer of integral, effectively converting the propagator into a contact interaction. In the context of effective field theory, this is familiar. When integrating out a heavy particle $\sigma$, its propagator $\sim \ev{\sigma(x) \sigma(y)}$ is essentially replaced by $\sim \delta[4](x-y)/\qty(m_\sigma^2 \sqrt{-g})$, which motivates us to consider the large-mass limit. However, it is also known that a simple contact interaction cannot generate parity-violating trispectrum \cite{Liu:2019fag}. Thus, we must keep some $k_s$ dependence to produce parity violation. The equation of motion of a massive spin-1 particle coupled to a chemical potential induced by a rolling field is known \cite{Liu:2019fag, Lu:2019tjj, Wang:2020ioa}
\begin{equation}
	\qty[\square - m_Z^2 - 2 hc H \frac{k_s}{a}] Z^{\mu}_{\vb{k}_s,(h)} = 0, \quad \square \defeq a^{-4} \qty(- \partial_\tau a^2 \partial_\tau - a^2 k_s^2), 
\end{equation}
in which we took the transverse component of the massive spin-1 field $Z$ with helicity $h = \pm 1$, $k_s$ is the co-moving momentum of the propagator, and $a$ is the scale factor. When the chemical potential $Hc$ is comparable to $m_Z$ but $k \ll m_Z$, we take the long wavelength limit, drop the Laplacian operator, and obtain an equation that the Green's function must satisfy
\begin{equation}
	\qty(m_Z^2 + 2 hc H \frac{k_s}{a}) \ev{Z^{\mu}_{(h)}(x) Z^{\nu}_{(h)}(y)} = \frac{a^2 \delta[4](x - y)}{\sqrt{-g}} {\varepsilon_{(h)}^{\mu}}^* \varepsilon_{(h)}^{\nu}.
\end{equation}
Without the time derivative, the propagator of $Z$ essentially becomes a contact interaction on the same time slice. Thus, the radial-dependent part of the propagator looks like
\begin{equation}
	\begin{aligned}
		D_{\pm\pm}(\tau_1, \tau_2; k) 
		\approx& \frac{\delta(\tau_1 - \tau_2)}{a^2 \mu_Z^2 H^2} - \delta(\tau_1 - \tau_2) \frac{2 hc k}{a^3 \mu_Z^4 H^3} + \order{\frac{k^2}{\mu_Z^6 H^4}},
	\end{aligned}
	\label{eqn:simplifiedDProp}
\end{equation}
where the subscripts on $D$ denote the Schwinger-Keldysh indices \cite{Chen:2017ryl}, $\mu_Z \defeq m_Z/H$ is the dimensionless mass of the massive spin-1 particle, and the tensor structure of the propagator reads
\begin{equation}
    {\varepsilon_{(h)}}^*_i {\varepsilon_{(h)}}_j = \frac{1}{2} \qty(\delta_{ij} - \hat{k}_i \hat{k}_j - \ii h \epsilon_{ijl} \hat{k}_l),
    \label{eqn:spin1PropTensorPart}
\end{equation}
in which $\hat{k}$ denotes the unit vector along the propagator momentum direction.%
\footnote{The $(\pm\mp)$-type propagators vanish in this limit as they solve the homogeneous equation of motion, with the source term being 0.}
As we sum over two helicity states $h = \pm 1$, only the $\order{h^0}$ or $\order{h^2}$ term in the full propagator survives. Therefore, the large-mass limit can be intuitively depicted as 
\begin{equation}
    \lim_{m_Z \gg k}
    \begin{tikzpicture}[baseline=-3pt] \begin{feynman}[small]
		\vertex (v) at (0,0); 
		\node[empty dot, fill = none, label = \(i\)] (i) at (-1.25,0);
		\node[empty dot, fill = none, label = \(j\)] (f) at (1.25,0);
		\diagram*{
			(i) -- [photon, edge label = \(\varepsilon_{(h)}\)] (v) -- [photon, edge label = \(\varepsilon_{(h)}^*\)] (f);
		};
	\end{feynman} \end{tikzpicture}
    \approx 
    \begin{tikzpicture}[baseline=-3pt] \begin{feynman}[small]
		\node[empty dot, fill = none] (i) at (-0.01,0);
		\node[empty dot, fill = none] (f) at (0.01,0);
		\diagram*{
			(i) -- [edge label = \(\delta_{ij}\)] (f);
		};
	\end{feynman} \end{tikzpicture}
    +
    \begin{tikzpicture}[baseline=-3pt] \begin{feynman}[small]
		\node[style = crossed dot, label = \( \hat{k}_{l} \epsilon_{ijl}\)] (v) at (0,0);
		\node[empty dot, fill = none, label = \(i\)] (i) at (-0.75,0);
		\node[empty dot, fill = none, label = \(j\)] (f) at (0.75,0);
		\diagram*{
			(i) -- [photon, momentum' = \(k\)] (v) -- [photon, momentum' = \(\ll m_H\ \)] (f);
		};
	\end{feynman} \end{tikzpicture}
    + \ldots.
    \label{eqn:propLargeMassPicture}
\end{equation}
When taking the large-mass limit, the $Z$ propagator becomes an almost contact interaction analogous to the Minkowski propagator $(p^2 + m_Z^2)^{-1} \approx m_Z^{-2}$. The leading term ($\sim \order{(Hck/m_Z^2)^0}$) in $D_{\pm\pm}$, then, has no parity violation. However, the presence of the chemical potential introduces a parity-violating term in the next-to-leading order. The $\sim \order{Hck/m_Z^2}$ term has helicity dependence and can introduce parity violation. Note that the parity-violating signal vanishes as propagator momentum $k$ approaches zero, this echoes the no-go theorem that a parity-violating signature in cosmological 4PCFs needs a nonlocal propagation effect. Note that we are not taking a rigorous EFT limit here as we dropped the time derivatives for simplicity, hence the name ``EFT-like toy shape". A true EFT limit includes contact interactions only and will lead to a vanishing parity-violating contribution to the trispectrum as required by the no-go theorem. This derivation aims to motivate a new kind of toy shapes that are numerically easy to evaluate.

To contract the tensor structure of the propagator with external momenta of inflaton fluctuations, we may introduce some operators of the form
\begin{equation}
    \begin{aligned}
		\Lag \propto \phi' \partial_i \phi Z^i.
	\end{aligned}
    \label{eqn:spin1InteractionLag}
\end{equation}
One possible model for this interaction comes from introducing two higher-dimensional operators between the $\U{1}$ sector and the inflaton
\begin{equation}
    \begin{aligned}
		\Lag =& -\sqrt{-g} \qty[\frac{c_1}{\Lambda} \partial_\mu \phi \qty(\Sigma^* D^\mu \Sigma) + \frac{c_2}{\Lambda^2} \qty(\partial \phi)^2 \abs{\Sigma}^2 + \text{h.c.}] \\
		\supset & \underbrace{\Im{ \frac{c_1 \dot{\phi}_0 m_Z}{\Lambda} }}_{\defeq -\rho_{1,Z}} \frac{a^2}{\dot{\phi}_0} \eta^{\mu\nu} \partial_\mu \phi Z_{\nu} h + a^3 \underbrace{\frac{c_2 v \dot{\phi}_0}{\Lambda^2}}_{\defeq \rho_2} \phi' h \\
        \to& -\ii a \qty(\frac{\rho_{1,Z} \rho_2}{\dot{\phi}_0 \mu_H^2}) \frac{1}{H^2} \eta^{\mu\nu} \phi' \partial_\mu \phi Z_\nu. 
	\end{aligned}
    \label{eqn:spin1InteractionModelLag}
\end{equation}
in which we integrated out the heavy Higgs boson by assuming that the external momenta are significantly smaller than the Higgs mass $m_H \defeq \mu_H H$, and $\phi'$ denotes the derivative of inflaton fluctuation $\phi$ with respect to the conformal time $\tau$. Combining the interaction vertex with the propagator and Wick-rotating the conformal time $\tau$ to Euclidean time $t_E$, we may obtain the following \emph{parity-violating part} of the trispectrum of the curvature perturbation
\begin{equation}
    \begin{aligned}
        & \qty[- \qty(2\pi)^4 P_\zeta^2 \frac{c}{2} \frac{\qty(4\pi)^{7/2}}{\sqrt{2}} ]^{-1} \qty(\frac{\rho_{1,Z}}{\dot{\phi}_0} \frac{\rho_2}{H} \frac{1}{\mu_Z^2 \mu_H^2})^{-2} \ev{\zeta^4}' \\
        \supset& - \qty(4\pi)^{-2} \iso{111}(\hat{\vb{k}}_1, \hat{\vb{k}}_3, \hat{\vb{k}}_s) \int_0^{\infty} \dd t_E \; \frac{t_E^3}{6} \qty[ \prod_{i=1}^4 \frac{e^{-k_i t_E}}{k_i} ] \frac{1 + k_1 t_E}{k_1} \frac{1 + k_3 t_E}{k_3} k_s + \text{23 perms}.
    \end{aligned}
\end{equation}
It is worth remarking that the parity-preserving part from the first term in \cref{eqn:simplifiedDProp} vanishes; hence, the leading contribution to the trispectrum from a heavy spin-1 exchange with chemical potential is parity-violating. Here, we also factored out a dimensionless overall magnitude $\mu_A$ of the correlation function for the EFT-like spin-1 model defined as 
\begin{equation}
    \mu_{A,\text{ Toy spin-1}} \defeq -\qty(2\pi)^4 P_\zeta^2 \frac{c}{2} \frac{(4\pi)^{7/2}}{\sqrt{2}} \qty(\frac{\rho_{1,Z}}{\dot{\phi}_0} \frac{\rho_2}{H} \frac{1}{\mu_Z^2 \mu_H^2})^2.
\end{equation}
This overall magnitude is factored out so that we may separate the shape of the parity-violating trispectrum from the model-specific factors. In performing the analysis against the BOSS data shown in \cref{sec:results}, we treat $\mu_A$ as a free parameter to be fitted with uncertainties based on the \textsc{MultiDark-Patchy} mocks for the BOSS survey.

\subsection{Full Models: Spin-1 and Spin-2 Exchange \label{sec:fullModels}}
In this subsection, we provide momentum trispectra from more realistic models based on cosmological collider physics. Specifically, we provide trispectra that involve the exchange of a massive spin-1 or spin-2 particle enhanced by a chemical potential. As most of the results are known here, we will direct readers to appropriate previous works that present the full analysis and shall only discuss salient points pertinent to this study.

\subsubsection{Spin-1 Exchange \label{sec:fullSpin1}}
Most of the result for this trispectrum is known in the literature \cite{Qin:2022fbv}. We present the result here both for completeness and in a form that will be convenient to our numerical routine. The Lagrangian of this model is the same as that shown previously in \cref{eqn:lagSpin1}. The Schwinger-Keldysh propagators for a massive spin-1 particle with chemical potential is
\begin{equation}
	D^{(h)}_{\pm\mp}(\tau_1, \tau_2;k)
	= \frac{e^{-h\pi c}}{2k} \W_{\mp \ii hc, \ii \tilde{\nu}_z}(\mp 2\ii k\tau_1) \W_{\pm \ii hc, \ii \tilde{\nu}_z}(\pm 2\ii k\tau_2),
    \label{eq:spin1Prop}
\end{equation}
in which $\tilde{\nu}_z \defeq \sqrt{m_Z^2/H^2 - 1/4}$  is the oscillation frequency of the cosmological collider signal and $\W_{c,\mu}(z)$ denotes the Whittaker $\W$ function. The tensor structure of the $Z$ propagator remains the same as that shown in \cref{eqn:spin1PropTensorPart}, and the $(\pm\pm)$-type propagators are
\begin{equation}
	D_{\pm\pm}(\tau_1, \tau_2) = D_{\mp\pm} \Theta(\tau_1 - \tau_2) + D_{\pm\mp} \Theta(\tau_2 - \tau_1), 
    \label{eq:spin1TOrdering}
\end{equation}
in which $\Theta(x)$ denotes the Heaviside step function. The explicit $e^{-h \pi c}$ chemical-potential dependence shows the exponential enhancement in the $h = -1$ state that leads to a large cosmological collider signal.\footnote{This provides an estimate of the cosmological collider signal as discussed in Ref.~\cite{Wang:2019gbi}. However, strictly speaking, the Whittaker function also encodes additional momentum-dependent enhancement or suppression in the presence of a chemical potential.}  Allowing the spin-1 particle to interact with the inflaton via \cref{eqn:spin1InteractionLag} and applying appropriate Wick rotation to the time integral, we get the following \emph{parity-violating part} of the full spin-1 exchange trispectrum model:
\begin{equation}
	\begin{aligned}
		& \qty(\frac{\rho_{1,Z}}{\dot{\phi}_0} \frac{\rho_2}{H \mu_H^2})^{-2} \qty[  \qty(2\pi)^4 P_\zeta^2 \frac{\qty(4\pi)^4 \sqrt{2}}{96}]^{-1} \ev{\zeta^4}'\\
		=& \iso{11(1)10}(\hat{\vb{k}}_1, \ldots, \hat{\vb{k}}_4) \int_0^\infty \dd t_L \dd t_R \; \qty[ \frac{e^{-k_1 t_L} \qty(1 + k_1 t_L)}{k_1^2} ] \qty[ e^{-k_2 t_L} ] \qty[ \frac{e^{-k_3 t_R} \qty(1 + k_3 t_R)}{k_3^2}] \qty[ \frac{e^{-k_4 t_R}}{k_4} ] \\
		& \times \qty(4\pi)^{-2} \frac{he^{- \pi hc}}{k_s^2} \left[ \W_{-\ii hc, \ii \tilde{\nu}_z}(2k_s t_L) \W_{\ii hc, \ii \tilde{\nu}_z}(2k_s t_R) \right. \\
        & + \Theta(t_L - t_R) \W_{-\ii hc, \ii \tilde{\nu}_z}(2k_s t_L) \W_{\ii hc, \ii \tilde{\nu}_z}(-2k_s t_R-\ii \epsilon) \\
		& + \left. \Theta(t_R - t_L) \W_{-\ii hc, \ii \tilde{\nu}_z}(-2k_s t_L + \ii \epsilon) \W_{\ii hc, \ii \tilde{\nu}_z}(2k_s t_R) + \text{h.c.} \right] + \text{23 perms},
	\end{aligned}
 \label{eq:full_spin1}
\end{equation}
where $\ii \epsilon$ shifts are introduced to avoid the branch cuts of the Whittaker $\W$ function. The expression of the isotropic basis function is given by~\cref{eq:P111}. Here, the model-dependent overall magnitude $\mu_A$ factored out from the trispectrum is 
\begin{equation}
    \mu_{A,\text{ Spin-1}} \defeq \qty[ \qty(2\pi)^4 P_\zeta^2 \frac{\qty(4\pi)^4 \sqrt{2}}{96} ]^{-1} \qty(\frac{\rho_{1,Z}}{\dot{\phi}_0} \frac{\rho_2}{H \mu_H^2})^{-2}.
\end{equation}
We should also clarify that this prefactor is not the only source of model dependence. In particular, the propagator of the exchanged spin-$1$ particle contains additional momentum-dependent information about the mass and chemical potential of the exchanged particle. Thus, the full spin-1 model is expected to imprint more model-specific information on the parity-violating part of the trispectrum apart from its overall shape.

\subsubsection{Spin-2 Exchange \label{sec:fullSpin2}}
For the massive spin-2 exchange, we consider the following model
\begin{equation}
    \begin{aligned}
        \Lag =& \sqrt{-g} \left[ \frac{1}{2} \nabla_\mu h^{\mu \rho} \nabla^{\nu} h_{\nu \rho} - \frac{1}{4} \nabla_{\mu}h_{\nu\rho} \nabla^{\mu} h^{\nu\rho} + \frac{1}{4} \nabla^\mu h \nabla_\mu h - \frac{1}{2} \nabla^\mu h^\mu_{\ \nu} \nabla_\mu h \right. \\
        & \left. - \frac{1}{2} H^2 \qty(h_{\mu\nu} h^{\mu\nu} + \frac{1}{2} h^2) 
        + \frac{1}{4} m^2 \qty(h_{\mu\nu} h^{\mu\nu} - h^2) \right] + \frac{\phi}{2\Lambda_c} \epsilon^{\mu\nu\rho\sigma} \nabla_\mu h_{\nu\lambda} \nabla_{\rho} h_{\sigma\lambda},
    \end{aligned}
\end{equation}
in which a spin-2 particle with a Fierz-Pauli mass and chemical potential term is present. A detailed analysis of this model has been done in Ref.~\cite{Tong:2022cdz}. The propagator for the $h=\pm 2$ mode of such particle is 
\begin{equation}
    H^{(h)}_{\pm\mp}(\tau_1, \tau_2; k) = \frac{e^{-h\kappa \pi/2}}{4H^2 k \tau_1 \tau_2} \W_{\mp\ii h\kappa/2, \ii \mu_h}(\mp 2\ii k \tau_1) \W_{\pm \ii h\kappa/2, \ii \mu_h}(\pm 2\ii k \tau_2)
\end{equation}
with a tensor structure 
\begin{equation}
    {\varepsilon_{(h)}}^*_{ij} {\varepsilon_{(h)}}_{mn} = \frac{1}{2} \qty(\delta_{im} - \hat{k}_i \hat{k}_m - \frac{\ii h}{2} \epsilon_{iml} \hat{k}_l) \qty(\delta_{jn} - \hat{k}_j \hat{k}_n - \frac{\ii h}{2} \epsilon_{jnl} \hat{k}_l),
    \label{eqn:spin2PropTensorPart}
\end{equation}
in which $\mu_h \defeq \sqrt{m^2/H^2 - 9/4}$ denotes the dimensionless mass of the spin-2 particle and $\kappa \defeq \dot{\phi}_0 / \Lambda_c$ denotes its dimensionless chemical potential. For this study, it suffices to consider the exchange of the mode with helicity $h = -2$ because the propagator of the spin-2 particle can enjoy a $\sim e^{-\pi \kappa h/2}$ enhancement due to the chemical potential. With the interaction term 
\begin{equation}
	\Lag \supset \frac{1}{M} h_{ij} \partial_i \phi \partial_j \phi. 
\end{equation}
and appropriate Wick rotation, one can acquire the following \emph{parity-violating} part of the full spin-2 trispectrum 
\begin{equation}
    \begin{aligned}
        & \qty[\qty(2\pi)^4 P_\zeta^2 \frac{1}{32} \frac{1}{15} \sqrt{\frac{2}{5}} ]^{-1} \qty(\frac{H}{M})^{-2} \ev{\zeta^4}' \\
        =& \int_0^\infty \frac{\dd t_L}{t_L} \frac{\dd t_R}{t_R} \; \qty[ \frac{\qty(1 + k_1 t_{L}) e^{-k_1t_{L}}}{k_1} \frac{\qty(1 + k_2 t_{L}) e^{-k_2t_{L}}}{k_2} \frac{\qty(1 + k_3 t_{R}) e^{-k_3t_{R}}}{k_3} ] \frac{\qty(1 + k_4 t_{R}) e^{-k_4t_{R}}}{k_4^3} \frac{1}{k_s^4} \\
        & \times \qty(4\pi)^{-2} he^{- \kappa h \pi / 2} \left[ \W_{-\ii \kappa h / 2, \ii \mu_h}(2k_s t_L) \W_{\ii \kappa h / 2, i \mu_h}(2k_s t_R) \right. \\
        & - \Theta(t_L - t_R) \W_{-\ii \kappa h/2,\ii \mu_h}(2k_s t_L) \W_{\ii \kappa h/2,\ii \mu_h}(-2k_s t_R - \ii \epsilon) \\
        & \left. - \Theta(t_R - t_L) \W_{-\ii \kappa h/2,\ii \mu_h}(-2k_s t_L + \ii \epsilon) \W_{\ii \kappa h/2,\ii \mu_h}(2k_st_R) + \text{h.c.} \right] \\
        & \times \qty[ k_1 \qty( 2 \iso{12(2)20} + \iso{32(2)20} ) - k_2 \qty( 2\iso{21(2)20} + \iso{23(2)20} ) ](\hat{\vb{k}}_1, \ldots, \hat{\vb{k}}_4).
    \end{aligned}
    \label{eq:full_spin2}
\end{equation}
where the isotropic basis functions respectively stand for
\begin{align}
\iso{12(2)20}(\hat{\vb{k}}_1, \hat{\vb{k}}_2, \hat{\vb{k}}_3, \hat{\vb{k}}_4) ={} & \frac{3\sqrt{5} \ii}{2^{1/2}(4\pi)^2} \hat{\vb{k}}_1 \cdot(\hat{\vb{k}}_2 \times \hat{\vb{k}}_3)(\hat{\vb{k}}_2 \cdot \hat{\vb{k}}_3) \\
\iso{32(2)20}(\hat{\vb{k}}_1, \hat{\vb{k}}_2, \hat{\vb{k}}_3, \hat{\vb{k}}_4) ={} & \frac{15\sqrt{5}\ii}{2^{3/2}(4\pi)^2}  \hat{\vb{k}}_3\cdot ( \hat{\vb{k}}_1 \times  \hat{\vb{k}}_2)\left[(\hat{\vb{k}}_1\cdot \hat{\vb{k}}_2)(\hat{\vb{k}}_1\cdot \hat{\vb{k}}_3)-\frac{1}{5}\hat{\vb{k}}_1\cdot \hat{\vb{k}}_3\right]\\
\iso{21(2)20}(\hat{\vb{k}}_1, \hat{\vb{k}}_2, \hat{\vb{k}}_3, \hat{\vb{k}}_4) ={} &\iso{12(2)20}(\hat{\vb{k}}_2, \hat{\vb{k}}_1, \hat{\vb{k}}_3, \hat{\vb{k}}_4)\\
\iso{23(2)20}(\hat{\vb{k}}_1, \hat{\vb{k}}_2, \hat{\vb{k}}_3, \hat{\vb{k}}_4) ={}& \iso{32(2)20}(\hat{\vb{k}}_2, \hat{\vb{k}}_1, \hat{\vb{k}}_3, \hat{\vb{k}}_4)
\end{align}
Here, the model-dependent overall magnitude $\mu_A$ is defined as 
\begin{equation}
    \mu_{A,\text{ Spin-2}} \defeq \qty[\frac{\qty(2\pi)^4 P_\zeta^2}{480} \sqrt{\frac{2}{5}} ] \qty(\frac{H}{M})^{2}
\end{equation}

\section{From Trispectra to Galaxy-Survey Data\label{sec:4PCFGeneralStrategy}}

We have explored the general features of models with parity-violating trispectra in momentum space. To compare these models to the observed galaxy survey data, we need to follow three main steps: (i) translating the momentum-space trispectra to the position-space 4PCF of curvature perturbations. The translation requires performing a high-dimensional Fourier transform, which is computationally demanding. This challenge can be addressed by converting the template into a factorizable form, though this often necessitates auxiliary integrals; alternatively, we could approximate the template as a sum of fully factorizable functions, though this approach may cost a considerable reduction in accuracy; (ii) connecting the 4PCF of curvature perturbation with the 4PCF of the galaxy number overdensities (galaxy 4PCF). This step requires incorporating structure growth and projection effects using numerical simulations or semi-analytical methods; (iii) inferring the parameters of the galaxy 4PCF templates from the galaxy survey data.

Significant progress has been made in developing an analysis pipeline that links the trispectrum to observational data, with particular advances in analyzing the parity-violating 4PCF from the BOSS data~\cite{Hou:2022wfj, Philcox:2022hkh, Cabass:2022oap}. Below, we summarize the key components of the existing analysis pipeline that help address the challenges in each of the three main steps. We focus especially on procedures that enable efficient computation of a position-space template for a full exchange-type 4PCF, a class of models that is not easily factorizable.

A few steps are required to convert the curvature fluctuation into a galaxy number overdensity: (i) perform the linear perturbation to convert the curvature perturbation to matter overdensity, (ii) relate matter overdensity to galaxy overdensity with the linear bias relation, and (iii) take redshift-space distortion into account. Here, we ignored nonlinear effects as the 4PCF observation extracted from BOSS has a minimum separation of $20\; h^{-1} \text{Mpc}$, significantly longer than the nonlinear scale $1/k_{NL} \approx 2\; h^{-1} \text{Mpc}$ around $z = 0.57$ at which the BOSS data was taken. It is, however, worth noticing that for smaller-scale observables, nonlinearity becomes important, and our treatment with a linear transfer function is insufficient. The galaxy number overdensity in position space is related to the curvature trispectrum by
\begin{align}
		\zeta_g^4(\vb{r}_1, \vb{r}_2, \vb{r}_3, \vb{r}_4) = 	
		\int \qty[ \prod_{i = 1}^{4} \dbar^3 k_i \; Z_1(\hat{\vb{k}}_i, z) M(k_i, z) ] 
		\ev{\zeta^4}'(\vb{k}) \nonumber\\
        \exp(\sum_{i = 1}^4 \ii\vb{k}_i \cdot \vb{r}_i) \qty(2\pi)^3 \delta[3](\sum_{i=1}^4 \vb{k}_i),
\label{eq:fourier}
\end{align}
where $Z_1$ is the Kaiser redshift-space distortion factor and $M$ is the transfer function, both derived from standard $\Lambda$CDM cosmology. The redshift-space distortion, along the line-of-sight direction $\hat{\vb{n}}$, is given by \beq
Z_1(\hat{\vb{k}}, z) = b_1(z) + f(z) (\hat{\vb{k}}\cdot\hat{\vb{n}})^2,
\label{eq:rsd}
\eeq
where $b_1(z)$ is the linear bias and $f(z)$ is the logarithmic growth factor (see e.g.~\cite{Hamilton:1997zq} for a review). Both $M(k, z)$ and $f(z)$ can be computed by Boltzmann solvers once and for all under linear perturbation theory. However, the challenge remains in performing the Fourier transform with the presence of the delta function. 

To mitigate this challenge, we leverage the properties of the galaxy survey data. Due to homogeneity, one position in the 4PCF, $\zeta_g^4(\vb{r}_1, \vb{r}_2, \vb{r}_3, \vb{r}_4)$, can be set at the origin (\textit{e.g.}, $\vb{r}_4=\vb{0}$) without loss of generality. Symmetry argument (see \cref{sec:intro}) suggests that the 4PCF should be described by six free parameters: three angular variables, which exhibit co-rotation invariance due to isotropy, and three radial distances. The angular and radial components of the  $\{\vb{r}_i\}$ variables can be separated by expanding the 4PCF in terms of the 3-point isotropic basis functions $\iso{\ell_1, \ell_2, \ell_3}(\hat{\vb{r}}_1, \hat{\vb{r}}_2, \hat{\vb{r}}_3)$ which are manifestly co-rotation invariant, i.e.,
\begin{equation}
    \zeta_g^4(\vb{r}_1, \vb{r}_2, \vb{r}_3) = \sum_{\ell_1, \ell_2, \ell_3} \zeta_{\ell_1, \ell_2, \ell_3}(r_1, r_2, r_3) \iso{\ell_1, \ell_2, \ell_3}(\hat{\vb{r}}_1,\hat{\vb{r}}_2,\hat{\vb{r}}_3),
\end{equation}
where the coefficients $\zeta_{\ell_1, \ell_2, \ell_3}(r_1, r_2, r_3)$ encode the 4PCF information with the $r_i$'s representing the radial distances. 

Under the isotropic basis function formalism, we adopt the following strategies to manage the Fourier transform: 
\begin{enumerate}
\item Split the Fourier kernel and the momentum-conserving delta functions into radial and angular parts using the plane wave expansions; 
\item Reduce the angular integrals analytically to Wigner symbols using identities of spherical harmonics; 
\item Perform the lower-dimensional radial integrals numerically by splitting them into subdiagrams; 
\item Integrate over time variables, which takes care of the time ordering in the propagator; 
\item Sum over all indices of angular momentum number to obtain the final result. 
\end{enumerate}
Step 3 and 4 can be naturally applied for all exchange-type trispectra, which take the general form 
\begin{equation}
    \begin{multlined}
        \ev{\zeta^4}'(\{\vb{k}_i\}) = \sum_{l_1, l_2, l', l_3, l_4} c_{l_1, l_2 (l') l_3, l_4} \iso{l_1, l_2 (l') l_3, l_4}(\hat{\vb{k}}_1, \hat{\vb{k}}_2, \hat{\vb{k}}_3, \hat{\vb{k}}_4) \\
        \times \int \dd t_L \dd t_R \; \mathcal{J}_L^{l_1, l_2 (l') l_3, l_4}(k_1, k_2, k_s, t_L) \Pi(k_s, t_L, t_R) \mathcal{J}_R^{l_1, l_2 (l') l_3, l_4}(k_3, k_4, k_s, t_R) 
        + \text{23 perms}
    \end{multlined}
\end{equation}
where $c_{l_1, l_2 (l') l_3, l_4}$ denotes the numerical prefactor and $\iso{l_1, l_2 (l') l_3, l_4}(\hat{\vb{k}}_1, \ldots, \hat{\vb{k}}_4)$ is the 4-point isotropic basis function~\cref{eq:4pt}.

After obtaining the galaxy 4PCF, the next step before comparing it to the data is to account for binning. Binning is necessary for inferring the underlying galaxy distributions from the noisy observed survey data. To match the binning of the data, the binning effects need to be incorporated into the theoretical prediction, which is performed for all six degrees of freedom. The angular binning is naturally achieved through different combinations of the three angular momentum numbers $\{\ell_1, \ell_2, \ell_3\}$. For the radial distance variables $\{r_1, r_2, r_3\}$,  each is binned into $n_r$ evenly-spaced radial bins within the range from $R_{\min}$ to $R_{\max}$. The  boundaries of the $b_i$-th radial bin are defined as
\beq
r_{b_i, \min}= R_{\min} + b_i \frac{R_{\max} -R_{\min}}{n_r}, \quad r_{b_i, \max}=  R_{\min} + (b_i+1) \frac{R_{\max} -R_{\min}}{n_r}.
\eeq
To incorporate radial binning in the theoretical prediction, we update the correlation coefficients by replacing the distance variable $r_i$  with the corresponding bin indices $b_i$, giving $\zeta_{\ell_1, \ell_2, \ell_3}(r_1, r_2, r_3) \to \zeta_{\ell_1, \ell_2, \ell_3}(b_1, b_2, b_3)$.
Additionally, we replace the Bessel function with the radial distance argument with the averaged spherical Bessel function,
$j_{\ell_i}  (k r_i) \to     \bar j_\ell(k, b_i) \defeq {\int_{r_{\min}}^{r_{\max}} \dd r\; r^2 j_{\ell_i} (k r_i)}/{\int_{r_{\min}}^{r_{\max}}  \dd r\; r^2},
\label{eq:avg_j}$
to compute the correlation coefficients (see \cref{app:detail4PCF} for more details).

Incorporating all the elements together, we obtain the following expression for the galaxy 4PCF coefficients for a general exchange-type model (detailed derivations are presented in \cref{app:detail4PCF})
\begin{keyeqn}
\begin{equation}
    \begin{aligned}
		& \zeta_{\ell_1, \ell_2, \ell_3}(b_1,b_2,b_3) \\
		=& \qty(4\pi)^{11/2} \sum_{\sigma, \ell_\sigma,L,j,\lambda,\red{l}} \Phi_\sigma \qty(-\ii)^{\ell_{1}+\ldots+\ell_{3}} \qty(-1)^{L'} \ii^{L_1 + \ldots + L_4} \qty(-1)^{\lambda_1 + \ldots + \lambda_4} \qty[\prod_{i=1}^4 Z_{j_i} \qty(2j_i + 1)] \\
		& \times \qty(2j'+1) \qty[\prod_{i=1}^4 \qty(2L_i + 1)] \qty(2L'+1) \mqty(j_1 & j_2 & j' \\ 0 & 0 & 0)  \mqty(j_3 & j_4 & j' \\ 0 & 0 & 0) \mqty(L_1 & L_2 & L' \\ 0 & 0 & 0) \mqty(L_3 & L_4 & L' \\ 0 & 0 & 0) \\
        & \times \qty[ \prod_{i=1}^4 \mqty( \ell_{\sigma_i} & j_i & \lambda_i \\ 0 & 0 & 0 ) \mqty( \red{l_i} & L_i & \lambda_i \\ 0 & 0 & 0 )]
		\qty{ \mqty{ \ell_{\sigma_1} & \ell_{\sigma_2} & \ell' \\ j_1 & j_2 & j' \\ \lambda_1 & \lambda_2 & \lambda' } } 
		\qty{ \mqty{ \ell' & \ell_{\sigma_3} & \ell_{\sigma_4} \\ j' & j_3 & j_4 \\ \lambda' & \lambda_3 & \lambda_4 } } 
		\qty{ \mqty{ \red{l_1} & \red{l_2} & \red{l'} \\ L_1 & L_2 & L' \\ \lambda_1 & \lambda_2 & \lambda' } } 
		\qty{ \mqty{ \red{l'} & \red{l_3} & \red{l_4} \\ L' & L_3 & L_4 \\ \lambda' & \lambda_3 & \lambda_4 } } \\
        & \times \red{c_{l_1,l_2(l')l_3,l_4}} \qty[\prod_{i=1}^4 \qty(2\lambda_i + 1)] \qty(2\lambda'+1) \qty[\prod_{i=1}^4 \sqrt{2\ell_{\sigma_i} + 1}] \sqrt{2\ell'+1} \red{\qty[\prod_{i=1}^4 \sqrt{2l_i + 1}] \sqrt{2l'+1}} \\
		& \times \int \dd \tau_L \dd \tau_R \int \frac{\dd k_s\, k_s^2}{2\pi^2} \; \red{\Pi(k_s, \tau_L, \tau_R)} \\
		& \times \qty[ \int \prod_{i = 1}^{2} \frac{\dd k_i \, k_i^2}{2\pi^2} \; M(k_i) \bar{j}_{\ell_{\sigma_i}}(k_i, b_{\sigma_i}) ] f_{L_1,L_2,L'}(k_1,k_2,k_s) \red{\mathcal{J}^{l_1,l_2(l')l_3,l_4}_L(k_1, k_2, k_s, \tau_L)}\\
		& \times \qty[ \int \prod_{i = 3}^{4} \frac{\dd k_i \, k_i^2}{2\pi^2} \; M(k_i) \bar{j}_{\ell_{\sigma_i}}(k_i, b_{\sigma_i}) ] f_{L_3,L_4,L'}(k_3,k_4,k_s)  \red{\mathcal{J}^{l_1,l_2(l')l_3,l_4}_{R}(k_3, k_4, k_s, \tau_R)}.
	\end{aligned}
	\label{eqn:strategy}
\end{equation}
\end{keyeqn}
Here, the red scripts denote model-dependent inputs that come from the trispectrum model, $\sigma$ refers to the 24 permutations of the four labels $\{1,2,3,4\}$, and the curly brackets denote the Wigner's $9j$ symbols. Some auxiliary functions or variables are defined as follows:
\begin{gather}
    \Phi_\sigma \defeq 
    \begin{dcases}
        (-1)^{\ell_1 + \ell_2 + \ell_3}, & \text{odd permutation }\sigma, \\
        1, & \text{even permutation }\sigma,
    \end{dcases}, ~\quad~ \ell_4 = 0, \\
    Z_{j_i} \defeq \qty(b + \frac{f}{3}) \delta_{0,j_i} + \frac{2f}{15} \delta_{2,j_i}, \label{eq:rsd}\\
    f_{L_1,L_2,L_3}(k_1,k_2,k_3) 
	\defeq \int_0^\infty \dd x\; x^2 j_{L_1}(k_1 x) j_{L_2}(k_2 x) j_{L_3}(k_3 x),
\end{gather}
with $\delta_{i,j}$ representing Kronecker delta.%
\footnote{In the numerical implementation, we fix the fiducial cosmology to that similar to Planck 2018 \cite{Planck:2018jri}, $\Omega_m = 0.307115$, $\Omega_\Lambda = 0.692885$, $\Omega_b = 0.048$, $\sigma_8 = 0.8288$, $h = 0.6777$ with $\Delta_\zeta^2 = 4.1\times 10^{-8}$ and $n_s = 1$. As BOSS data are mostly taken at $z = 0.57$, the corresponding linear bias is $b = 2.0$ with a growth factor $f \approx 0.777$.}
Conveniently, the 24 permutations on external momenta are performed explicitly during numerical evaluation.  In numerical implementation, we improved the numerical stability of evaluation of the auxiliary function $f_{L_1, L_2, L_3}(k_1, k_2, k_3)$. More discussions about these numerical technicalities can be found in \cref{sec:tripleJ}.

\section{Results \label{sec:results}}

In this section, we first investigate the features of the galaxy 4PCF templates by computing the galaxy 4PCF coefficients for a number of toy models and full models presented in Sec.~\ref{sec:theoreticalModels}. We then compare the templates with the BOSS survey data, using the correlation matrix provided by Ref.~\cite{Philcox:2022hkh}.

We consider the local-type and equilateral-type toy models with local or equilateral trispectra multiplying $\{\iso{111}, \iso{122}, \ldots \iso{333}\}(\hat{\vb k}_1, \hat{\vb k}_2, \hat{\vb k}_s)$ (see~\cref{sec:canonicalModels}). Besides, we consider toy spin-1, i.e., exchange spin-1 particle under the large mass limit (see~\cref{sec:largeMassModels}). Finally, we consider the full spin-1 and spin-2 exchange model with the chemical potential enhancement. We pick $\tilde{\nu}_z = c = 4$ and $\mu_h = \kappa = 4$ as the benchmark for spin-1 and spin-2 scenarios respectively.

In the numerical analysis, we use $L_{\max} = 10$ for the spin-1 template and $L_{\max} = 8$ for all other templates for the summation of $L$ in~\cref{eqn:strategy}. See discussion on the $L_{\max}$ choice in~\cref{app:Lmax}. Besides, we adopt the radial distance binning scheme from~\cite{Philcox:2022hkh}, with parameters $R_\text{min} = 20\; h^{-1} \text{Mpc}$, $R_\text{max} = 160\; h^{-1} \text{Mpc}$ and $n_r = 10$. This means the binning index for $i$-th radial distance $b_i = 0, ..., 9$. Due to permutation symmetry of radial indices (see \cref{eqn:strategy}, in particular, the sum over permutation $\sigma$), we can demand that $b_1 < b_2 < b_3$ without loss of generality \cite{Cahn:2020axu}. Note that we do not include bins with $b_1 = b_2$ or $b_2 = b_3$ due to concerns about late-time small-scale effects~\cite{Philcox:2022hkh}. Under the restrictions, we have 120 different radial bins. We expect the features we find below will not significantly change if we use a different radial binning scheme for a similar distance range.  

For canonical toy models, including both local- and equilateral-type, we split the temporal integral into $50$ log-even grids ranging from $\ln \tau = \ln 10^{-4}$ to $\ln \tau = \ln 10^4$ and the momentum integral into $1000$ linear grids from $k = 0.03\; h \text{Mpc}^{-1}$ to $k = 3\; h\text{Mpc}^{-1}$. The results cost about $3\times 10^2$ CPU-hours for each local toy model and $4\times10^3$ CPU-hours for each equilateral toy model. For the EFT-like toy shapes, the computation for one model costs about $2\times 10^3$ CPU-hours. For the full spin-$1$ model, the computation becomes more resource-demanding due to an additional layer of temporal integral and a more complicated propagator shape. (For instance, in performing the $\ii \epsilon$ shift, we set $\epsilon = 10^{-9}$ to avoid the branch cut of the Whittaker function in the propagator.) The more demanding computation takes a total of $4\times10^4$ CPU-hours for the full template. For the spin-$2$ model, because of two different momentum dependences in the two sets of isotropic basis functions as shown in \cref{eq:full_spin2}, the computational cost for evaluating the full spin-$2$ position-space template is about $7\times 10^4$ CPU-hours.

\subsection{Features of Position-space Templates \label{sec:templateFeatures}}

After computing the galaxy 4PCF coefficients $\zeta_{\ell_1, \ell_2,\ell_3}(b_1,b_2,b_3)$ for various models, we identify interesting characteristics in the angular and radial distribution, which are closely tied to the input trispectra. We first explore these features with toy models to illustrate angular and radial dependencies, followed by a comprehensive analysis of the full models displaying both types of features.

\subsubsection{Angular Distribution}
\begin{figure}[!t]
    \centering
    \includegraphics[width=0.96\linewidth]{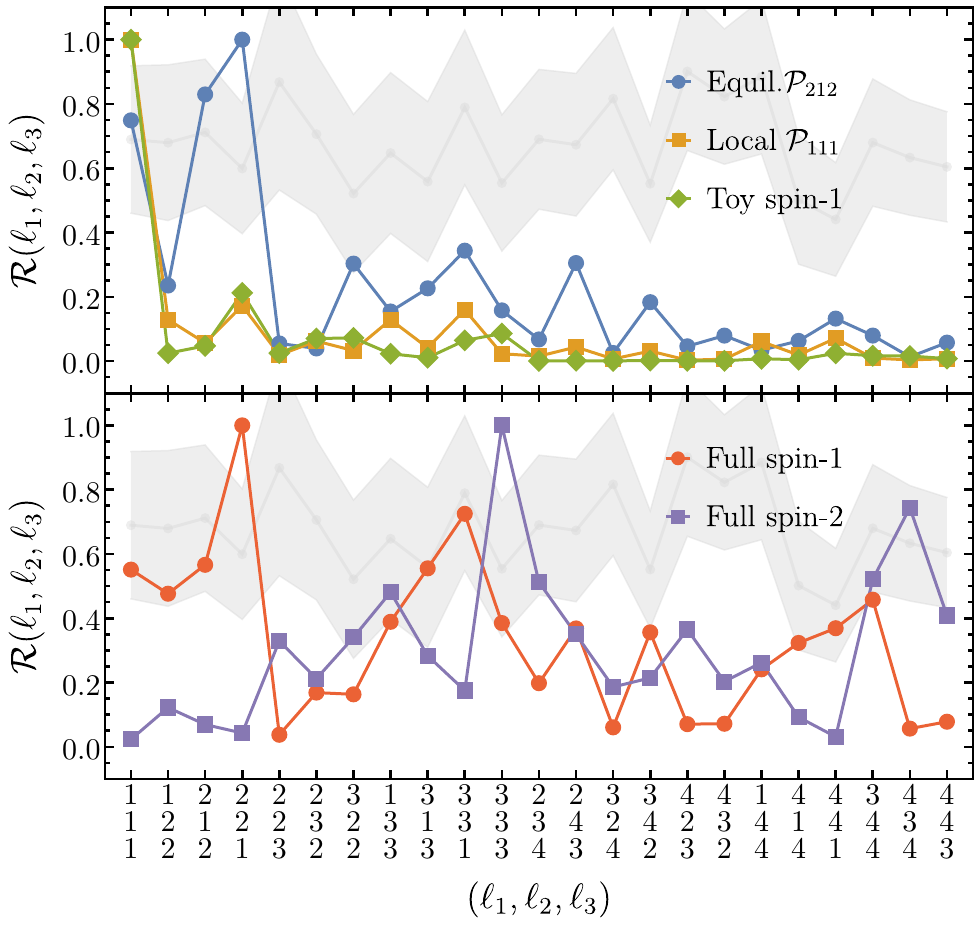}
    \caption{Angular dependence of parity-violating galaxy 4PCF coefficients for the toy and full models discussed in Sec.~\ref{sec:theoreticalModels}. For each model, we show the relative size $\mathcal R (\ell_1, \ell_2, \ell_3)$ defined by~\cref{eq:relative}. Contact-like toy models, equil.$\times \mathcal P_{212}$, local$\times \mathcal P_{111}$, and toy spin-1, are shown as blue, yellow, green lines respectively in the upper panel, while full models, spin-1 and  spin-2, are red and purple lines in the lower panel. The coefficient for contact-like models concentrates in a few angular bins, whereas a full model exhibits nontrivial angular correlations distributed across a broader range of angular bins. The gray line indicates the averaged BOSS CMASS data with respect to the North and South Galactic Caps, along with the $1\sigma$ error band due to the diagonal entries in the covariance matrix estimated according to \textsc{MultiDark-Patchy} mocks.}
    \label{fig:fullVsToy_ang_dep}
\end{figure}

Fig.~\ref{fig:fullVsToy_ang_dep} shows the angular dependence of galaxy 4PCF coefficients across different models. For each model, we calculate the relative size of various angular bins $\mathcal{R}$,
\beq
\mathcal{R}(\ell_1, \ell_2, \ell_3) \defeq  \frac{\sum_{(b_1,b_2, b_3)} \left|\zeta_{\ell_1,\ell_2, \ell_3} (b_1, b_2, b_3)\right| }
{ \max_{(\ell_1,\ell_2, \ell_3)} \left\{\sum_{(b_1,b_2, b_3)} \left|\zeta_{\ell_1,\ell_2, \ell_3} (b_1,b_2, b_3)\right|\right\} },
\label{eq:relative}
\eeq
defined as the absolute value of the summed coefficients over all radial bins within each angular bin, then normalized by the highest sum across angular bins. This normalized sum captures the typical correlation strength within each angular bin. Fig.~\ref{fig:fullVsToy_ang_dep} shows that the tensor structures in the input primordial trispectra impact the galaxy 4PCF. Generally, trispectra with specific combination of multipole moment indices of the tensor structure lead to peak correlations (or peak anti-correlations) around the corresponding angular bin multipoles of the galaxy 4PCF, e.g.,  local$\times \iso{111}(\vb{k}_1, \vb{k}_3, \vb{k}_s)$ has the largest coefficient at $(\ell_1, \ell_2, \ell_3) =(1,1,1)$ while equil.$\times \mathcal P_{212}(\vb{k}_1, \vb{k}_3, \vb{k}_s)$ has the largest coefficient at $(\ell_1, \ell_2, \ell_3) =(2,2,1)$.\footnote{Additional angular dependence in the galaxy 4PCF, such as that introduced by redshift distortion (see \cref{eq:rsd}), can influence the correspondence between the angular bins of the trispectrum and those of the galaxy 4PCF.} Thus, observing large correlations in specific angular bins ($\ell_1$, $\ell_2$, $\ell_3$) hints us to consider primordial interaction models with tensor structure $\iso{L_1,L_2,L_3}(\hat{\vb{k}}_1, \hat{\vb{k}}_3, \hat{\vb{k}}_s)$ (or similar tensor structures by permuting over $L_i$'s).

Another interesting observation is that toy models from contact interactions concentrate their correlations within a few angular bins. For instance, the 4PCF of local$\times \iso{111}(\vb{k}_1, \vb{k}_3, \vb{k}_s)$ (see~\cref{eq:local-1}) shows strong correlations only in the angular bin $(\ell_1, \ell_2, \ell_3) = (1,1,1)$, with much smaller values elsewhere. Other toy models exhibit similar behavior, as seen in the upper panel of \cref{fig:fullVsToy_ang_dep}. By contrast, both full spin-1 and spin-2 models display large (anti-)correlations across many angular bins, with values more widely distributed than the toy models, which generally have lower values outside a few peaks.

\subsubsection{Radial Distribution}

\begin{figure}[!bp]
    \centering
    \includegraphics[width=0.96\linewidth]{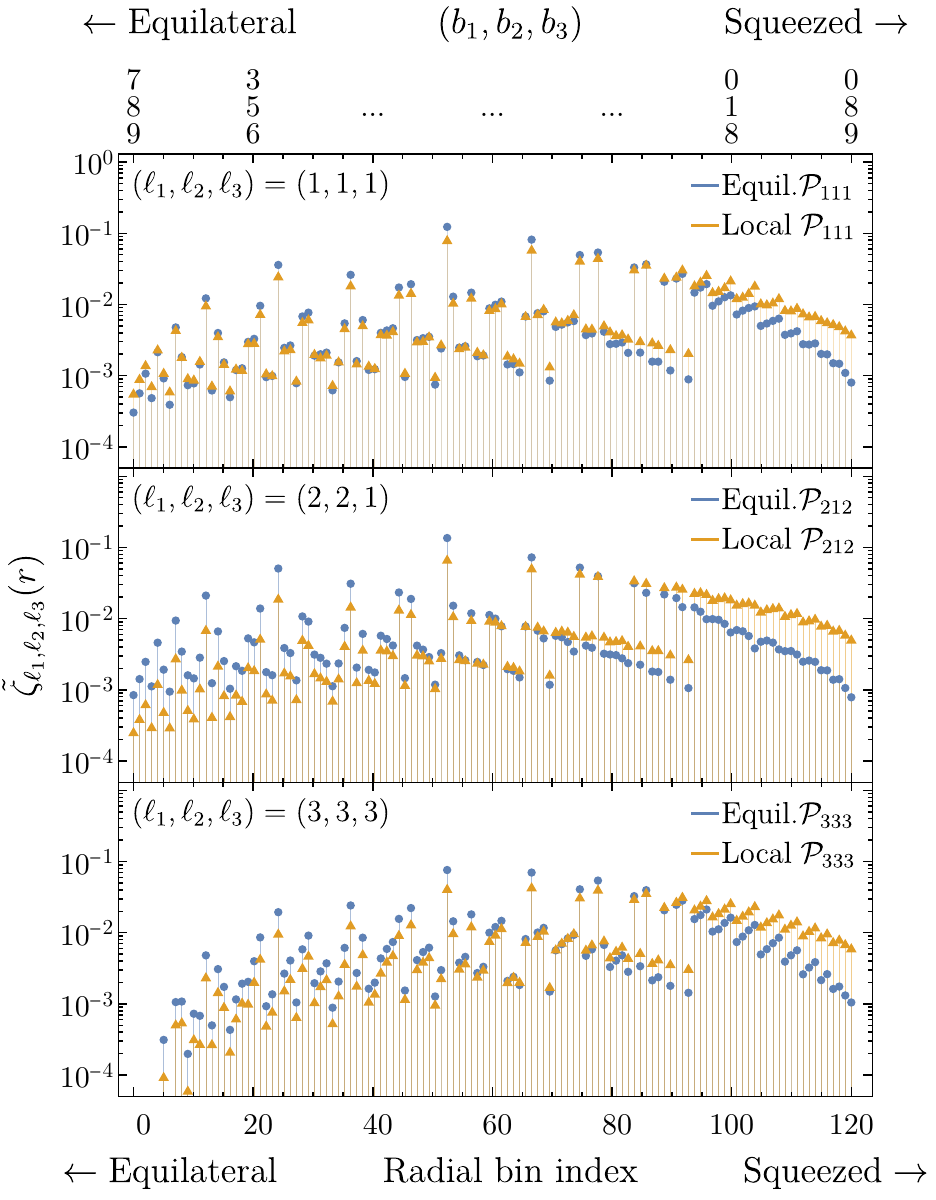}
    \caption{The radial dependence of 4PCF templates for three pairs of equilateral vs. local toy models with angular-dependent factor $\mathcal{K} \propto \iso{111}(\hat{\vb{k}}_1, \hat{\vb{k}}_3, \hat{\vb{k}}_s)$, $\iso{212}(\hat{\vb{k}}_1, \hat{\vb{k}}_3, \hat{\vb{k}}_s)$, and $\iso{333}(\hat{\vb{k}}_1, \hat{\vb{k}}_3, \hat{\vb{k}}_s)$ from the top to the bottom, respectively. The angular bin with the largest signal for both local and equilateral templates is selected ($\zeta_{111}$, $\zeta_{221}$ and $\zeta_{333}$) and we normalize the correlation coefficients according to~\cref{eq:normalized-coeff}. The radial bin index is ordered from a more equilateral configuration with a small index to a more squeezed configuration with a large index. Appropriate negative signs are introduced when plotting mainly anticorrelated signals so that they appear in the logarithmic scale. }
    \label{fig:toyModel_rad_dep}
\end{figure}

Galaxy 4PCFs not only provide insights into tensor structures but also provide the shape information of the trispectra through its radial distribution. Recall that we  select radial bins $(b_1, b_2, b_3)$ with the restriction $b_1 < b_2 < b_3$ ($b_i = 0, ..., 9$). We sort these bins by the ratios $(b_3 + b_2 + 2)/(b_1 + 1)$ and $(b_3 + 1)/(b_2 + 1)$, with lower ratios representing the 4PCF tetrahedron configuration close to equilateral ($b_1 \sim b_2 \sim b_3$) and higher ratios representing squeezed configuration ($b_1 \ll b_2 \ll b_3$). Specifically, the bins are ordered as $(b_1, b_2, b_3) = (7,8,9), (6, 7, 8)\cdots (0,7,9), (0,8,9)$. The averaged radii for all the bin configurations are reported in the first panel of~\cref{fig:LMAXDep}.

In \cref{fig:toyModel_rad_dep}, we compared several equilateral-shape toy models with local-shape ones, each multiplied by an angular-dependent factor $\mathcal K$: $\iso{111}(\hat{\vb{k}}_1, \hat{\vb{k}}_3, \hat{\vb{k}}_s)$, $\iso{212}(\hat{\vb{k}}_1, \hat{\vb{k}}_3, \hat{\vb{k}}_s)$, and $\iso{333}(\hat{\vb{k}}_1, \hat{\vb{k}}_3, \hat{\vb{k}}_s)$. We report the coefficients for the three angular bins $(\ell_1, \ell_2, \ell_3) = (1,1,1), (2,2,1), (3,3,3)$ for the three $\mathcal K$'s respectively, where both local and equilateral models have peak (anti-)correlations. We further normalize these coefficients in that angular bin by its area under the curve,
\beq
    \tilde{\zeta}_{\ell_1, \ell_2, \ell_3}(\vb{r}) \defeq \frac{\zeta_{\ell_1,\ell_2, \ell_3} (b_1, b_2, b_3)}{ \sum_{(b_1,b_2, b_3)} \abs{\zeta_{\ell_1,\ell_2, \ell_3} (b_1,b_2, b_3)} }.
    \label{eq:normalized-coeff}
\eeq

Comparing the local-shape and equilateral-shape models, we observe the distributions in the coefficients of the local-shape models and the equilateral-shape models generally follow each other. The difference lies in that the local-shape models generally produce an enhanced correlation when one galaxy pair distance is much larger than others, reflecting a squeezed configuration. In contrast, equilateral-shape 4PCFs show a milder enhancement when the three radial distances are similar. 

\begin{figure}[!bp]
    \centering
    \includegraphics[width=0.96\linewidth]{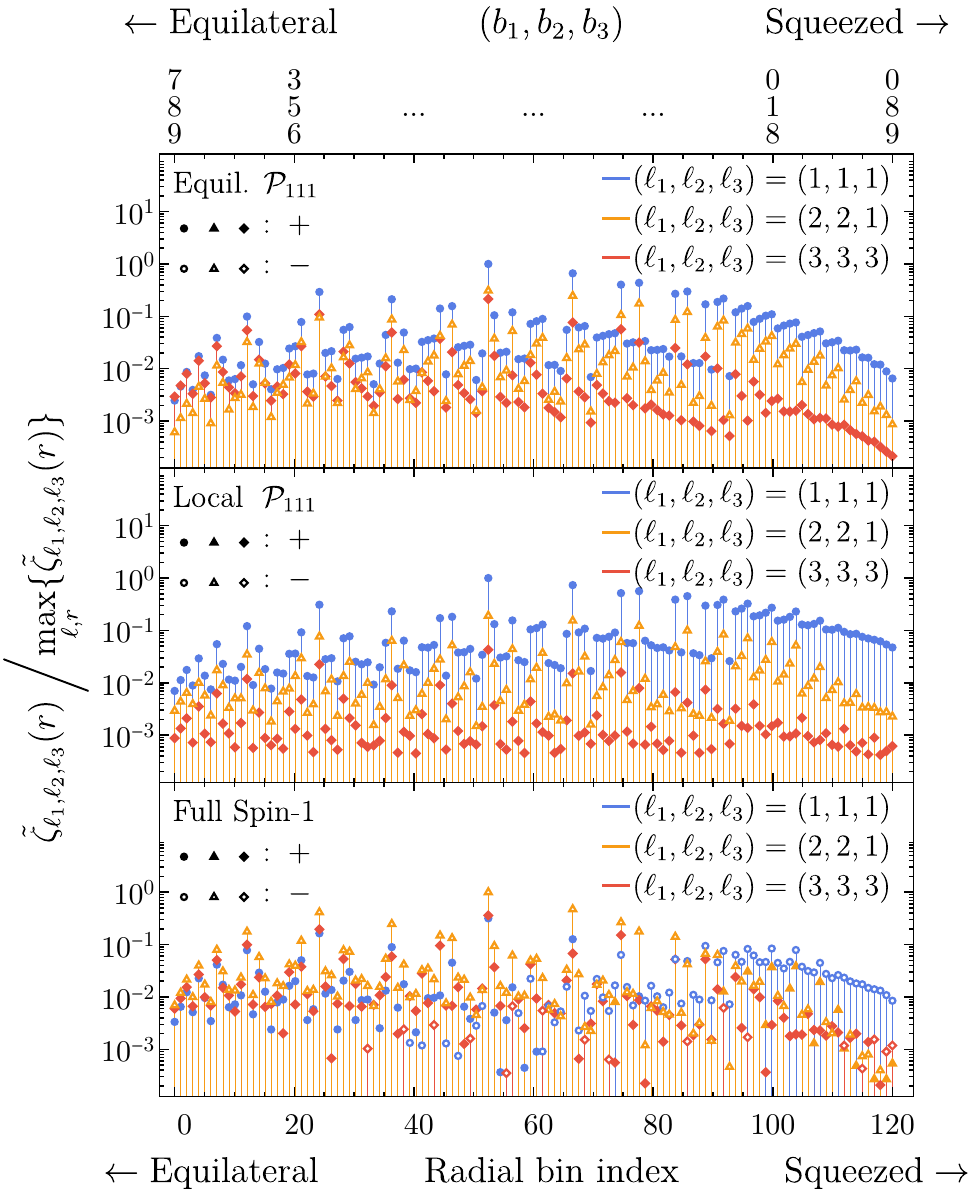}
    \caption{The radial dependence of 4PCF templates between two canonical toy models and the full spin-1 model: The vertical axis and the radial bins are the same as that in \cref{fig:toyModel_rad_dep}. However, we used solid markers to indicate positive correlations and open markers to indicate negative correlations. Also, note that three panels are now organized by templates instead of angular bin index $(\ell_1, \ell_2, \ell_3)$. We normalize each panel by the maximal correlation coefficient shown here for each template so that a comparison across angular bins can also be inferred.}
    \label{fig:toyvsFull_rad_dep}
\end{figure}

\subsubsection{Features of the Full Models}

Beyond the toy models, the full models show more complex patterns in the angular and radial distributions of the coefficients as shown in~\cref{fig:fullVsToy_ang_dep} and~\cref{fig:toyvsFull_rad_dep}, respectively. 
\begin{itemize}

\item \textbf{Angular distribution.} In~\cref{fig:fullVsToy_ang_dep}, we found two distinctions:  (1) Comparing the two spins, we see that significant coefficients of the spin-2 model are distributed at higher-$\ell$ bins than those of the spin-1 model. (2) Full model coefficients exhibit strong (anti-)correlations across a broader range of angular bins, while the contact-interaction toy model only shows strong (anti-)correlations in a few angular bins. 

The distinction (1) arises because the spin-2 trispectra contains isotropic basis functions with higher $\ell$'s than that of the spin-1 trispectra, as shown in~\cref{eq:full_spin1} and~\cref{eq:full_spin2}. The distinction (2) arises because, in contact interactions, the dynamical part of the trispectra (see \cref{eqn:dynamical}) generally has a simple form:
\begin{equation}
	\mathcal{J}(k_1, \ldots, k_s) = \int \dd \tau_L \dd \tau_R\; \mathcal{J}_L(k_1, k_2, k_s, \tau_L) f(k_s) \delta(\tau_L - \tau_R) \mathcal{J}_R(k_3, k_4,k_s, \tau_R),
\end{equation}
where $f(k_s)$ is at most a power function of $k_s$, and $\delta(\tau_L - \tau_R)$ enforces the contact interaction to be local, \textit{i.e.} on the same time slice. This leads to a straightforward angular dependence of the external momenta $k_i$, resulting in a simple angular dependence of the galaxy 4PCF. In contrast, the full model often includes more intricate propagators, such as Hankel or Whittaker functions in $\Pi(k_s, \tau_L, \tau_R)$ and Heaviside step functions in $\tau_L$ and $\tau_R$. These complex structures introduce a more complicated dynamical part as a function of $k_s$. After integrating over $k_s = \sqrt{k_1^2 + k_2^2 + 2 \vb{k}_1 \cdot \vb{k}_2}$, we obtain a complex angular dependence on the external momenta $k_i$, resulting in a nontrivial angular dependence of galaxy 4PCF. 

Thus, observing significant (anti-)correlations across multiple angular bins may signal primordial particle production, while concentration on a few angular bins suggest contact interaction. For the former scenario, the peak value of the distribution hints the possible spin of the exchanged massive particle.

\item \textbf{Radial distribution.} As shown in~\cref{fig:toyvsFull_rad_dep}, we again observed that the radial distribution of the \emph{absolute} coefficients of the full spin-1 model largely follows those of the canonical toy models. However, unlike the toy models, whose coefficients are predominantly either negative or positive, the coefficients of the full spin-1 model exhibit fluctuations between positive and negative values. These fluctuations may be attributed to the oscillatory propagators of the trispectrum. 

While quantitatively comparing the difference between toy and full templates in position space is challenging, one may compare the dynamical part $\mathcal{J}(k_1, \ldots, k_4, k_s)$ of trispectra directly in momentum space as shown in \cref{fig:compare_trispectra}. This was previously challenging to perform as the analytical result for the full model is usually not known. Fortunately, this has been computed exactly in a recent study \cite{Qin:2022fbv}. We fix $k_2 = k_3 = k_4 = (80 h^{-1} \text{Mpc})^{-1}$ and $k_s = (100 h^{-1} \text{Mpc})^{-1}$ to show the $k_1$ dependence only. Nonetheless, one notices that the two toy models peak at different $k_1$ while other external and propagator momenta are fixed. More importantly, the oscillatory feature in the full model due to heavy $Z$ exchange cannot be easily reproduced by either toy model. This further illustrates the complementarity between evaluating the full template and performing the quick canonical toy template fitting.

\begin{figure}
    \centering
    \includegraphics[width=0.9\linewidth]{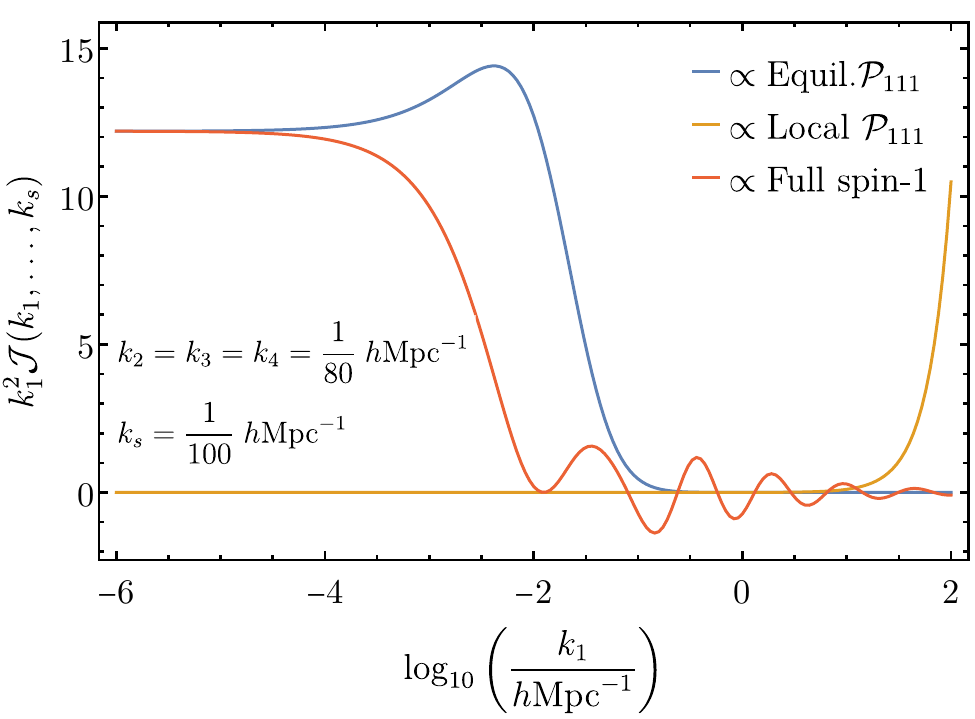}
    \caption{Comparison of the dynamical part of the trispectra: The vertical axis is proportionate to $k_1^2 \mathcal{J}(k_1, \ldots, k_s)$ in which $\mathcal{J}(k_1)$ is the radial part of the trispectra. The $k_1^2$ scaling and overall scaling factors are chosen so that $\mathcal{J}(k_1, \ldots, k_s)$ is visible on the plotted range. We fix $k_2 = k_3 = k_4 = (80 h^{-1} \text{Mpc})^{-1}$ and $k_s = (100 h^{-1} \text{Mpc})^{-1}$ to show the $k_1$ dependence only. Note that, in the squeezed limit, the unique oscillatory feature in the full model comes from the exchange of heavy particles (spin-1 gauge boson) during inflation.}
    \label{fig:compare_trispectra}
\end{figure}

Furthermore, as we go to higher $\ell$ bins, a substantial suppression of the squeezed-limit coefficients occurs for the Equil.$\times \mathcal{P}_{111}$ and the full spin-1 model, while the equilateral-limit coefficients largely remain unaffected by the change in $\ell$. See the first and third panels of~\cref{fig:toyvsFull_rad_dep}. Such a feature does not occur for the Local$\times \mathcal{P}_{111}$ model, where all the coefficients are uniformly suppressed regardless of radial configuration in higher $\ell$ bins. See the second panel of~\cref{fig:toyvsFull_rad_dep}. Similar features in the radial distributions are observed for the full spin-2 model.
\end{itemize}

Our analysis shows that the 4-point correlation function (4PCF) coefficients' distributions -- both angular and radial -- provide insights into the primordial trispectrum's angular ($\mathcal{K}$) and radial ($\mathcal{J}$) components, helping us understand potential primordial interactions. It will be an interesting study to see how the distributions of the 4PCF coefficient change under different mass parameters and chemical potentials. Given the limited computational resources, we will leave it for future exploration. 

\subsection{Comparison with BOSS Data \label{sec:compareWithBOSS}}
Each theoretical model considered in \cref{sec:theoreticalModels} has a model-dependent overall magnitude, $\mu_A$, which will be inferred by comparing the scaled templates to the observed BOSS data.%
\footnote{
Here, we are treating the overall magnitude $\mu_A$ as an arbitrary freedom of the template shape to be fitted. However, once a statistically significant shape is identified, one may attempt to further interpret $\mu_A$ as the model-dependent prefactor of the trispectrum. For instance, as shown in \cref{eq:full_spin1}, the amplitude of the parity-violating part of the trispectrum for a full spin-$1$ exchange is proportionate to $\qty[\rho_{1,Z} \rho_2/(\dot{\phi}_0 H \mu_H^2)]^2$ which depends on the inflaton-$Z$ couplings shown in \cref{eqn:spin1InteractionModelLag}. Since the trispectrum depends on the non-negative squared couplings, one may even expect the sign of $\mu_A$ to provide information about whether the positive or negative helicity state of the spin-$1$ particle is enhanced, hence probing the sign of the chemical potential. We leave its precise connection to model parameters for a future study. More detail about how $\mu_A$ is defined for each model can be found in \cref{sec:theoreticalModels}.
}
For the comparison, we adopt the existing compressed $\chi^2$ analysis pipeline and covariance matrix from Refs.~\cite{Philcox:2022hkh, Hou:2022wfj}. The latter is estimated using \textsc{MultiDark-Patchy} mock galaxy catalogs \cite{Kitaura:2015uqa, Rodriguez-Torres:2015vqa}. To perform Markov chain Monte Carlo on $\mu_A$, we compress the data by projecting the galaxy 4PCF templates and the observed data onto the first $N_\text{eig} \sim \order{10 - 100}$ eigenvectors of the covariance matrix with the lowest eigenvalues, corresponding to basis functions combinations with smallest uncertainties.

\begin{figure}[!t]
    \centering
    \includegraphics[width=0.96\linewidth]{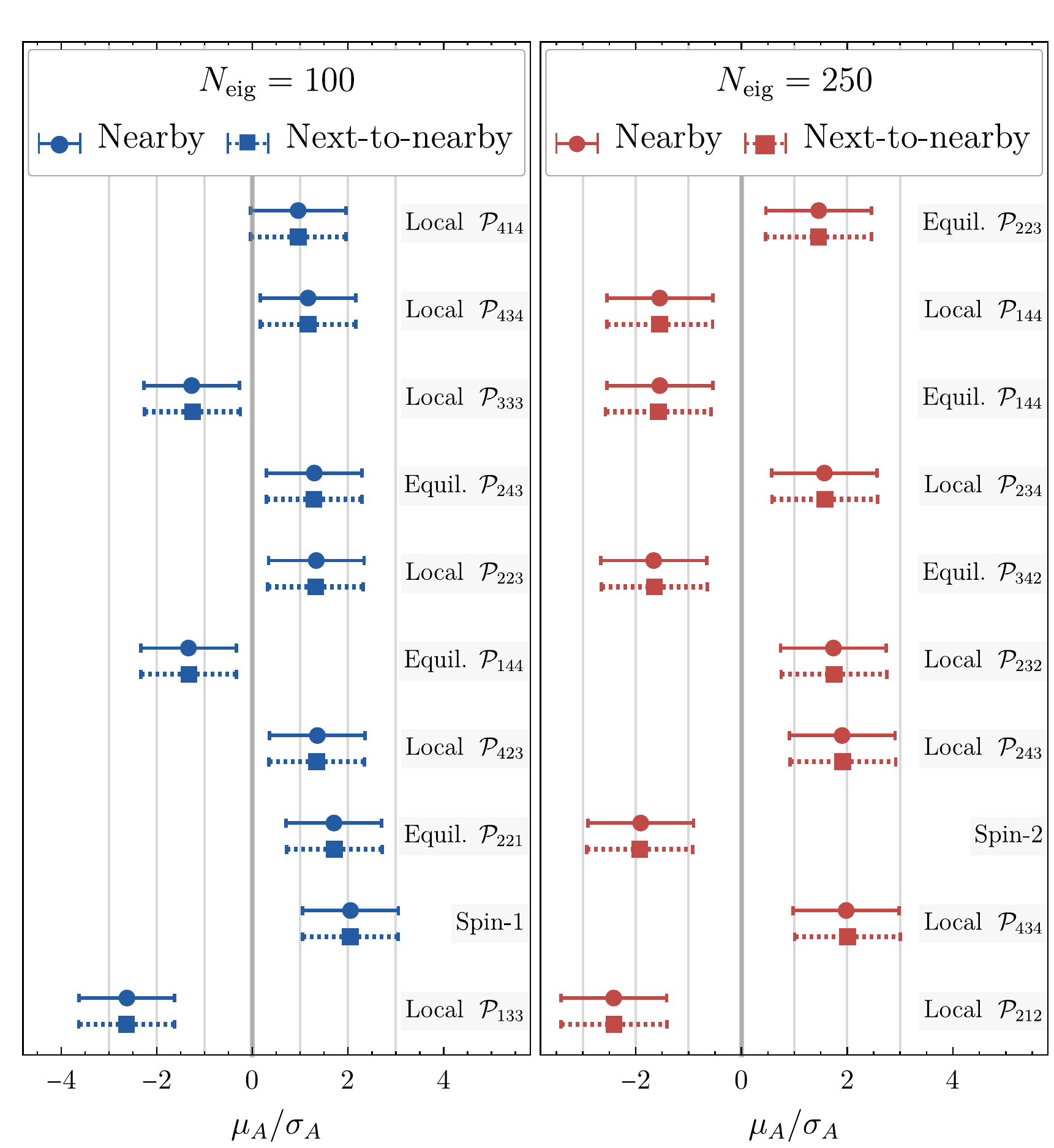}
    \caption{Significance of various galaxy 4PCF templates against the observed BOSS data. The error bars are fixed to be 1 to provide a visual guide. We show ten models with the most significant nonzero overall magnitude over its standard derivation for each compression strategy. However, no detection beyond the $3\sigma$ level has been found. \textbf{Left}: Comparison between nearby radial filtering (solid bars) and next-to-nearby radial filtering (dotted bars) schemes with $N_\text{eig} = 100$. \textbf{Right}: The same as left but with $N_\text{eig} = 250$.}
    \label{fig:sigBOSS}
\end{figure}

\begin{figure}[!t]
 \centering
    \includegraphics[width=0.48\linewidth]{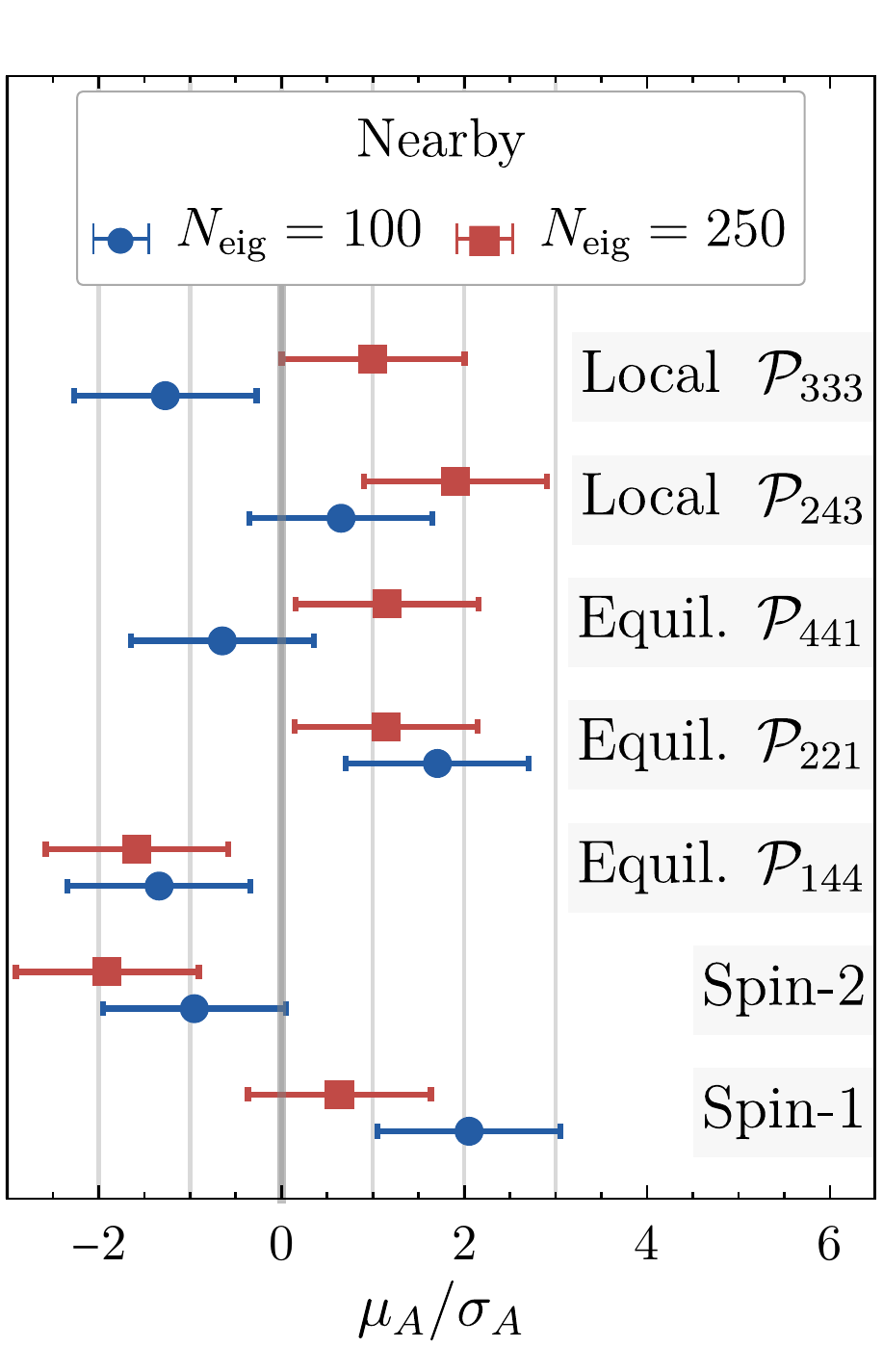}
     \caption{Comparison of $\mu_A/\sigma_A$ between nearby radial filtering compression with $N_\text{eig} = 100$ and that with $N_\text{eig} = 250$ for the same model. 
     }
\label{fig:sigBOSS2}
\end{figure}

We considered four compression schemes: (1) $N_\text{eig} = 100$ with nearby radial filtering, (2) $N_\text{eig} = 100$ with next-to-nearby radial filtering, (3) $N_\text{eig} = 250$ with nearby radial filtering, and (4) $N_\text{eig} = 250$ with next-to-nearby radial filtering. Increasing $N_\text{eig}$ yields more data points for fitting but may introduce additional noise bias in sample covariance estimation~\cite{Philcox:2021hbm}. As for the radial filtering, recall from the previous section that we demand $b_1 < b_2 < b_3$ as a canonical ordering of radial bins of 4PCF, leaving $120$ radial bins per angular bin. This is what we will call the ``nearby radial filtering". To further minimize late-time small-scale effects, Ref.~\cite{Philcox:2022hkh} suggests enforcing $b_1 < b_2 - 1 < b_3 - 2$, creating 56 radial bins per angular bin. We call this the ``next-to-nearby radial filtering".

We measure the ratio of the fitted central value of the overall magnitude to its standard deviation, $\sigma_A$, as an indicator of detection significance. Among the 49 theoretical models investigated, \cref{fig:sigBOSS} highlights the ten models exhibiting the largest deviations from zero for each compression strategy and \cref{tab:sigBOSS} shows the $\mu_A/\sigma_A$ values for the top five models in each case explicitly. No strong deviation ($|\mu_A/\sigma_A|\geq 3$) from the null hypothesis is observed in the BOSS data. As seen in the figure, different radial filtering schemes do not significantly affect the inferred significance. However, when we compare the same model and radial filtering scheme but use different eigenvalue compression schemes, some models display a significant discrepancy in the inferred significance, as shown in \cref{fig:sigBOSS2}. This discrepancy may arise because removing eigenvectors associated with larger eigenvalues can discard substantial information about parity violation contained within them~\cite{Philcox:2022hkh}. It highlights the need for a more robust data analysis approach. Nevertheless, through the exercise, we demonstrate that full spin-1 and spin-2 model templates can be effectively compared to observed galaxy survey data.

 \begin{table}
     \centering
     \begin{tabular}{c | *{4}{c} | *{2}{c} }
         \hline
        \multicolumn{7}{c}{$N_\text{eig} = 100$, Nearby radial filtering} \\
         \hline
         Model & Local $\iso{133}$ & Equil. $\iso{211}$ & Local $\iso{423}$ & Equil. $\iso{144}$ & Spin-1 & Spin-2 \\
         $\abs{\mu_A/\sigma_A}$ & $2.62$ & $1.71$ & $1.36$ & $1.34$ & $2.05$ & $0.95$ \\
         \hline 
         \multicolumn{7}{c}{$N_\text{eig} = 100$, Next-to-nearby radial filtering} \\
         \hline
         Model & Local $\iso{133}$ & Equil. $\iso{211}$ & Local $\iso{423}$ & Equil. $\iso{144}$ & Spin-1 & Spin-2 \\
         $\abs{\mu_A/\sigma_A}$ & $2.63$ & $1.72$ & $1.34$ & $1.33$ & $2.05$ & $0.94$ \\
         \hline 
         \hline
         \multicolumn{7}{c}{$N_\text{eig} = 250$, Nearby radial filtering}\\
         \hline
         model & Local $\iso{212}$ & Local $\iso{434}$ & Local $\iso{243}$ & Local $\iso{232}$ & Spin-1 & Spin-2 \\
         $\abs{\mu_A/\sigma_A}$ & $2.42$ & $1.98$ & $1.90$ & $1.74$ & $0.63$ & $1.90$ \\
         \hline 
         \multicolumn{7}{c}{$N_\text{eig} = 250$, Next-to-nearby radial filtering}\\
         \hline
         model & Local $\iso{212}$ & Local $\iso{434}$ & Local $\iso{243}$ & Local $\iso{232}$ & Spin-1 & Spin-2 \\
         $\abs{\mu_A/\sigma_A}$ & $2.41$ & $2.01$ & $1.92$ & $1.75$ & $0.65$ & $1.92$ \\
         \hline
     \end{tabular}
     \caption{Significance of various position-space templates of 4PCF against the observed BOSS data: Four toy models whose fitted overall magnitude deviates from zero the most in units of its standard deviation along with the two full models are presented. However, there is no strong tension for these models beyond $3\sigma$ level according to BOSS data. } 
     \label{tab:sigBOSS}
 \end{table}

\section{Conclusion and Discussion \label{sec:conclusion}}
In this work, we studied various theoretical models that may produce interesting parity-violating trispectra, from canonical trispectra models, such as local-shape-like or equilateral-shape-like models, to full spin-1 and spin-2 exchange models with chemical potential enhancement. We also introduced a new class of models by replacing the propagators of the exchanged particle in the large-mass limit to produce a contact-interaction-like trispectrum. Translating these trispectra into position-space templates is usually difficult because of the high-dimensional Fourier transform and handling of singular momentum-conserving delta functions. We presented a general formalism that enables the evaluation of these Fourier transforms of tree-level exchange processes, by splitting the exchange-type trispectra using time integrals, carrying out analytical angular integrals, and performing lower-dimensional numerical integrals on radial-dependent factors to maximally utilize the factorizability of the target trispectrum. Such a pipeline shows how one may fit higher-point LSS survey data to theoretical models of primordial interactions. We also discussed qualitative features of galaxy 4PCF correlation coefficients for a given primordial trispectrum and how the observed distribution of the correlation coefficients can be used to guide further model buildings of these parity-violating signals. While the BOSS data does not provide strong evidence of detecting any theoretical models considered in this study, we look forward to new models and a more robust data analysis that may shed light on the reported parity-violating signals. 

From this study, we also found that the massive spinning particle production with chemical potential enhancement is a promising target for future primordial non-Gaussianity searches. Their trispectra can be both sizable and parity-violating, and they can be naturally expressed in factorizable form for computing the galaxy 4PCF. We anticipate the upcoming LSS missions will place stronger constraints on such models or even uncover potential evidence of their distinctive signatures. 

As extensions to this study, one may consider carrying out a parameter scan on the particle mass and chemical potential of the full model to see if BOSS data prefers other shapes of the full model. From a theoretical perspective, a full model with chemical potential enhancement should also show up in other correlation functions, such as in the bispectrum. Therefore, a joint analysis of 3PCF, parity-preserving 4PCF, and parity-violating 4PCF on model parameters may also be interesting. We also mentioned that full models seem to encode nontrivial angular dependence beyond contact-interaction-like toy models. This raises the question of whether there are criteria for determining when full-model template fitting is indispensable, as opposed to the more efficient fitting that uses templates based on fully factorizable functions approximating the full model. The answer lies in how to quantify the ``goodness of fit" of the full model and its approximation. In the future, it may be interesting to study whether it is feasible to ``unfold" the noisy data from position space back to the distributions in momentum space which typically interfaces with phenomenological model building better using techniques such as machine learning.

\acknowledgments
We thank Jiamin Hou, Wayne Hu, Austin Joyce, Hayden Lee, Oliver Philcox, and Xi Tong for useful discussions. This work was completed with resources provided by the University of Chicago’s Research Computing Center. YB and LTW are supported by the Department of Energy grant DE-SC0013642. ZX is supported by NSFC under Grants No.\ 12275146 and No.\ 12247103, the National Key R\&D Program of China (2021YFC2203100), and the Dushi Program of Tsinghua University. YZ is supported by the GRF Grants No. 11302824 and No. 11310925 from the Hong Kong Research Grants Council and the Grant No. 9610645 and No. 7020130 from the City University of Hong Kong. YZ acknowledges the Aspen Center for Physics, which is supported by NSF grant PHY-2210452, for hospitality during the initial phase of the work.

\section*{Data Availability Statements}
Some codes and data corresponding to the finding of this paper are available at the GitHub repository \cite{GitHub_repo}, and others are cited in the paper. 

\appendix

\section{Detailed Computation for the 4PCF \label{app:detail4PCF}}

The inflaton fluctuation $\phi$ with wave vector $\vb{k}$ is related to the galaxy correlation $\zeta_g$ at redshift $z$ by 
\begin{equation}
	\zeta_g(\vb{k}, z) = -  Z_1(\hat{\vb{k}}, z) M(k, z) \frac{H}{\dot{\phi}_0} \phi(\vb{k}),
\end{equation}
in which $Z_1$ is the Kaiser redshift-space distortion~\cref{eq:rsd}, $M$ is the transfer function (linear region), $H$ is the Hubble scale during inflation, and $\dot{\phi}_0 \sim 3600 H^2$ is the rolling speed of the inflaton background. See,~e.g.,~\cite{dodelson2020modern} for a review on galaxy clustering and growth of structure.

The position-space galaxy 4PCF is related to the momentum-space trispectrum, $\ev{\phi^4}'$,  by a Fourier transformation
\begin{equation}
	\begin{multlined}
		\zeta_g^4(\vb{r}_1, \vb{r}_2, \vb{r}_3, \vb{r}_4) = \frac{H^4}{\dot{\phi}_0^4} 
		\int \qty[ \prod_{i = 1}^{4} \dbar^3 k_i \; Z_1(\hat{\vb{k}}_i, z) M(k_i, z) ] 
		\ev{\phi^4}'(\vb{k}) \\
		\times \exp(\sum_{i = 1}^4 \ii\vb{k}_i \cdot \vb{r}_i) \qty(2\pi)^3 \delta[3](\sum_{i=1}^4 \vb{k}_i),
	\end{multlined}
\label{eq:fourier}
\end{equation}
where $\dbar^3 k_i \defeq \dd^3 k_i / \qty(2\pi)^3$. As discussed in Sec.~\ref{sec:4PCFGeneralStrategy}, we adopt the isotropic basis function formalism in computing the galaxy 4PCF. 

The derivation below largely follows Refs.~\cite{Cahn:2020axu,Philcox:2021hbm, Philcox:2021bwo, Cahn:2021ltp, Philcox:2022hkh, Hou:2022wfj}. Here, we show the derivation for the readers' convenience. Under the formalism, the correlation function can be expressed as 
\begin{equation}
    \zeta_g^4(\vb{r}_1, \vb{r}_2, \vb{r}_3) = \sum_{\ell_1, \ell_2, \ell_3} \zeta_{\ell_1, \ell_2, \ell_3}(r_1, r_2, r_3) \iso{\ell_1, \ell_2, \ell_3}(\hat{\vb{r}}_1,\hat{\vb{r}}_2,\hat{\vb{r}}_3),
\end{equation}
where $\iso{\ell_1, \ell_2, \ell_3}(\hat{\vb{r}}_1,\hat{\vb{r}}_2,\hat{\vb{r}}_3)$ denotes the isotropic $3$-point basis function for the unit vector $\hat{\vb{r}}_{1,2,3}$ that is defined by~\cref{eq:3pt}. The information of the 4PCF is encoded in the coefficient of the basis functions, $\zeta_{\ell_1, \ell_2, \ell_3}(r_1, r_2, r_3)$. 

For the isotropic basis function, the parity operation, $\mathbb P$, is a reflection of $\hb r_{1,2,3}$ to $-\hb r_{1,2,3}$ on top of a co-rotation, which the isotropic basis function is invariant under. Under the parity $\mathbb P$, the isotropic basis function satisfies the relations, 
\begin{align}
& \mathbb P [\iso{\ell_1, \ell_2, \ell_3}(\hat{\vb{r}}_1,\hat{\vb{r}}_2,\hat{\vb{r}}_3)]=\iso{\ell_1, \ell_2, \ell_3}(-\hat{\vb{r}}_1,-\hat{\vb{r}}_2,-\hat{\vb{r}}_3) \nonumber \\
={}&(-1)^{\ell_1+\ell_2+\ell_3}\iso{\ell_1, \ell_2, \ell_3}(\hat{\vb{r}}_1, \hat{\vb{r}}_2,\hat{\vb{r}}_3) = \iso{\ell_1, \ell_2, \ell_3}^*(\hat{\vb{r}}_1,\hat{\vb{r}}_2,\hat{\vb{r}}_3),
\end{align}
where the second lines follow the properties of spherical harmonics. 
Generically, one can split a generic $\zeta_g^4(\vb{r}_1, \vb{r}_2, \vb{r}_3)$ into an even sum and odd sum of $\ell_1+\ell_2+\ell_3$, i.e.,
\begin{align}
\zeta_g^4(\vb{r}_1, \vb{r}_2, \vb{r}_3) ={}& \sum_{\ell_1+\ell_2+\ell_3 = \text{even}} \zeta_{\ell_1, \ell_2, \ell_3}(r_1, r_2, r_3) \iso{\ell_1, \ell_2, \ell_3}(\hat{\vb{r}}_1,\hat{\vb{r}}_2,\hat{\vb{r}}_3) \nonumber\\
&+ \sum_{\ell_1+\ell_2+\ell_3=\text{odd}} \zeta_{\ell_1, \ell_2, \ell_3}(r_1, r_2, r_3) \iso{\ell_1, \ell_2, \ell_3}(\hat{\vb{r}}_1,\hat{\vb{r}}_2,\hat{\vb{r}}_3),
\end{align}
where the first and second lines correspond to the parity-preserving and -violating part of $\zeta_g^4(\vb{r}_1, \vb{r}_2, \vb{r}_3)$, i.e. $\mathbb P[\zeta_g^4(\vb{r}_1, \vb{r}_2, \vb{r}_3)_\text{even/odd}] = \pm \zeta_g^4(\vb{r}_1, \vb{r}_2, \vb{r}_3)_\text{even/odd}$. We will focus on the parity-violating part of 4PCF. Under $\ell_1+\ell_2+\ell_3=\text{odd}$, $\iso{\ell_1, \ell_2, \ell_3}(\hat{\vb{r}}_1,\hat{\vb{r}}_2,\hat{\vb{r}}_3)$ is imaginary while $\zeta_g^4(\vb{r}_1, \vb{r}_2, \vb{r}_3)$ is real. Hence the coefficient for the parity-violating 4PCF, $\zeta_{\ell_1,\ell_2, \ell_3}$ is also imaginary.

Using the completeness relation of isotropic basis functions, one can obtain the correlation coefficients via 
\begin{equation}
	\zeta_{\ell_1, \ell_2, \ell_3}(r_1, r_2, r_3) = \int \qty[\prod_{i = 1}^3 \dd \hat{\vb{r}}_i] \; \zeta_g^4(\vb{r}_1, \vb{r}_2, \vb{r}_3) \iso{\ell_1, \ell_2, \ell_3}^*(\hat{\vb{r}}_1,\hat{\vb{r}}_2,\hat{\vb{r}}_3),
 \label{eq:coefficient}
\end{equation}
where $\dd \hat{\vb{r}}_i$ denotes the integration measure of the solid angle associated with the $i$\textsuperscript{th} coordinates. Substituting (\ref{eq:fourier}) to (\ref{eq:coefficient}), the 4PCF coefficient is related to the trispectrum by
\begin{equation}
	\begin{multlined}
		\zeta_{\ell_1, \ell_2, \ell_3}(r_1, r_2, r_3) = \frac{H^4}{\dot{\phi}_0^4} 
		\int \qty[\prod_{i = 1}^3 \dd \hat{\vb{r}}_i] 
		\qty[ \prod_{i = 1}^{4} \dbar^3 k_i \; Z_1(\hat{\vb{k}}_i, z) M(k_i, z) ] \\
		\times \iso{\ell_1, \ell_2, \ell_3}^*(\hat{\vb{r}}_1,\hat{\vb{r}}_2,\hat{\vb{r}}_3)
		\ev{\phi^4}'(\vb{k}) \exp(\sum_{i = 1}^4 \ii\vb{k}_i \cdot \vb{r}_i) \qty(2\pi)^3 \delta[3](\sum_{i=1}^4 \vb{k}_i),
	\end{multlined}
\label{eq:4PCF_key}
\end{equation}
where we impose $\vb{r}_4 = \vb{0}$ in the complex exponential. 

In \cref{sec:theoreticalModels}, we will discuss different trispectra from various models. Generally speaking, the trispectrum $\ev{\phi^4}'$ for a local exchange is usually a function of external momenta $\{\vb{k}_{1}, \ldots, \vb{k}_4\}$ as well as a propagator momentum $\vb{k}_s$. Moreover, if only one species of intermediate field is exchanged during inflation, we can frequently separate each trispectrum diagram into a radial part and an angular part, \textit{i.e.}
\begin{equation}
	\ev{\phi^4}'(\vb{k}_1, \ldots, \vb{k}_4, \vb{k}_s) = \mathcal{J}(k_1, \ldots, k_4, k_s) \mathcal{K}(\hat{\vb{k}}_1, \ldots, \hat{\vb{k}}_4, \hat{\vb{k}}_s) + \qty(\text{other diagrams}),
\end{equation}
in which the angular factor $\mathcal{K}$ is determined by the interaction vertex and the polarization vector of the propagator. It is convenient to treat the external field as distinct so that Wick contractions result in summation over $4!=24$ different momentum combination, \textit{i.e.}
\begin{equation}
	\ev{\phi^4}' = \sum_{\sigma \in S_4} \mathcal{J}(k_{\sigma_1}, \ldots, k_{\sigma_4}, k_s) \mathcal{K}(\hat{\vb{k}}_{\sigma_1}, \ldots, \hat{\vb{k}}_{\sigma_4}, \hat{\vb{k}}_s),
\end{equation}
in which the summation runs over all 24 maps in the $S_4$ group and $\sigma_i \defeq \sigma(i)$ for each index $i$ ranging from $1$ to $4$.  Correspondingly, the correlation coefficients can be evaluated via 
\begin{align}
		\zeta_{\ell_1, \ell_2, \ell_3}(r_1, r_2, r_3) & = \frac{H^4}{\dot{\phi}_0^4} \sum_{\sigma}
		\int \qty[\prod_{i = 1}^3 \dd \hat{\vb{r}}_{\sigma_i}] 
		\qty[ \prod_{i = 1}^{4} \dbar^3 k_i \; Z_1(\hat{\vb{k}}_i, z) M(k_i, z) ] \iso{\ell_1, \ell_2, \ell_3}^*(\hat{\vb{r}}_{\sigma_1},\hat{\vb{r}}_{\sigma_2},\hat{\vb{r}}_{\sigma_3}) \nonumber\\
		&\times
		\mathcal{J}(k_1, \ldots, k_4, k_s) \mathcal{K}(\hat{\vb{k}}_1, \ldots, \hat{\vb{k}}_4, \hat{\vb{k}}_s)
		\exp(\sum_{i = 1}^4 \ii\vb{k}_i \cdot \vb{r}_{\sigma_i}) \qty(2\pi)^3 \delta[3](\sum_{i=1}^4 \vb{k}_i).
 \label{eq:4PCF_key_2}
\end{align}
The general strategy to compute $\zeta_{\ell_1, \ell_2, \ell_3}(r_1, r_2, r_3)$ is to express $\dbar^3 k_i = \dd \hat{\vb{k}}_i \dd k_i \; k_i^2 / \qty(2\pi)^3$ and perform the angular integrals analytically while computing the radial integrals numerically. To do so,  we need to first expand the angular-relevant parts of~\cref{eq:4PCF_key_2}, i.e., the redshift distortion $\prod_{i=1}^4 Z_1(\hat{\vb{k}}_i, z)$, the complex exponential $ \exp(\sum_{i = 1}^4 \ii\vb{k}_i \cdot \vb{r}_i)$, and the momentum-conserving $\delta$-function $\qty(2\pi)^3 \delta[3](\sum_{i=1}^4 \vb{k}_i)$ into isotropic basis functions and then perform the integral over the products of the isotropic basis functions.

\subsection{Expanding into the Angular-relevant Part Isotropic Basis Functions}
We start with the kernel of the Fourier transform. By using the plane-wave expansion formula
\begin{equation}
	e^{\ii \vb{K} \cdot \vb{R}} = 4\pi \sum_{L = 0}^{\infty} \sum_{m = -L}^{L} \ii^{L} j_{L}(KR) Y_{L}^m(\hat{\vb{K}}) Y_{L}^{m*}(\hat{\vb{R}})
\end{equation}
one can separate the angular part of a complex exponential from its radial part as 
\begin{equation}
	\exp( \sum_{i} \ii\vb{k}_i \cdot \vb{r}_{\sigma_i} )
	= \qty(4\pi)^4 \prod_{i = 1}^{4} \sum_{\ell_i = 0}^{\infty} \sum_{m_i = -\ell_i}^{\ell} \ii^{\ell_i} j_{\ell_i}(k_i r_{\sigma_i}) Y_{\ell_i}^{m_i *}(\hat{\vb{k}}_i) Y_{\ell_i}^{m_i}(\hat{\vb{r}}_{\sigma_i}).
    \label{eq:FourierKernalPlaneWaveExp}
\end{equation}
At this point, it is useful to introduce the following identity $\ell_4 = m_4 = 0$ so that $1 = \int \dd \hat{\vb{r}}_4 \; Y_{l_4}^{m_4*}(\hat{\vb{r}}_4) \qty(4\pi)^{-1/2}$ can be inserted into the evaluation of $\zeta_{\ell}$. Now, one can perform the four $\dd \hat{\vb{r}}_i$ integrals to obtain
\begin{equation}
	\begin{aligned}
		& \qty(4\pi)^{-1/2} \sum_{\sigma} \int \qty[\prod_{i = 1}^4 \dd \hat{\vb{r}}_{\sigma_i}] \; \iso{\ell_1, \ell_2, \ell_3}^*(\hat{\vb{r}}_1, \hat{\vb{r}}_2, \hat{\vb{r}}_3) Y_{\ell_4}^{m_4*}(\hat{\vb{r}}_4) \prod_{i=1}^4 Y_{\ell_i'}^{m_i'}(\hat{\vb{r}}_{\sigma_i})\\
		=& \qty(4\pi)^{-1/2} (-1)^{\ell_1 + \ldots + \ell_3} \sum_{\sigma} \int \qty[\prod_{i = 1}^4 \dd \hat{\vb{r}}_{\sigma_i}] \; \sum_{\ell'} \iso{\ell_1, \ell_2, (\ell'), \ell_3, \ell_4}(\hat{\vb{r}}_1, \ldots, \hat{\vb{r}}_4) \delta_{\ell', \ell_3} \prod_{i=1}^4 Y_{\ell_i'}^{m_i'}(\hat{\vb{r}}_{\sigma_i})\\
		=& \qty(4\pi)^{-1/2} (-1)^{\ell_1 + \ldots + \ell_3} \sum_{\sigma} \int \qty[\prod_{i = 1}^4 \dd \hat{\vb{r}}_{\sigma_i}] \; \sum_{\ell'} \iso{\ell_{\sigma_1}, \ell_{\sigma_2}, (\ell'), \ell_{\sigma_3}, \ell_{\sigma_4}}(\hat{\vb{r}}_{\sigma_1}, \ldots, \hat{\vb{r}}_{\sigma_4}) \prod_{i=1}^4 Y_{\ell_i'}^{m_i'}(\hat{\vb{r}}_{\sigma_i})\\
		=& \qty(4\pi)^{-1/2} \sum_{\sigma} \sum_{\ell,m} \qty(-1)^{\ell_1 + \cdots + \ell_4} \Phi_\sigma C_{m}^{\ell} \prod_{i = 1}^{4} \delta_{\ell_i',\ell_{\sigma_i}} \delta_{m_i', m_{\sigma_i}}
	\end{aligned}
\label{eq:complex}
\end{equation}
where $\Phi_\sigma$ is the symmetry factor due to permutations of three-$j$ symbols, and $C_{m}^{\ell}$ denotes decomposition coefficient of isotropic function in terms of spherical harmonics, \textit{i.e.} $\iso{\ell} = \sum_{m,\ell} C_{m}^{\ell} Y_{\ell}^{m*}$. $\iso{\ell_1, \ell_2, (\ell'), \ell_3, \ell_4}(\hat{\vb{r}}_1,\hat{\vb{r}}_2,\hat{\vb{r}}_3, \hat{\vb{r}}_4)$ is the isotropic 4-pt function defined in (\ref{eq:4pt}) where the subscript $\ell$ with (without) bracket represents the primary (intermediate) angular momentum number. Substituting (\ref{eq:complex}) and (\ref{eq:FourierKernalPlaneWaveExp}) into (\ref{eq:4PCF_key_2}) yields
\begin{equation}
	\begin{aligned}
		\zeta_{\ell_1, \ell_2, \ell_3}(r_1,r_2,r_3) =&
		\frac{H^4}{\dot{\phi}_0^4} \qty(4\pi)^{7/2} \sum_{\sigma, \ell_\sigma} \Phi_\sigma \qty(-\ii)^{\ell_{1}+\ldots+\ell_{3}}
		\qty[ \int \prod_{i = 1}^{4} \dbar^3 k_i \; Z_1(\hat{\vb{k}}_i) M(k_i) j_{\ell_{\sigma_i}}(k_i r_{\sigma_i}) ] \\
  \times& \mathcal{J}(k_1, \ldots, k_4, k_s) \mathcal{K}(\hat{\vb{k}}_1, \ldots, \hat{\vb{k}}_4, \hat{\vb{k}}_s)\\
		\times &
		\iso{\ell_{\sigma_1},\ell_{\sigma_2}(\ell')\ell_{\sigma_3},\ell_{\sigma_4}}(\hat{\vb{k}}_1, \ldots, \hat{\vb{k}}_4) \qty(2\pi)^3 \delta[3](\sum_{i=1}^4 \vb{k}_i).
	\end{aligned}
\end{equation}
Next, the momentum conserving $\delta$-function can be expressed as 
\begin{equation}
	\qty(2\pi)^3 \delta[3](\sum_{i=1}^4 \vb{k}_i) = \int \dd^3 x\; e^{ \sum_{i=1}^4 \ii \vb{k}_i \cdot \vb{x}} = \int \dbar^3 k_s \int \dd^3 x \dd^3 y \; e^{\ii \qty(\vb{k}_1 + \vb{k}_2 + \vb{k}_s) \cdot \vb{x}} e^{\ii \qty(\vb{k}_3 + \vb{k}_4 - \vb{k}_s) \cdot \vb{y}}
\end{equation}
with the help of one ($\vb{x}$) or two ($\vb{x}$ and $\vb{y}$) auxiliary variables. Applying the plane-wave expansion to the two complex exponential yields 
\begin{equation}
	\begin{aligned}
		& \int \dbar^3 k_s \dd^3 x \dd^3 y\; e^{\ii \qty(\vb{k}_1 + \vb{k}_2 + \vb{k}_s) \cdot \vb{x}} e^{\ii \qty(\vb{k}_3 + \vb{k}_4 - \vb{k}_s) \cdot \vb{y}} \\
		=& \qty(4\pi)^5 \int \frac{\dd k_s }{2\pi^2} \;  k_s^2  \sum_{L,m} \ii^{L_1 + \ldots + L_5 + L_6} \qty(-1)^{L_6} \\
  & \times \qty[\int \dd x\; x^2 j_{L_1}(k_1 x) j_{L_2} (k_2 x) j_{L_5}(k_s x)] \qty[\int \dd y\; y^2 j_{L_3}(k_3 y) j_{L_4} (k_4 y) j_{L_6}(k_s y)] \\
		& \times \int \dd \hat{\vb{x}} \; Y^{m_1}_{L_1}(\hat{\vb{x}}) Y^{m_2}_{L_2}(\hat{\vb{x}}) Y^{m_5}_{L_5}(\hat{\vb{x}}) \times \int \dd \hat{\vb{y}} \; Y^{m_3}_{L_3}(\hat{\vb{y}}) Y^{m_4}_{L_4}(\hat{\vb{y}})\; Y^{m_6}_{L_6}(\hat{\vb{y}}) \\
		& \times \int \dd \hat{\vb{k}}_s
		Y^{m_1*}_{L_1}(\hat{\vb{k}}_1) 
		Y^{m_2*}_{L_2}(\hat{\vb{k}}_2) 
		Y^{m_3*}_{L_3}(\hat{\vb{k}}_3) 
		Y^{m_4*}_{L_4}(\hat{\vb{k}}_4)  Y^{m_5*}_{L_5}(\hat{\vb{k}}_s) Y^{m_6*}_{L_6}(\hat{\vb{k}}_s).
	\end{aligned}
  \label{eq:delta0}
\end{equation}
Using the Gaunt integral identity, 
\begin{equation}
		\int \dd \hat{\vb{x}} \; Y^{m_1}_{L_1}(\hat{\vb{x}}) Y^{m_2}_{L_2}(\hat{\vb{x}}) Y^{m_3}_{L_3}(\hat{\vb{x}}) 
		=  \sqrt{\frac{\prod_{i=1}^3 \qty(2L_i + 1)}{4\pi}} \mqty(L_1 & L_2 & L_3 \\ 0 & 0 & 0) \mqty(L_1 & L_2 & L_3 \\ m_1 & m_2 & m_3)
\end{equation}
and collect terms for the isotropic 3-pt basis functions~(\ref{eq:3pt}), we get
\begin{equation}
	\begin{aligned}
		&\qty(2\pi)^3 \delta[3](\sum_i \vb{k}_i) = \qty(4\pi)^4 \sum_{L} \qty(-1)^{L_6} \ii^{L_1 + \ldots + L_6} \sqrt{\prod_{i=1}^6 \qty(2L_i + 1)} \mqty(L_1 & L_2 & L_5 \\ 0 & 0 & 0)  \mqty(L_3 & L_4 & L_6 \\ 0 & 0 & 0) \\
		& \times \int \frac{\dd k_s}{2\pi^2} \; k_s^2 \qty[\int \dd x\; x^2 j_{L_1}(k_1 x) j_{L_2} (k_2 x) j_{L_5}(k_s x)] \qty[\int \dd y\; y^2 j_{L_3}(k_3 y) j_{L_4} (k_4 y) j_{L_6}(k_s y)]\\
		& \times \int \dd \hat{\vb{k}}_s \iso{L_1,L_2,L_5}(\hat{\vb{k}}_1, \hat{\vb{k}}_2, \hat{\vb{k}}_s)\iso{L_3,L_4,L_6}(\hat{\vb{k}}_3, \hat{\vb{k}}_4, \hat{\vb{k}}_s).
	\end{aligned}
 \label{eq:delta1}
\end{equation}
Note that the radial integrals have a common triple spherical Bessel integral, 
\begin{equation}
	f_{L_1,L_2,L_3}(k_1,k_2,k_3) 
	\defeq \int \dd x\; x^2 j_{L_1}(k_1 x) j_{L_2}(k_2 x) j_{L_3}(k_3 x).
 \label{eq:triplej}
\end{equation}
The integral can be evaluated numerically after changing the order of the $x$ and $k_i$ integrals from the Fourier transform (e.g. Ref~\cite{Philcox:2022hkh}). But this requires a fine sampling of the $k_{1,2,3,4,s}-x$ lattice given the high oscillatory nature of the spherical Bessel function and thus is computationally expensive. A better way to evaluate the triple spherical Bessel integral is by using its analytical solutions, e.g., Refs.~\cite{jackson1972integrals, Mehrem:1990eg}. The analytical solution of Ref.~\cite{Mehrem:1990eg} has been adopted in Ref.~\cite{Cabass:2022oap}. But the solution is numerically unstable if the ratio of the input momenta is large. Instead, we adopt the analytical solution of Ref.~\cite{jackson1972integrals}, which does not suffer the numerical instability and has fewer layers of summation than that of Ref.~\cite{Mehrem:1990eg}. We will compare the analytical solutions in detail in Sec.~\ref{sec:tripleJ}.

If there is no explicit $\hat{\vb{k}}_s$ dependence in $\mathcal{K}$, one can directly integrate out $\hb k_s$ in (\ref{eq:delta0}) by using the orthogonal relationship 
\beq
\int \dd \hat{\vb{k}}_s Y^{m_5*}_{L_5}(\hat{\vb{k}}_s) Y^{m_6 *}_{L_6}(\hat{\vb{k}}_s) = (-1)^{m_6} \delta_{L_5,L_6} \delta_{m_5,-m_6}
\eeq
and collecting terms for the isotropic 4-pt basis function \cref{eq:4pt} to obtain
\begin{equation}
	\begin{aligned}
		\qty(2\pi)^3 \delta[3](\sum_i \vb{k}_i) = &\qty(4\pi)^4 \sum_{L} \qty(-1)^{L'} \ii^{L_1 + \ldots + L_4} \sqrt{\qty(2L'+1) \prod_{i=1}^4 \qty(2L_i + 1)} \\
		& \times \mqty(L_1 & L_2 & L' \\ 0 & 0 & 0)  \mqty(L_3 & L_4 & L' \\ 0 & 0 & 0) \\
		& \times \int \frac{\dd k_s}{2\pi^2} \; k_s^2 f_{L_1, L_2, L'}(k_1, k_2, k_s)  f_{L_3, L_4, L'} (k_3, k_4, k_s)\\
		& \times \iso{L_1,L_2(L')L_3,L_4}(\hat{\vb{k}}_1, \ldots, \hat{\vb{k}}_4).
	\end{aligned}
 \label{eq:delta1}
\end{equation}

Next, one expands the redshift-space distortion factor into the isotropic basis functions. The Kaiser factor can be expressed as a function of $\hat{\vb{k}}$ and the line-of-sight direction $\hat{\vb{n}}$ 
\begin{equation}
	Z_1(\hat{\vb{k}}, \hat{\vb{n}}) = b + f \qty(\hat{\vb{n}} \cdot \hat{\vb{k}})^2 = b + \frac{f}{3} + \qty(4\pi) \cdot \frac{2f}{15} \sum_{m=-2}^{2} Y_{2}^{m*}(\hat{\vb{n}}) Y_{2}^{m}(\hat{\vb{k}}),
\end{equation}
It is then understood that the redshift-space distortion should be integrated over a solid angle of all possible lines of sight, \textit{i.e.}
\begin{equation}
	\begin{aligned}
	& \prod_{i=1}^4 Z_i(\hat{\vb{k}}_i) 
	= \int \frac{\dd \hat{\vb{n}}}{4\pi} \prod_{i=1}^4 Z_i(\hat{\vb{k}}_i, \hat{\vb{n}}) 
	= \qty(4\pi)^3 \int \dd \hat{\vb{n}} \; \sum_{j,m} \qty[\prod_{i = 1}^{4} Z_{j_i}] Y^{m_i}_{j_i}(\hat{\vb{n}}) Y^{m_i}_{j_i}(\hat{\vb{k}_i})\\
 = & \qty(4\pi)^{2} \sum_{j} \qty[\prod_{i=1}^4 Z_{j_i} \sqrt{2j_i + 1}] \sqrt{2j'+1} \qty(-1)^{j'} \mqty(j_1 & j_2 & j' \\ 0 & 0 & 0)  \mqty(j_3 & j_4 & j' \\ 0 & 0 & 0) \iso{j_1, j_2 (j') j_3, j_4}(\hat{\vb{k}}_1, \ldots, \hat{\vb{k}}_4)
 	\end{aligned}
\end{equation}
in which 
\begin{equation}
	Z_{j_i} \defeq \qty(b + \frac{f}{3}) \delta_{0,j_i} + \frac{2f}{15} \delta_{2,j_i}. 
\end{equation}
Since $j_i$ can only take even values, the three-$j$ symbol dictates that $j'$ is also an even number, implying that $\qty(-1)^{j'} = 1$. 

Finally, we comment on the angular part of the trispectrum $\mathcal K$. For the exchange-type trispectrum, $\mathcal K$ is a function of the unit vector $\{\vb{k}_{1}, \ldots, \vb{k}_4, \hb k_s\}$ and can be decomposed as the linear combination of the isotropic 5-pt basis functions, i.e.,
\begin{equation}
	\mathcal{K}(\hat{\vb{k}}_1, \ldots, \hat{\vb{k}}_4, \hb k_s) = \sum_{l} c_{l_1, l_2, (l'), l_3, (l''), l_4, l_5} \iso{l_1, l_2, (l'), l_3, (l''), l_4, l_5}(\hat{\vb{k}}_1, \ldots, \hat{\vb{k}}_4, \hb k_s). 
\end{equation}
where the isotropic 5-pt basis function is defined by
\begin{equation}
	\begin{aligned}
	&\iso{\ell_1, \ell_2, (\ell'), \ell_3, (\ell''),\ell_4, \ell_5}(\hat{\vb{r}}_1,\hat{\vb{r}}_2,\hat{\vb{r}}_3, \hat{\vb{r}}_4, \hb r_5) 
	\defeq \!\!\!\!\!\! \sum_{m_1, m_2, m_3, m_4, m_5} \!\!\!\!\!\!  \sqrt{(2\ell'+1)(2\ell''+1)} \sum_{m', m''} (-1)^{\ell'-m'+\ell''-m''}\\
 &\mqty(\ell_1 & \ell_2 & \ell' \\ m_1 & m_2 & -m')\mqty(\ell' & \ell_3 & \ell'' \\ m' & m_3 & m'')
 \mqty(\ell'' & \ell_4 & \ell_5 \\ m'' & m_4 & m_5)
  Y_{\ell_1}^{m_1*}(\hat{\vb{r}}_1) Y_{\ell_2}^{m_2*}(\hat{\vb{r}}_2) Y_{\ell_3}^{m_3*}(\hat{\vb{r}}_3)Y_{\ell_4}^{m_4*}(\hat{\vb{r}}_4)Y_{\ell_4}^{m_5*}(\hat{\vb{r}}_5).
 	\end{aligned}
 \label{eq:5pt}
\end{equation}
Note that the explicit $\hb k_s$ dependence in $\mathcal{K}(\hat{\vb{k}}_1, \ldots, \hat{\vb{k}}_4, \hb k_s)$ can be dropped by changing the variable based on the momentum conservation
\beq
\hb k_s = -(k_1 \hb k_1 + k_2 \hb k_2)/k_s \quad \text{or} \quad \hb k_s = (k_3 \hb k_3+ k_4 \hb k_4)/k_s.
\eeq\footnote{$\zeta_{\ell_1, \ell_2, \ell_3}$ will obtain a negative sign if we define $k_s = (k_1 \hb k_1 + k_2 \hb k_2)/k_s$ or $k_s = - (k_3 \hb k_3+ k_4 \hb k_4)/k_s$. $\vb{k}_s \to -\vb{k}_s$ is equivalent to a parity transform on all external momentum, which is preserved by the Fourier transform, i.e. under an inverse Fourier transform $\mathcal{F}^{-1}(f(-k)) = f(-x)$. Hence, the parity-preserving part of the 4PCF remains unchanged, while the parity-violating part of the 4PCF obtains a relative negative sign. This, however, is consistently done so across all parity-violating parts. As we report detection significance in terms of $\abs{ \sigma_A / \sigma_A }$, this sign change is not crucial for this study.} This operation simplifies the angular integral at the expense of rendering the trispectrum as a summation of a larger number of terms. 

\subsection{The General Expression for the 4PCF Coefficient}
Combing all the expansions together, we have the 4PCF coefficient
\begin{equation}
	\begin{aligned}
		& \zeta_{\ell_1, \ell_2, \ell_3}(r_1,r_2,r_3) \\
		=& \frac{H^4}{\dot{\phi}_0^4} \qty(4\pi)^{19/2} \sum_{\sigma, \ell_\sigma} \Phi_\sigma \qty(-\ii)^{\ell_{1}+\ldots+\ell_{3}} \sum_{L,j} \qty[\prod_{i=1}^4 Z_{j_i} \sqrt{2j_i + 1}] \sqrt{2j'+1} \mqty(j_1 & j_2 & j' \\ 0 & 0 & 0)  \mqty(j_3 & j_4 & j' \\ 0 & 0 & 0) \\
		& \times \qty(-1)^{L_6} \ii^{L_1 + \ldots + L_6} \qty[\prod_{i=1}^6 \sqrt{2L_i + 1}]  \mqty(L_1 & L_2 & L_5 \\ 0 & 0 & 0)  \mqty(L_3 & L_4 & L_6 \\ 0 & 0 & 0) \\
		& \times \int \frac{\dd k_s\, k_s^2}{2\pi^2} \; \qty[ \int \prod_{i = 1}^{2} \frac{\dd k_i \, k_i^2}{2\pi^2} \; M(k_i) j_{\ell_{\sigma_i}}(k_i r_{\sigma_i}) ] f_{L_1,L_2,L_5}(k_1,k_2,k_s) \mathcal{J}_L(k_1, k_2, k_s)\\
		& \times \qty[ \int \prod_{i = 3}^{4} \frac{\dd k_i \, k_i^2}{2\pi^2} \; M(k_i) j_{\ell_{\sigma_i}}(k_i r_{\sigma_i}) ] f_{L_3,L_4,L_6}(k_3,k_4,k_s)  \mathcal{J}_{R}(k_3, k_4, k_s)\\
		& \times \int \qty[\prod_{i=1}^{4} \dd \hat{\vb{k}}_i] \int \dd \hat{\vb{k}}_s \iso{L_1,L_2,L_5}(\hat{\vb{k}}_1, \hat{\vb{k}}_2, \hat{\vb{k}}_s)\iso{L_3,L_4,L_6}(\hat{\vb{k}}_3, \hat{\vb{k}}_4, \hat{\vb{k}}_s) \\
        & \times \iso{\ell_{\sigma_1},\ell_{\sigma_2}(\ell')\ell_{\sigma_3},\ell_{\sigma_4}}(\hat{\vb{k}}_1, \ldots, \hat{\vb{k}}_4)
		\iso{j_1, j_2 (j') j_3, j_4} (\hat{\vb{k}}_1, \ldots, \hat{\vb{k}}_4)
		\mathcal{K}(\hat{\vb{k}}_1, \ldots, \hat{\vb{k}}_4, \hat{\vb{k}}_s)\\
		=& \frac{H^4}{\dot{\phi}_0^4} \qty(4\pi)^{19/2} \sum_{\sigma, \ell_\sigma} \Phi_\sigma \qty(-\ii)^{\ell_{1}+\ldots+\ell_{3}} \sum_{L,j} \qty[\prod_{i=1}^4 Z_{j_i} \sqrt{2j_i + 1}] \sqrt{2j'+1} \mqty(j_1 & j_2 & j' \\ 0 & 0 & 0)  \mqty(j_3 & j_4 & j' \\ 0 & 0 & 0) \\
		& \times \qty(-1)^{L'} \ii^{L_1 + \ldots + L_4} \qty[\prod_{i=1}^4 \sqrt{2L_i + 1}] \sqrt{2L'+1} \mqty(L_1 & L_2 & L' \\ 0 & 0 & 0)  \mqty(L_3 & L_4 & L' \\ 0 & 0 & 0) \\
		& \times \int \frac{\dd k_s\, k_s^2}{2\pi^2} \; \qty[ \int \prod_{i = 1}^{2} \frac{\dd k_i \, k_i^2}{2\pi^2} \; M(k_i) j_{\ell_{\sigma_i}}(k_i r_{\sigma_i}) ] f_{L_1,L_2,L'}(k_1,k_2,k_s) \mathcal{J}_L(k_1, k_2, k_s)\\
		& \times \qty[ \int \prod_{i = 3}^{4} \frac{\dd k_i \, k_i^2}{2\pi^2} \; M(k_i) j_{\ell_{\sigma_i}}(k_i r_{\sigma_i}) ] f_{L_3,L_4,L'}(k_3,k_4,k_s)  \mathcal{J}_{R}(k_3, k_4, k_s)\\
		& \times \int \qty[\prod_{i=1}^{4} \dd \hat{\vb{k}}_i] \; \iso{\ell_{\sigma_1},\ell_{\sigma_2}(\ell')\ell_{\sigma_3},\ell_{\sigma_4}}
		\iso{L_1,L_2(L')L_3,L_4}
		\iso{j_1, j_2 (j') j_3, j_4}
		\mathcal{K}(\hat{\vb{k}}_1, \ldots, \hat{\vb{k}}_4).
	\end{aligned}
	\label{eqn:strategyOverview}
\end{equation}
Here, we expressed the trispectrum involving the exchange of an intermediate particle into the product of a left part $\mathcal{J}_{L}(k_1, k_2, k_s)$ and a right part $\mathcal{J}_{R}(k_3, k_4, k_s)$. In the second eqality, we applied~\cref{eq:delta1} assuming $\mathcal{K}$ has no explicit $\hat{\vb{k}}_s$ dependence.  The 1\textsuperscript{st} and 2\textsuperscript{nd} line of \cref{eqn:strategyOverview} can be regarded as some coupling between various angular momentum indices $\{\ell_{\sigma}, L, j\}$, and the 3\textsuperscript{rd} and 4\textsuperscript{th} line contain $\ell_\sigma$- and $L$-dependent radial integrals to be evaluated numerically. Lastly, the last (two) line can be evaluated analytically by manipulating the isotropic basis functions to provide additional angular couplings between $\{\ell_{\sigma}, L, j\}$.

In general, the factorization of $\mathcal{J} = \mathcal{J}_{L} \mathcal{J}_{R}$ is not always guaranteed. However, for an exchange-type diagram, using Schwinger-Keldysh formalism \cite{Chen:2017ryl}, one can almost always partition such a diagram into a left and right subdiagram at the cost of introducing additional conformal time integrals. This means that 
\begin{equation}
	\mathcal{J}(k_1, \ldots, k_s) = \int \dd \tau_L \dd \tau_R\; \mathcal{J}_L(k_1, k_2, k_s, \tau_L) \Pi(k_s, \tau_L, \tau_R) \mathcal{J}_R(k_3, k_4,k_s, \tau_R),
	\label{eqn:dynamical}
\end{equation}
in which $\Pi(\tau_L, \tau_R)$ denotes some additional conformal time dependence due to the propagator of the exchanged particle. In recapitulation, the general formalism for computing the correlation coefficients is
\begin{equation}
	\begin{aligned}
		& \zeta_{\ell_1, \ell_2, \ell_3}(r_1,r_2,r_3) \\
		=& \frac{H^4}{\dot{\phi}_0^4} \qty(4\pi)^{19/2} \sum_{\sigma, \ell_\sigma} \Phi_\sigma \qty(-i)^{\ell_{1}+\ldots+\ell_{3}} \sum_{L,j} \qty[\prod_{i=1}^4 Z_{j_i} \sqrt{2j_i + 1}] \sqrt{2j'+1} \mqty(j_1 & j_2 & j' \\ 0 & 0 & 0)  \mqty(j_3 & j_4 & j' \\ 0 & 0 & 0) \\
		& \times \qty(-1)^{L'} i^{L_1 + \ldots + L_4} \qty[\prod_{i=1}^4 \sqrt{2L_i + 1}] \sqrt{2L'+1} \mqty(L_1 & L_2 & L' \\ 0 & 0 & 0)  \mqty(L_3 & L_4 & L' \\ 0 & 0 & 0) \\
		& \times \int \dd \tau_L \dd \tau_R \int \frac{\dd k_s\, k_s^2}{2\pi^2} \; \Pi(k_s, \tau_L, \tau_R) \\
		& \times \qty[ \int \prod_{i = 1}^{2} \frac{\dd k_i \, k_i^2}{2\pi^2} \; M(k_i) j_{\ell_{\sigma_i}}(k_i r_{\sigma_i}) ] f_{L_1,L_2,L'}(k_1,k_2,k_s) \mathcal{J}_L(k_1, k_2, k_s, \tau_L)\\
		& \times \qty[ \int \prod_{i = 3}^{4} \frac{\dd k_i \, k_i^2}{2\pi^2} \; M(k_i) j_{\ell_{\sigma_i}}(k_i r_{\sigma_i}) ] f_{L_3,L_4,L'}(k_3,k_4,k_s)  \mathcal{J}_{R}(k_3, k_4, k_s, \tau_R)\\
		& \times \int \qty[\prod_{i=1}^{4} \dd \hat{\vb{k}}_i] \; \iso{\ell_{\sigma_1},\ell_{\sigma_2}(\ell')\ell_{\sigma_3},\ell_{\sigma_4}}
		\iso{L_1,L_2(L')L_3,L_4}
		\iso{j_1, j_2 (j') j_3, j_4}
		\mathcal{K}(\hat{\vb{k}}_1, \ldots, \hat{\vb{k}}_4).
	\end{aligned}
\end{equation}

\subsection{Handling Angular Integrals}
To reduce computational time, we will calculate the angular integrals analytically. First, we will decompose the kinematic part of the trispectrum onto isotropic basis functions as well
\begin{equation}
	\mathcal{K}(\hat{\vb{k}}_1, \ldots, \hat{\vb{k}}_4) = \sum_{l} c_{l_1,l_2(l')l_3,l_4} \iso{l_1,l_2(l')l_3,l_4}(\hat{\vb{k}}_1, \ldots, \hat{\vb{k}}_4). 
\end{equation}
Recall that isotropic functions are specific combinations of spherical harmonics; hence, products of isotropic functions can be reduced using angular momentum addition. (More detailed discussions can be found in Ref.~\cite{Cahn:2020axu}.) This leads to 
\begin{equation}
	\begin{aligned}
		& \int \qty[\prod_{i=1}^{4} \dd \hat{\vb{k}}_i] \; \iso{\ell_{\sigma_1},\ell_{\sigma_2}(\ell')\ell_{\sigma_3},\ell_{\sigma_4}}
		\iso{L_1,L_2 (L') L_3,L_4}
		\iso{j_1, j_2 (j') j_3, j_4}
		\iso{l_1,l_2 (l') l_3, l_4}(\hat{\vb{k}}_1, \ldots, \hat{\vb{k}}_4) \\
		=& \qty(4\pi)^{-4} \sum_{\lambda} \qty(-1)^{\lambda_1 + \ldots + \lambda_4} \qty[\prod_{i=1}^4 \qty(2\lambda_i + 1)] \qty(2\lambda'+1) \qty[\prod_{i=1}^4 \sqrt{2\ell_{\sigma_i} + 1}] \sqrt{2\ell'+1} \\
		& \times \qty[\prod_{i=1}^4 \sqrt{2j_i + 1}] \sqrt{2j'+1} \qty[\prod_{i=1}^4 \sqrt{2l_i + 1}] \sqrt{2l'+1} \qty[\prod_{i=1}^4 \sqrt{2L_i + 1}] \sqrt{2L'+1} \\
		& \times 
		\qty[ \prod_{i=1}^4 \mqty( \ell_{\sigma_i} & j_i & \lambda_i \\ 0 & 0 & 0 ) \mqty( l_i & L_i & \lambda_i \\ 0 & 0 & 0 )]
		\qty{ \mqty{ \ell_{\sigma_1} & \ell_{\sigma_2} & \ell' \\ j_1 & j_2 & j' \\ \lambda_1 & \lambda_2 & \lambda' } } 
		\qty{ \mqty{ \ell' & \ell_{\sigma_3} & \ell_{\sigma_4} \\ j' & j_3 & j_4 \\ \lambda' & \lambda_3 & \lambda_4 } } 
		\qty{ \mqty{ l_1 & l_2 & l' \\ L_1 & L_2 & L' \\ \lambda_1 & \lambda_2 & \lambda' } } 
		\qty{ \mqty{ l' & l_3 & l_4 \\ L' & L_3 & L_4 \\ \lambda' & \lambda_3 & \lambda_4 } }. 
	\end{aligned}
\end{equation}
Hence, angular integrals over external momenta can be done completely, and the result is a general coupling matrix between various angular indices so that 
\begin{equation}
    \begin{aligned}
		& \zeta_{\ell_1, \ell_2, \ell_3}(r_1,r_2,r_3) \\
		=& \frac{H^4}{\dot{\phi}_0^4} \qty(4\pi)^{11/2} \sum_{\sigma, \ell_\sigma,L,j,\lambda,l} \Phi_\sigma \qty(-i)^{\ell_{1}+\ldots+\ell_{3}} \qty(-1)^{L'} i^{L_1 + \ldots + L_4} \qty(-1)^{\lambda_1 + \ldots + \lambda_4} \\
		& \times \qty[\prod_{i=1}^4 Z_{j_i} \qty(2j_i + 1)] \qty(2j'+1) \qty[\prod_{i=1}^4 \qty(2L_i + 1)] \qty(2L'+1) \\
		& \times \mqty(j_1 & j_2 & j' \\ 0 & 0 & 0)  \mqty(j_3 & j_4 & j' \\ 0 & 0 & 0) \mqty(L_1 & L_2 & L' \\ 0 & 0 & 0)  \mqty(L_3 & L_4 & L' \\ 0 & 0 & 0) \\
        & \times \qty[ \prod_{i=1}^4 \mqty( \ell_{\sigma_i} & j_i & \lambda_i \\ 0 & 0 & 0 ) \mqty( l_i & L_i & \lambda_i \\ 0 & 0 & 0 )]
		\qty{ \mqty{ \ell_{\sigma_1} & \ell_{\sigma_2} & \ell' \\ j_1 & j_2 & j' \\ \lambda_1 & \lambda_2 & \lambda' } } 
		\qty{ \mqty{ \ell' & \ell_{\sigma_3} & \ell_{\sigma_4} \\ j' & j_3 & j_4 \\ \lambda' & \lambda_3 & \lambda_4 } } 
		\qty{ \mqty{ l_1 & l_2 & l' \\ L_1 & L_2 & L' \\ \lambda_1 & \lambda_2 & \lambda' } } 
		\qty{ \mqty{ l' & l_3 & l_4 \\ L' & L_3 & L_4 \\ \lambda' & \lambda_3 & \lambda_4 } } \\
        & \times c_{l_1,l_2(l')l_3,l_4} \qty[\prod_{i=1}^4 \qty(2\lambda_i + 1)] \qty(2\lambda'+1) \qty[\prod_{i=1}^4 \sqrt{2\ell_{\sigma_i} + 1}] \sqrt{2\ell'+1} \qty[\prod_{i=1}^4 \sqrt{2l_i + 1}] \sqrt{2l'+1} \\
		& \times \int \dd \tau_L \dd \tau_R \int \frac{\dd k_s\, k_s^2}{2\pi^2} \; \Pi(k_s, \tau_L, \tau_R) \\
		& \times \qty[ \int \prod_{i = 1}^{2} \frac{\dd k_i \, k_i^2}{2\pi^2} \; M(k_i) j_{\ell_{\sigma_i}}(k_i r_{\sigma_i}) ] f_{L_1,L_2,L'}(k_1,k_2,k_s) \mathcal{J}^{l_1,l_2(l')l_3,l_4}_L(k_1, k_2, k_s, \tau_L)\\
		& \times \qty[ \int \prod_{i = 3}^{4} \frac{\dd k_i \, k_i^2}{2\pi^2} \; M(k_i) j_{\ell_{\sigma_i}}(k_i r_{\sigma_i}) ] f_{L_3,L_4,L'}(k_3,k_4,k_s)  \mathcal{J}^{l_1,l_2(l')l_3,l_4}_{R}(k_3, k_4, k_s, \tau_R).
	\end{aligned}
	\label{eqn:strategyalmost}
\end{equation}

\subsection{Handling the Radial Bins}

Finally, we note that in practice, the radial distances $r_{1,2,3}$ are binned variables for the radial range of the survey, which are not continuous. Therefore the 4PCF coefficients should instead use the radial bin indices $b_{1,2,3}$ as its variables, i.e., $\zeta_{\ell_1,\ell_2,\ell_3} (b_1, b_2, b_3)$, or the mean radii of the bins, i.e., $\zeta_{\ell_1,\ell_2,\ell_3} (\bar r_{b_1}, \bar r_{b_2}, \bar r_{b_3})$. Correspondingly, the spherical Bessel function $j_{\ell} (k r)$ is replaced with the bin-integrated spherical Bessel function $\bar j_{\ell} (k,b)$~(c.f.~\cite{Philcox:2019hdi}), which is defined as
\beq
\bar j_\ell(k, b) \defeq \frac{\int_{r_{\min}}^{r_{\max}} \dd r r^2 j_\ell (k r)}{\int_{r_{\min}}^{r_{\max}}  \dd r r^2},
\eeq
with
\beq
r_{\min} = R_{\min} + b \frac{R_{\max} -R_{\min}}{n_r}, \quad r_{\max} =  R_{\min} + (b+1) \frac{R_{\max} -R_{\min}}{n_r}
\eeq
for a linear-even-binning. Here, $R_{\min}$, $R_{\max}$, $n_r$, and $b$ are the minimal and maximal radial distance, the total number of radial bins, and the bin index, respectively. Note that for $\vb r_4 =\vb 0$, we have $j_0(0)=1$. After replacing $j_\ell$ to $\bar j_\ell$ of \cref{eqn:strategyalmost}, we finally obtain~\cref{eqn:strategy}.

\section{Numerical Technicalities \label{app:numTechnicalities}}
In this appendix, we comment on a few important numerical technicalities we encountered and provide details on how they are handled.

\subsection{Numerical Stability of Integrals over Triple Products of Spherical Bessel Functions}
\label{sec:tripleJ}

Recall that the radial integral will always contain an auxiliary integral of the form
\begin{equation}
	f_{L_1,L_2,L_3}(k_1, k_2, k_s) = \int_0^\infty \dd x\; x^2 j_{L_1}(k_1 x) j_{L_2}(k_2 x) j_{L_3}(k_s x),
\end{equation}
in which $j_{L}(x)$ denotes the spherical Bessel function of the first kind. This type of integral has a closed-form expression as shown in \cite{jackson1972integrals} given by
\begin{equation}
    \begin{multlined}
        f_{L_1,L_2,L_3}(k_1, k_2, k_s) = \frac{\ii^{-\qty(L_1 + L_2 + L_3)} \pi \Delta}{4k_1 k_2 k_s} \mqty( L_1 & L_2 & L_3 \\ 0 & 0 & 0 )^{-1} \sum_{m=-\min\{L_1, L_2\}}^{\min\{L_1, L_2\}} \qty(-1)^m  \\
        \times\sqrt{\frac{\qty(L_1 - m)! \qty(L_2 + m)!}{\qty(L_1 + m)! \qty(L_2 - m)!}} \mqty(L_1 & L_2 & L_3 \\ m & -m & 0) P_{L_1}^{m}(\cos\theta_{13}) P_{L_2}^{-m}(\cos\theta_{23}),
    \end{multlined}
    \label{eqn:new3JInteg}
\end{equation}
where 
\begin{equation}
    \cos\theta_{13} \defeq \frac{k_2^2 - k_1^2 - k_s^2}{2k_1 k_s}, \quad 
    \cos\theta_{23} \defeq \frac{k_1^2 - k_2^2 - k_s^2}{2k_2 k_s},
    \quad 
    \Delta = 
        \begin{dcases}
            1, & \abs{\cos\theta_{13}} < 1, \\
            \frac{1}{2}, & \abs{\cos\theta_{13}} = 1, \\
            0, & \abs{\cos\theta_{13}} > 1. 
        \end{dcases}
\end{equation}

Note that this is a different strategy as the one described in Mehrem \textit{et al.}'s paper \cite{Mehrem:1990eg}, which is used in Ref.~\cite{Cabass:2022oap},
\begin{align}
  f_{L_1,L_2,L_3}(k_1, k_2, k_s) = &\f{\pi \Delta}{4 k_1 k_2 k_s} \ii^{L_1+L_2-L_3} (2L_3+1)^{1/2} \left(\f{k_1}{k_s}\right)^{L_3}  \begin{pmatrix}
L_1 &L_2 & L_3 \\
0 & 0 & 0
\end{pmatrix}^{-1} \nonumber\\
& \times \sum_{L_4 =0}^{L_3}\begin{pmatrix}
   2 L_3 \\
   2 L_4 
\end{pmatrix}^{1/2}
\left(\f{k_2}{k_1}\right)^{L_4} \sum_{\ell = |L_2 - L_4|}^{L_2+L_4} (2\ell+1)
\begin{pmatrix}
   L_1 &L_3-L_4 & \ell \\
   0 & 0 & 0
   \end{pmatrix}\nonumber\\
  &\times \begin{pmatrix}
      L_2 &L_4 & \ell \\
      0 & 0 & 0
      \end{pmatrix} 
      \begin{Bmatrix}
         L_1 &L_2 & L_3 \\
         L_4 & L_3-L_4 & \ell
         \end{Bmatrix}  P_\ell (\cos \theta_{12})
\label{eq:old3JInteg}
\end{align}
where 
\beq
\cos \theta_{12} \defeq \f{k_1^2 +k_2^2-k_s^2}{2 k_1 k_2},  \quad 
    \Delta = 
        \begin{dcases}
            1, & \abs{\cos\theta_{12}} < 1, \\
            \frac{1}{2}, & \abs{\cos\theta_{12}} = 1, \\
            0, & \abs{\cos\theta_{12}} > 1. 
        \end{dcases}
\eeq

We adopted (\ref{eqn:new3JInteg}) instead of (\ref{eq:old3JInteg}) for two main reasons. First, as \cref{eqn:new3JInteg} shows, only one layer of summation from $m = -\min\{L_1, L_2\}$ to $m = \min\{L_1, L_2\}$ is needed as opposed to two layers of summation needed in the old strategy. This significantly speeds up the computation. However, more importantly, this new strategy avoids numerical instabilities which we will discuss below. 

\begin{figure}[!t]
    \centering
    \includegraphics[width=0.96\textwidth]{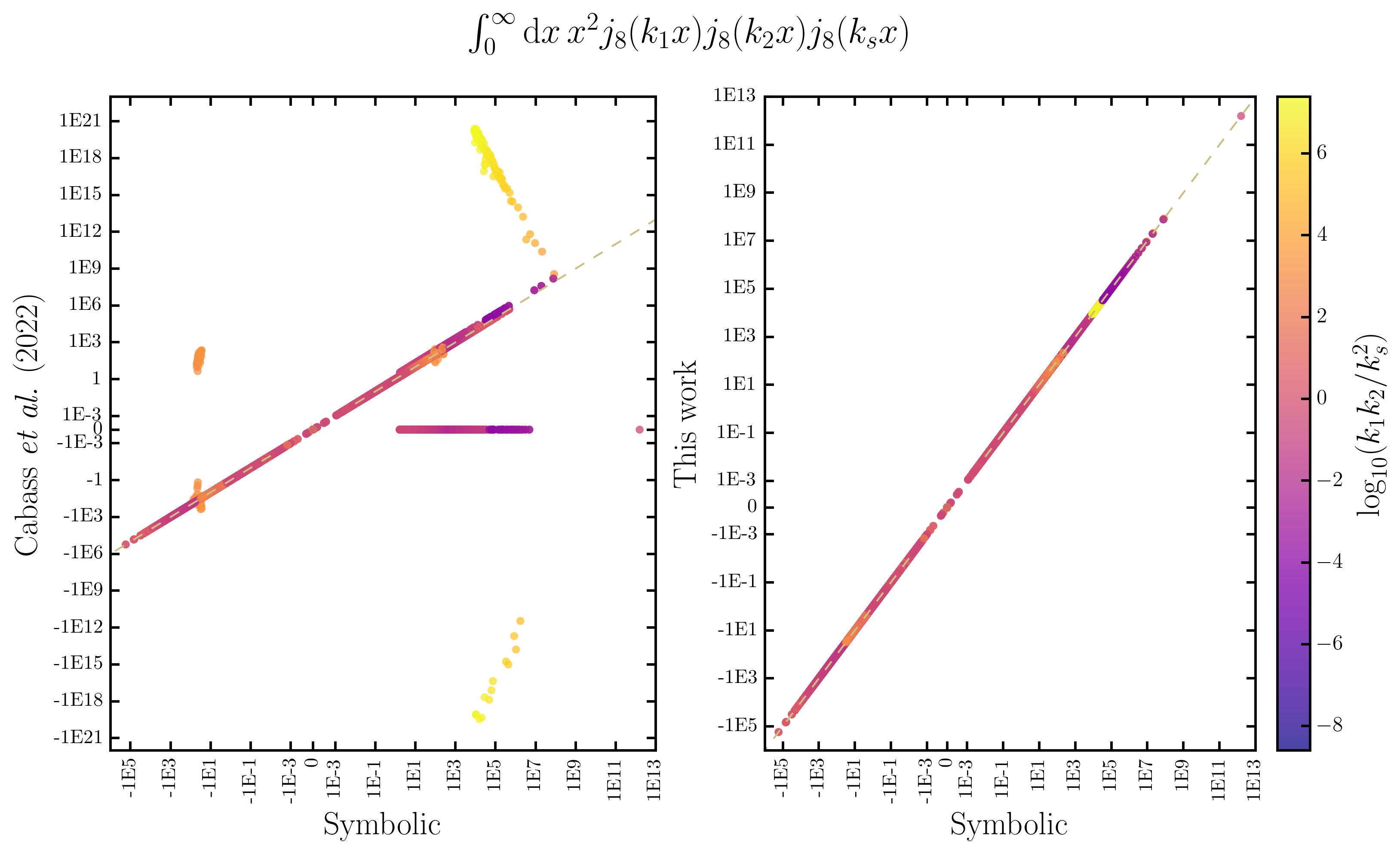}
    \caption{(Left) Numerical values for $f_{8,8,8}(k_1, k_2, k_s)$ for a given set of input momentum $(k_1,k_2, k_s)$ ($k_{1,2} =  \{5\times 10^{-5}, 5.05\times 10^{-3}\cdots,0.49505\}\,h/\text{Mpc}$ and $k_s =\{ 10^{-4}, 1.01\times 10^{-2}, \cdots, 0.9901\}\,h/\text{Mpc}$) from the symbolic computation \textit{vs.} those from Ref.~\cite{Cabass:2022oap} for the same momentum input. (Right) Numerical values from the symbolic computation vs. those from our \texttt{numpy} implementation for the same momentum input. In both panels, the dash lines indicate where the numerical values from both implementations are equal. Points off the dash lines indicate a discrepancy between the implementations. The colors indicate the value of $k_1 k_2 /k_s^2$ for a given momentum set.}
    \label{fig:threej_comp}
\end{figure}

As noted in their original paper \cite{Mehrem:1990eg}, Mehrem \textit{et al.}'s strategy can lead to unstable results when $k_s \ll k_1$ or $k_2$. This is a numerical issue rather than an analytical one. When using Mehrem \textit{et al.}'s formula, each term in the summation can be large since the summand contains factors similar to $\sim \qty(k_1 / k_s)^L$ or $\sim \qty(k_2 / k_s)^L$; however, the final result of such sum is usually orders-of-magnitude smaller due to cancellations among terms. This would not be an issue if the evaluation is symbolic, but when evaluating the summation numerically, one faces the issue of floating-point errors and finite numerical precision. For example, the 64-bit floating-point format, which is usually used for \texttt{numpy}, has a precision of $1.11\times 10^{-16}$. It is inadequate when the computation involves a large range of values of $k_{1,2,s}$  or a high values of $L_{1,2,3}$. Thus, in general, one expects larger $L_1$, $L_2$, and $L_3$ yields larger discrepancy between symbolic and numerical result. As a benchmark, we compared the symbolic result of $f_{8,8,8}(k_1, k_2, k_s)$ with its numerical result for $k_{1,2} =  \{5\times 10^{-5}, 5.05\times 10^{-3}\cdots,0.49505\}\,h/\text{Mpc}$ and $k_s =\{ 10^{-4}, 1.01\times 10^{-2}, \cdots, 0.9901\}\,h/\text{Mpc}$. The total number of comparisons is $10^6$. Note that this benchmark can be relevant for the computation in the main text when we set $L_\text{max}=8$. As shown in the left panel of \cref{fig:threej_comp}, the numerical result (ordinate) can differ by orders of magnitude from the symbolic result (abscissa) according to the strategy used in Ref.~\cite{Cabass:2022oap}, and such discrepancy usually happens when $k_1 k_2 / k_s^2 \gtrsim 10^2$. On the other hand, the new strategy shown in \cref{eqn:new3JInteg} uses angular variables that is generally robust again large ratios of $\sim k_1/k_s$ or $\sim k_2/k_s$, hence, bypassing the numerical instability. Speed-wise, \cref{eqn:new3JInteg} is much faster than its symbolic counterpart.

\begin{figure}[!t]
    \centering
    \includegraphics[width=0.96\textwidth]{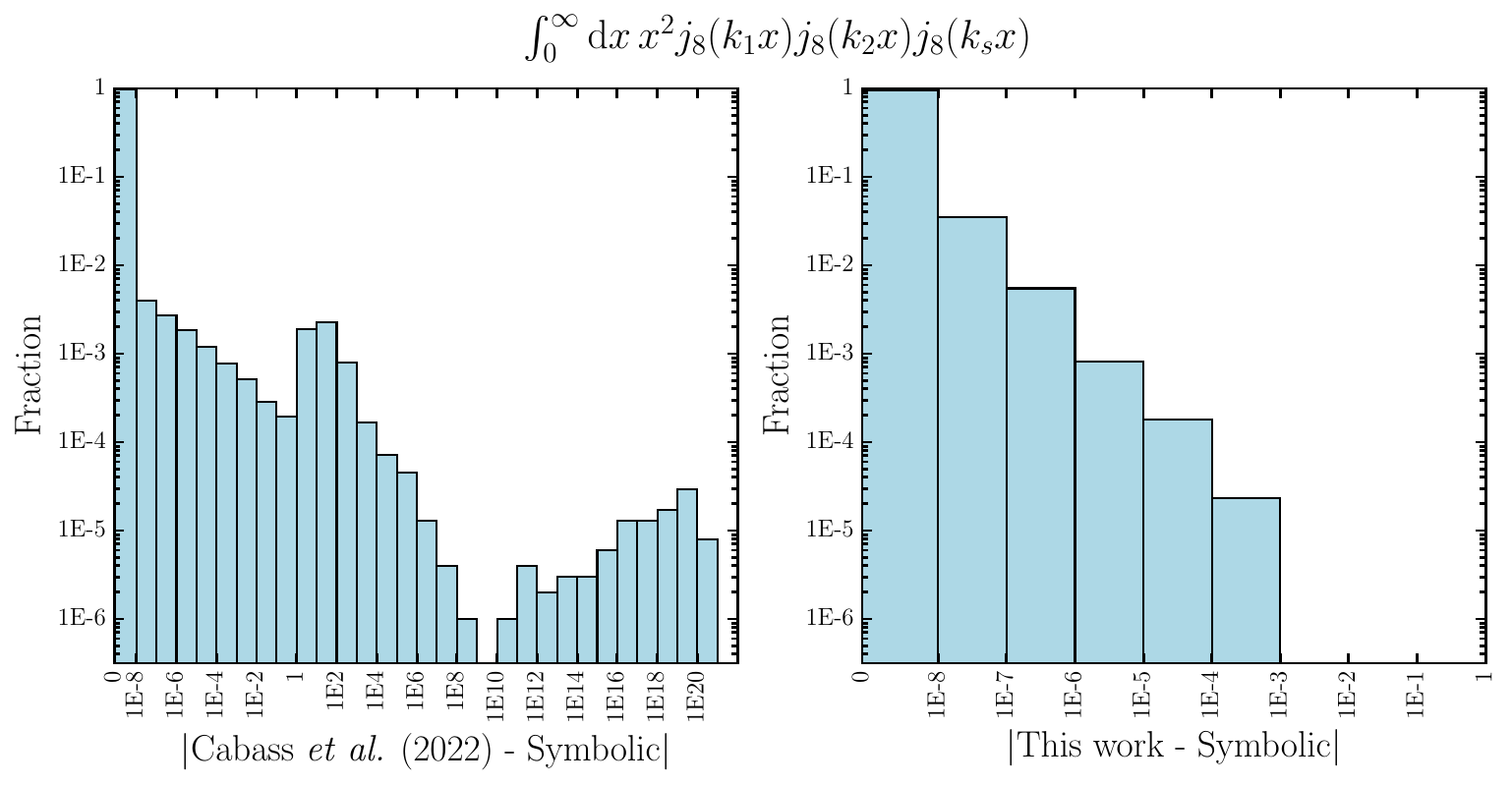}
    \caption{(Left) The normalized histogram for the absolute value difference of $I(L_1= 8, L_2 =8, L_p =8, k_1, k_2, k_s)$ between those from the symbolic computation and those from Ref.~\cite{Cabass:2022oap} for the same $(k_1,k_2, k_s)$ input. The total number of comparisons (normalization) is $10^6$. (Right) The normalized histogram for the absolute value difference of $I(L_1= 8, L_2 =8, L_p =8, k_1, k_2, k_s)$ between those from the symbolic computation and those from our \texttt{numpy} implementation for the same $(k_1,k_2, k_s)$ input.}
    \label{fig:threej_detail}
\end{figure}

We also note that the numerical result sometimes evaluates to zero while the symbolic result is nonzero, as shown in the horizontal line in the left panel of \cref{fig:threej_comp}. This comes from the numerical evaluation of the $\Delta$ factor, which appears in both (\ref{eq:old3JInteg}) and (\ref{eqn:new3JInteg}). This $\Delta$ factor is effectively a step function that selects combinations of $k_1$, $k_2$, and $k_s$ such that they can form a triangle. However, the numerical comparison between $\abs{\cos\theta_{13}}$ and $1$ for (\ref{eqn:new3JInteg}) (or $\abs{\cos\theta_{12}}$ and $1$ for (\ref{eq:old3JInteg})) can be inaccurate because of floating-point errors. However, the floating-point error is expected near the machine epsilon $\sim 10^{-15}$ while $\min\{\cos\theta_{13}\}\sim 10^{-6}$ for our sampling over $(k_1, k_2, k_s)$ space. It is, thus, generally sufficient to chop off small values at the level of $10^{-10}$ and compare $\abs{1 - \cos\theta_{13}}$ with zero. With both the instability due to large $k_1k_2/k_s^2$ ratio and the issue of $\Delta$ patched up, the new strategy becomes more stable and agrees well with the symbolic computation as shown in the right panel of \cref{fig:threej_comp}. In \cref{fig:threej_detail}, we plotted the absolute difference between the numerical and symbolic implementation of the two strategies for the benchmark $f_{8,8,8}(k_1,k_2,k_s)$. While Ref.~\cite{Cabass:2022oap}'s strategy leaves a $\sim 10^{-3}$ fraction of the phase space that yields numerically unstable integrals, the strategy introduced above patched up these instabilities so that the numerical and symbolic results agree at least at the level of $\sim 10^{-3}$. 

\subsection{Truncation in Angular Sums \label{app:Lmax}}
While evaluating the Fourier transform shown in \cref{eqn:strategy}, we used a plane-wave expansion to expand some momentum-conserving delta functions in Fourier space. Such expansion usually requires a summation over an angular index $L$ from $0$ to infinity. In actual numerical computation, this summation must be truncated at contain $L_\text{max}$. The choice of this truncation order should be reflected by the desired accuracy in the angular bins. In particular, if one desires to obtain the correlation coefficients $\zeta_{4,4,3}(\vb{r})$ instead of $\zeta_{1,1,1}(\vb{r})$, we must truncate at a larger $L_\text{max}$. This is because the plane-wave expansion essentially encodes the angular dependence of the propagator momentum. More precisely, it encodes how the propagator momentum couples to some particular configuration of external momenta, showing up as Wigner symbols in \cref{eqn:strategy}. For higher angular bins, we must keep couplings from higher angular momentum modes propagating through the propagator. 

\begin{sidewaysfigure}
    \centering
    \includegraphics[width=1\linewidth]{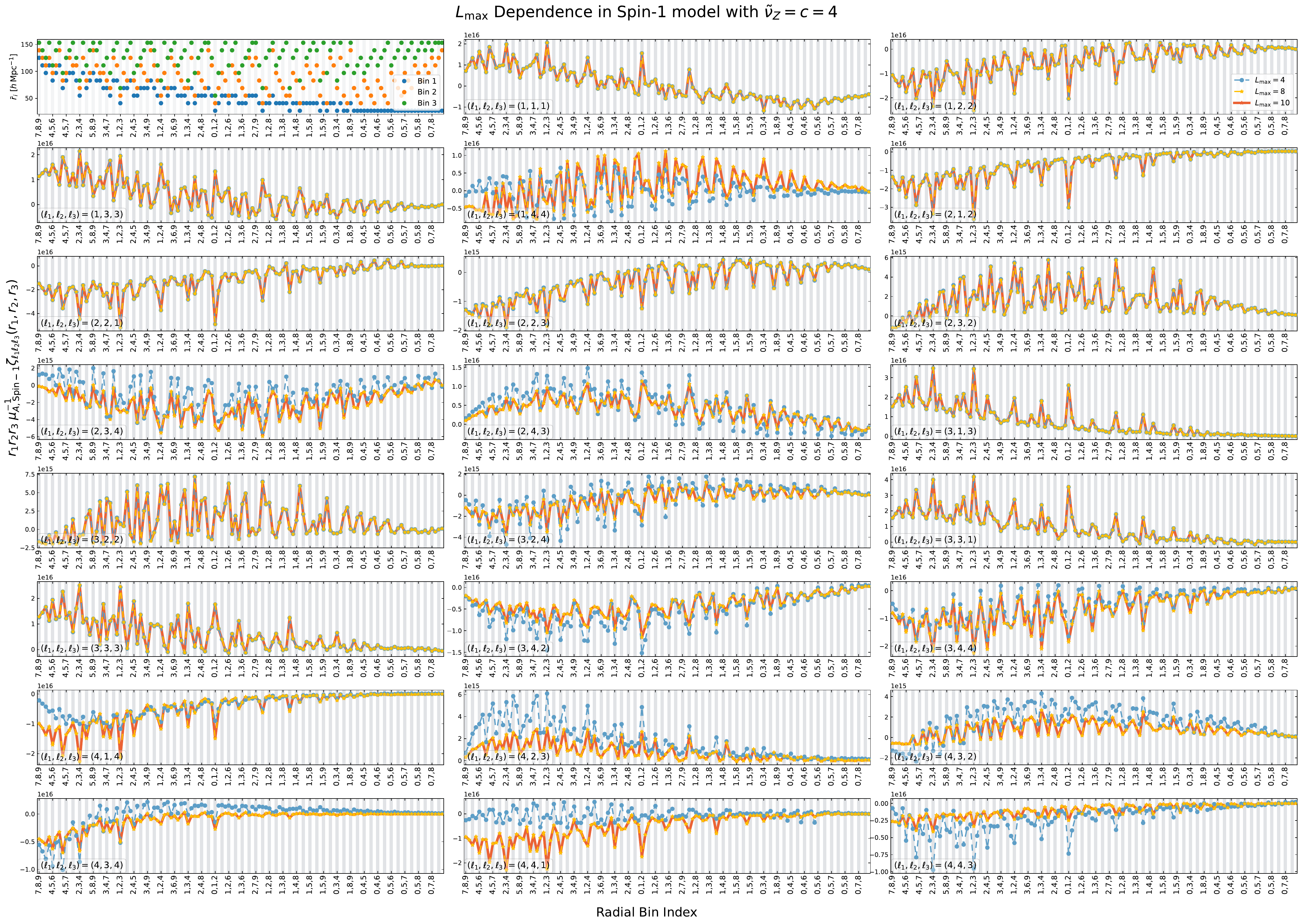}
    \caption{Comparison plot to illustrate $L_\text{max}$ dependence in correlation coefficients for the full spin-1 model. The blue, green, and red dots show results with $L_{\max} = 4, 8, 10$ respectively. While for low $L_\text{max}$, correlation coefficients $\zeta_{\ell_1, \ell_2, \ell_3}$ with lower angular index $\ell_i$ is not too sensitive to the choice of $L_\text{max}$. However, for angular bins with higher $\ell_i$, smaller truncation parameter $L_\text{max}$ affects the result significantly as some angular correlations encoded in summation over Wigner symbols of $L$ with other angular momentum labels, such as $\ell_i$, are lost.}
    \label{fig:LMAXDep}
\end{sidewaysfigure}

As a concrete demonstration, we computed the 4PCF for the full spin-$1$ model with three different truncation parameters $L_\text{max} = 4, 8, \text{ or } 10$. Its reweighted correlation coefficients $(r_1 r_2 r_3) \zeta_{\ell_1, \ell_2, \ell_3}$ is shown in \cref{fig:LMAXDep}. For small angular bins such as $\zeta_{1,1,1}$, we note that even the lowest truncation parameter $L_\text{max} = 4$ can yield a satisfactory match with that of the highest $L_\text{max}$. However, as the angular bin index increases, $L_\text{max} = 4$ begins to drop too many Wigner symbols in the angular coupling, and the resulting correlation coefficients for $L_\text{max} = 4$ deviate significantly from those of $L_\text{max} = 8$ or $10$. Fortunately, the results for $L_\text{max} = 8$ and $L_\text{max} = 10$ agree decently well that we may confidently say that correlation coefficients indeed converge as $L_\text{max}$ increase to a moderate value for numerical computation. In computing our templates for the full model, we chose $L_\text{max} = 8$ for the full spin-2 model. As for the full spin-1 model, we used $L_\text{max} = 10$ in the showcase and in the fitting analysis to BOSS data. For the survey of all toy models, we decided to use $L_\text{max} = 4$ for computation efficiency.

\section{Spin-2 Exchange in Large-mass Limit \label{app:spin2LargeMass}}
Similar to the spin-1 exchange in large-mass limit presented in~\cref{sec:largeMassModels}, one can compute the trispectrum for the exchange of a massive spin-2 particle with a mass $m$ and a chemical potential $\kappa$ whose Lagrangian reads \cite{Tong:2022cdz}
\begin{equation}
	\Lag = \sqrt{-g} \qty[ \MPl^2 R + \frac{1}{4} m^2 \qty(h_{\mu\nu} h^{\mu\nu} - h^2) ] + \frac{\theta}{2\Lambda_c} \epsilon^{\mu\nu\rho\sigma} \nabla_\mu h_{\nu\lambda} \nabla_{\rho} h_{\sigma}^{\ \lambda},
\end{equation}
in which $g_{\mu\nu} = \bar{g}_{\mu\nu} + \MPl^{-1} h_{\mu\nu}$, and $\nabla$ denotes the covariant derivative with respect to the dS metric $\bar{g}_{\mu\nu}$. Here, the dimensionless chemical potential is $\kappa \defeq \dot{\theta}/(\Lambda_c H)$ introduced by the rolling of the $\theta$ field. In the large-mass limit, the propagator for the spin-2 mode with helicity $h$ can be found as 
\begin{equation}
	\begin{multlined}
	    H_{++(--)}(\tau_1, \tau_2; p) 
	   \approx \frac{\delta(\tau_1 - \tau_2)}{\qty(m^2 - 2H^2)} - \delta(\tau_1 - \tau_2) \frac{\kappa H h p}{a \qty(m^2 - 2H^2)^2}.
	\end{multlined}
\end{equation}
Comparing with \cref{eqn:simplifiedDProp}, one realizes that the effective propagator for a spin-$2$ particle can be found via $c \to \kappa$ and $\tilde{\nu}_Z^2 \to \qty(\mu_h^2 - 2)$. Following Ref.~\cite{Tong:2022cdz}, massive spin-$2$ particle may couple to the inflaton via 
\begin{equation}
	\Lag \supset \frac{1}{M} h_{ij} \partial_i \phi \partial_j \phi. 
\end{equation}
After contracting the external momenta to the polarization tensors from the massive spin-2 propagator, we obtain the resulting trispectrum
\begin{equation}
	\begin{aligned}
		& \qty[\qty(\frac{H}{M})^2 \qty(\frac{H^2}{m_h^2 - 2H^2})^2 \qty(\frac{2}{15} \sqrt{\frac{2}{5}} \qty(4\pi)^{3/2}) H^3 \kappa \frac{H^4}{\dot{\phi}_0^4}]^{-1} \ev{\zeta^4}'_{\text{spin-}2} \\
        =& \qty[\int_{0}^{\infty} \dd t_E \; t_E
		\qty(\frac{1 + k_1 t_E}{k_1} e^{-k_1 t_E})
		\qty(\frac{1 + k_2 t_E}{k_2^3} e^{-k_2 t_E})
		\qty(\frac{1 + k_3 t_E}{k_3} e^{-k_3 t_E})
		\qty(\frac{1 + k_4 t_E}{k_4^3} e^{-k_4 t_E}) k_s ] \\ 
		& \times \qty(2 \mathcal{P}_{221}(\hat{\vb{k}}_1, \hat{\vb{k}}_3, \hat{\vb{k}}_s) + \mathcal{P}_{223}(\hat{\vb{k}}_1, \hat{\vb{k}}_3, \hat{\vb{k}}_s)),
	\end{aligned}
\end{equation}
where the isotropic basis functions are given by
\begin{align}
\mathcal{P}_{221}(\hat{\vb{k}}_1, \hat{\vb{k}}_3, \hat{\vb{k}}_s) ={}& -\frac{3\sqrt{5}\ii}{2^{1/2} (4\pi)^{3/2}} \hat{\vb{k}}_1 \cdot (\hat{\vb{k}}_3 \times \hat{\vb{k}}_s) (\hat{\vb{k}}_1\cdot \hat{\vb{k}}_3),\\
\mathcal{P}_{223}(\hat{\vb{k}}_1, \hat{\vb{k}}_3, \hat{\vb{k}}_s)={}& -\frac{15\sqrt{5}\ii}{2^{3/2}(4\pi)^{3/2}} \hat{\vb{k}}_1 \cdot (\hat{\vb{k}}_3 \times \hat{\vb{k}}_s) \left[(\hat{\vb{k}}_1 \cdot \hat{\vb{k}}_s)(\hat{\vb{k}}_3 \cdot \hat{\vb{k}}_s) -\frac{1}{5}\hat{\vb{k}}_1 \cdot \hat{\vb{k}}_3\right].
\end{align}

\section{Full Results of 4PCF Templates \label{app:fullresults}}
\begin{sidewaysfigure}[!t]
    \centering
    \includegraphics[width=1\linewidth]{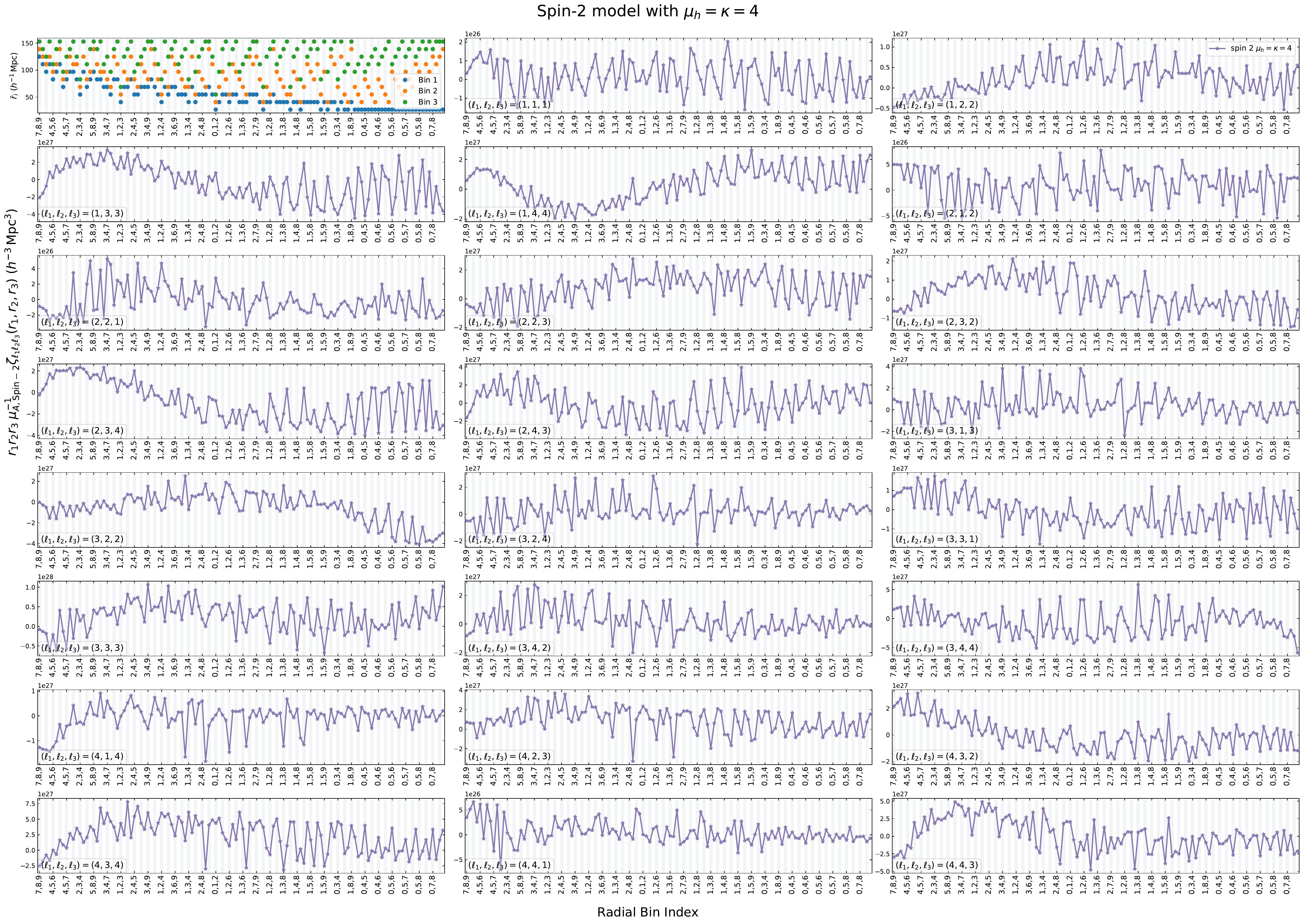}
    \caption{Correlation coefficients $r_1 r_2 r_3 \zeta$ for the full spin-2 model as a function of the angular bin index. }
    \label{fig:fullSpin2}
\end{sidewaysfigure}

\begin{sidewaysfigure}[!t]
    \centering
    \includegraphics[width=1\linewidth]{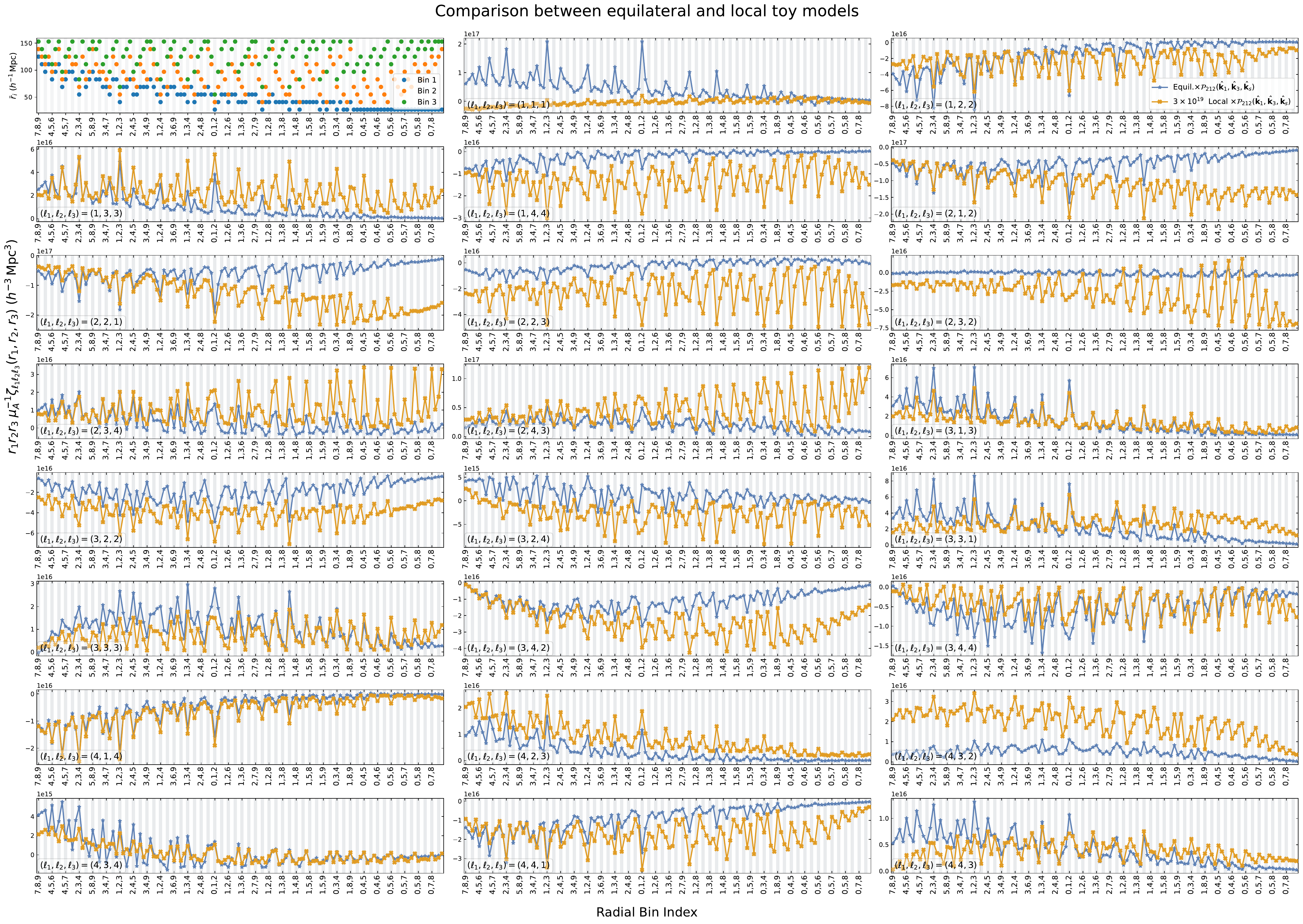}
    \caption{Comparison of the coefficient coefficients $r_1 r_2 r_3 \zeta$ for two toy models: equilateral shape times $\iso{2,1,2}(\hat{\vb{k}}_1, \hat{\vb{k}}_3, \hat{\vb{k}}_s)$ vs. local shape times $\iso{2,1,2}(\hat{\vb{k}}_1, \hat{\vb{k}}_3, \hat{\vb{k}}_s)$. Here, we rescaled the local template by a factor of $3\times 10^{19}$ so that the correlation coefficients of the two models are comparable in size.}
    \label{fig:toyCompare}
\end{sidewaysfigure}

In addition to the full 4PCF coefficients of the full spin-1 model shown in~\cref{app:Lmax}, here we provide the full coefficients for two example position-space templates: the full spin-2 models (\cref{sec:fullSpin2}) and a comparison between equilateral and local shapes multiplied by the angular-dependent factor $\iso{133}(\hat{\vb{k}}_1, \hat{\vb{k}}_3, \hat{\vb{k}}_s)$ (\cref{sec:canonicalModels}). They are shown in \cref{fig:fullSpin2,fig:toyCompare}.

\clearpage
\bibliography{P_odd_4PCF.bib}

\end{document}